\documentclass[12pt]{article}
\usepackage[utf8]{inputenc}
\usepackage{amsmath}
\usepackage{geometry}
\usepackage{graphicx}
\usepackage{comment}
\usepackage{CJKutf8}
\usepackage{amsfonts}
\usepackage{amssymb}
\usepackage{xcolor}
\usepackage{booktabs}
\usepackage{colortbl}
\usepackage{microtype}
\usepackage{setspace}
\usepackage{mathtools}
\usepackage{caption}
\usepackage{authblk}
\usepackage{subcaption}
\usepackage[hidelinks]{hyperref}
\usepackage[bottom]{footmisc}
\usepackage{float}
\usepackage[backend=biber, style=authoryear, sorting=nyt, uniquename = false]{biblatex}

\title{Volatility Spillovers in China’s Real Estate Crisis: \\
A Network Approach}
\author{Julia Manso\thanks{Correspondence: Julia Manso, Department of Statistics, 24-29 St Giles', Oxford OX1 3LB, United Kingdom. \newline Email: jumanso@stats.ox.ac.uk. I thank Steve Bond, Frank Windmeijer, and David Steinsaltz for their advice and feedback on this paper. }}
\affil{\small{\textit{Department of Statistics and Nuffield College, University of Oxford, Oxford, U.K.}}}
\date{}
\date{}

\begin{document}

\maketitle
\singlespacing
\vspace{-10mm}
\begin{abstract}
    Over the past six years, sentiment towards the Chinese real estate sector has deteriorated sharply following the introduction of financing constraints in 2020 with the ``three red lines." Forcing developers to restructure their debt, the policy triggered a cascade of financing troubles, defaults, and reduced housing demand, ultimately culminating in a prolonged real estate crisis. This paper utilizes a network approach in line with \textcite{demirer_estimating_2018} and \textcite{diebold_network_2014} to measure daily time-varying connectedness in the stock return volatilities of major Chinese real estate developers throughout the crisis. Particularly focusing on spillover between real estate companies as reflected by market perception, this paper examines how connectedness evolves over time across firms with different regional exposures and state-ownership statuses, filling a gap in the literature to elucidate where property demand and real estate firm trustworthiness have deteriorated most. An event-study analysis of four key moments of the crisis outlines distinct phases of market sentiment: with the introduction of the three red lines, connectedness primarily reflects shared exposure to the sector and a uniform shock to the market. Then, the early unrest surrounding Evergrande exposes strong regional differentiation, with firms concentrated in less developed regions receiving significant spillover. By one year into the crisis, previously stable regions receive higher levels of spillover, and there is evidence of a substitution effect towards private developers. Two years into the crisis, the market has much less homogeneity in effects across regions and state-ownership status: major shocks induce minimal network changes, reflecting how investors have already priced in their beliefs. This paper also offers one of the most extensive timelines of the Chinese real estate crisis to date, and a new R package, \texttt{GephiForR}, was created for the network visualization in this paper. 
\end{abstract}
\noindent \textbf{Keywords\textemdash} Chinese real estate, connectedness, volatility, spillover, networks, variance decomposition

\vspace{-7mm}
\doublespacing

\onehalfspacing

\section{Introduction}

In March of 2019, Evergrande\textemdash China's largest real estate developer and one of the largest companies in the world\textemdash was riding high, like much of the Chinese real estate sector. It had just announced that its 2018 core profit rose 93.3 percent compared to the previous year, a ``record high" as the company delivered more properties while cutting costs. Some investors were concerned about the company's leverage ratio and other performance indicators, but the outlook was largely positive (\cite{jim_china_2019}). Fast-forwarding to 29 January 2024, after losing 99\% of its share value and struggling to make debt payments, China Evergrande was ordered to wind up by a Hong Kong court\textemdash a far cry from the rosy expectations of just five years earlier (\cite{hoskins_evergrande_2021}; \cite{jim_china_2024}).

While Evergrande has been the most prominent Chinese developer to struggle on the global stage, problems in the Chinese real estate market run much deeper: government policies limiting annual debt growth have handicapped the borrowing-dependent industry, creating a massive credit crunch that threatens to expose even the most financially reticent parties. As trouble built and investors realized the extent of their exposure, developers were hit by share selloffs, declines in credit ratings, and fewer willing lenders, further impeding their ability to raise cash. Yet, these massive and decisive share selloffs also offer an opportunity to examine spillover effects in the industry and map the pattern of contagion.

The existing literature on the Chinese real estate crisis focuses on spillover from the real estate sector to the broader economy (\cite{hu_risk_2024}), the financial sector (\cite{huangfu_research_2024}; \cite{ouyang_interconnected_2023}; \cite{xu_risk_2021}; \cite{nong_financial_2024}), the industrial/manufacturing sector (\cite{xie_characteristics_2025}), and the energy market (\cite{xie_dynamic_2025}). This paper thus fills an important gap in the literature, analyzing spillover patterns among developers themselves across different events/phases of the real estate crisis. Following the work of \textcite{demirer_estimating_2018} and \textcite{diebold_network_2014}, this paper uses a network approach to measure the daily time-varying connectedness of major real estate companies’ stock return volatilities. Examining how spillover patterns differ by region and state-ownership status, it maps how firms with specific regional focuses are impacted, working to understand where demand for property and real estate firm trustworthiness have deteriorated most\textemdash as well as whether state ownership impacts these perceptions. 

In terms of methodology, this paper follows \textcite{demirer_estimating_2018}, first fitting a rolling window vector autoregression (VAR) to stock return volatility data for the Chinese real estate market. After using the elastic net for shrinkage and selection and estimating the VAR with a 100-day rolling window to capture the dynamism of developer relationships, a variance decomposition is then performed to determine how much of a firm's forecast error variance is attributable to shocks from other developers, summing these to obtain the directional Diebold-Yilmaz measures of connectedness. These directional connectedness measures are then plotted as a network to understand how connections between firms evolve across different events. 

An event study is performed on four key events that signal different phases of the crisis: the announcement of the ``three red lines," the first release of insider news about Evergrande's cash crunch, the initial suspension of share trading by Chinese developer Kaisa, and a key profit warning issued by Country Garden. 

Indeed, I ultimately discern how investor sentiment has evolved as the real estate crisis progressed. First illustrating evidence of an external shock to the network with the announcement of the three red lines, connectedness primarily reflects shared exposure to real estate and an experience of a uniform, external shock.

Then, analyzing the network's behavior surrounding the release of news about Evergrande's cash crunch reveals investors' immediate reactions to market exposure: when the news broke, investors' instincts were linked with the regional development pattern of China, as they expected stronger contractions in less developed regions and shifted their investments away. There is also evidence that state-owned firms are seen as slightly more insulated from market shocks than their privately-owned counterparts during this early period. 

One year into the crisis, network behavior during the suspension of share trading by Kaisa offers updated insight into investors' beliefs: now, regions that were regarded as relatively stable earlier in the crisis are increasingly at risk as the downturn continues. The network also reflects a change in sentiment surrounding the relative stability of state-owned enterprises, exhibiting a substitution effect towards private developers. 

Then, fast-forwarding to nearly two years later, by the time of Country Garden's profit warning in August 2023, the network captures just how much investor sentiment has changed. Unlike in earlier periods, there is much less homogeneity across regions and state-ownership status, likely reflecting how developers have adapted to investing amidst the crisis. Now, when a ``shocking" event occurs, the network change is minimal because investors have already priced in their beliefs. Nevertheless, there is evidence of a substitution effect towards state-owned enterprises\textemdash the reverse of that seen under the Kaisa case\textemdash reflecting the market sentiment that state-owned developers are now ``safer" bets for investment.  

In the process of this analysis, I also develop one of the most comprehensive timelines of the Chinese real estate crisis to date, as well as an R package for graphing networks using Gephi-based layout orientations (\cite{manso_gephiforr_2024-1}).\footnote{Gephi is a popular network visualization software. The R package (\texttt{GephiForR}) can be found at \href{https://CRAN.R-project.org/package=GephiForR}{\texttt{https://CRAN.R-project.org/package=GephiForR}}.} 

This paper first offers a brief primer on the Chinese real estate market, discussing state ownership in China (Section \ref{background}). Next, after introducing the data (Section \ref{sec: data}), as well as the methodology and visualization process (Section \ref{sec: methodology}), Section \ref{sec: results} discusses the results. Section \ref{sec: conclusion} concludes. The comprehensive timeline of the real estate crisis can be found in Appendix \ref{Appendix: Dated Timeline}.

\section{Background} \label{background}

The Chinese real estate market is unique not only given the depth of State ownership, but also because it is a largely nascent industry that boomed within the past 30 years. Most importantly, though, housing in China is first and foremost a commodity, and the property market is distinctly speculative, with property developers wielding significant market power and housing prices increasing exponentially since 2000 (\cite{zhao_playing_2017}). 

While incredibly complex with many local disparities, the real estate market is the interaction of four groups who collectively influence price movement: local governments, real estate developers, banks, and speculators (\cite{liu_urban_2018}). First, local governments gain ``land revenue" by selling commercial and residential land use rights (\cite{wu_evaluating_2016}). High land costs are subsequently transferred to high real estate prices, and real estate developers often engage in property hoarding and price discrimination to further increase real estate prices and create artificial scarcity (\cite{zhou_real_2021}). Local governments then borrow heavily from banks by mortgaging land use rights and using the funds to support infrastructure development, which further drives up its potential for land revenue (\cite{liu_formation_2022}). Developers also borrow heavily, not only obtaining funds from banks and other financial institutions, but from shadow banks, private financing, and even illegal fundraising channels (\cite{liu_formation_2022}). Another key source of developer income is presales, wherein buyers pay for properties before their construction has concluded (sometimes even before construction has started), often through mortgages. Lending to all parties involved, banks offer excessive credit support to the real estate industry; this behavior in turn drives prices higher as developers borrow more to build, and buyers borrow more to pay. Indeed, buyers often act as speculators, purchasing property and selling it after prices rise. They too help support a cycle wherein all parties involved have an incentive to increase land value as much as possible, creating an overheating market and a property bubble.

Prior to 2020, the State's attempts to curb this cycle had been half-hearted: property has been one of the driving forces of China's economy, and to truly curb rising house prices would harm this growth (\cite{tan_effect_2022}). Further, the onus had traditionally been on local governments to reform their policies and lower land sale prices\textemdash a difficult thing for them to do when 59.8 percent of revenue from property sales goes to the government, and combined revenues from land transfers and special taxes on real estate account for 37.6 percent of local government revenue as of 2020 (\cite{zhang_high_2024}; \cite{ren__2021}). When the central government shifted towards persistent tightening with the ``three red lines" in 2020, speculators, developers, and banks all thought the government would back down from its measures. Yet, as the real estate crisis began to spiral during 2021, the public did not realize how deep that commitment to tightening\textemdash and its implications\textemdash ran.

When the three red lines policy was imposed, it was done as a pilot program for 12 developers and essentially restricted their ability to borrow capital/grow their debt subject to meeting three limits (``red lines") on the liability-to-asset ratio, the liability-to-equity ratio, and the cash-to-short-term-debt ratio. The maximum allowable annual growth in debt was restricted to 15\% if no lines were crossed, dropping by 5\% based on the number of lines violated. If a developer failed to meet all three criteria, it could not grow its debt annually at all. Realistically, the structure of the market ensured that almost all developers were violating the first red line on the liability-to-asset ratio, but others like Evergrande were in violation of all three at the time of announcement.

The three red lines pressured firms to internally restructure in order to meet government guidelines, and stress percolated into the market as internal documents\textemdash often describing just how badly firms were violating the red lines\textemdash were leaked to the public. As the situation escalated and the three red lines were expanded to all developers in January 2021, any actions by developers that hinted at financial trouble were met with sharp tumbles in stock price (\cite{white_chinese_2021}). Developers thus took great pains to project images of financial stability and security, often liquidating assets behind the scenes to make bond payment deadlines. By early 2022, the situation was untenable, with leading firm Evergrande suspending trading of its shares, citing its inability to produce audited results (\cite{jim_china_2022}; \cite{stevenson_china_2022}).\footnote{On the Stock Exchange of Hong Kong, listed companies can suspend for up to 18 consecutive months, by which time they must supply results and be relisted or be delisted from the exchange altogether. Suspension is also a type of stop-gap mechanism that prevents the stock from falling in value while giving the company some time to rectify its struggling financial position, such that it can hopefully return to trading in a much stronger position (\cite{leung_rare_2024}). } Several other struggling firms\textemdash China Aoyuan, Kaisa, Fantasia, Modern Land, and Sunac\textemdash followed almost immediately. Since then, shock after shock has impacted the market\textemdash and trouble continues still to this day. A comprehensive timeline of the events of the real estate crisis, from the three red lines to Evergrande's liquidation in early 2024, is included in the Appendix (\ref{Appendix: Dated Timeline}), offering one of the most comprehensive timelines on the topic to date. 

Also note that one additional factor further complicates firm behavior in this analysis\textemdash the prevalence of State ownership. Following \textcite{chow_evergrande_2024}, a State-owned enterprise (SOE) is defined here as a company whose largest shareholder is the State. Conversely, privately-owned enterprises (POEs) are understood as companies where the largest shareholder is a private company or individual. In China, these SOEs are legally recognized as corporate entities (rather than government entities) and are managed by State-owned Assets Supervision and Administration Commissions (SASACs). The literature suggests that Chinese SOEs are numerous and powerful enterprises, but less effective and profitable than their POE peers, limited by their ability to engage with and respond to market forces (and therein increase profitability). See, for instance, \cite{mei_fortune_2022}.\footnote{However, viewing POEs as independent of state influence is essentially a false dichotomy. \textcite{milhaupt_beyond_2015} argue that in China, virtually all large and successful firms have ``close connections to state actors and agencies, access to state largesse, and a role in carrying out the policies of the ruling political party"; in essence, no firm is wholly autonomous (p. 668). Indeed, the POEs who most embrace the State's involvement often become the largest and most successful firms because State support can increase market access and yield business advantage through proximity to state power, among other benefits. Yet, clever real estate firms have also tapped into additional dimensions of State support: the State can provide stability amid high risk-taking and has a deep-seated interest in protecting its own enterprises\textemdash and deep coffers to accompany it. }

\section{Data} \label{sec: data}

In line with \textcite{diebold_network_2014}, I use high-frequency stock market returns and return volatilities to estimate connectedness. This approach is quite intuitive, as those with the most knowledge about the connections under investigation are those financially involved. The data encompass real estate companies domiciled and operating in China from the Shanghai and Shenzhen stock exchanges; data collection and handling procedures are described in Appendices \ref{Appendix: Data Cleaning Procedures} and \ref{Appendix: Region Assignment and State-Ownership Status}. With this raw data, volatility\textemdash the ``dispersion from an expected value, price or model" (\cite{daly_financial_2008}, p. 2379)\textemdash is calculated via the estimator developed by \textcite{garman_estimation_1980}, which applies Brownian motion principles to stocks to estimate daily stock return volatility as a function of the natural logarithms of daily high, low, opening, and closing prices for stock $i$ on day $t$. This specification is used broadly in the literature (as in \cite{ji_dynamic_2019}, \cite{diebold_financial_2015}, and \cite{longstaff_how_2011}) and is ``nearly as efficient as realized volatility based on high-frequency [five-minute] intraday sampling" while still being robust under several conditions including microstructure noise (\cite{demirer_estimating_2018}, \cite{alizadeh_range-based_2002}). 

\section{Methodology} \label{sec: methodology}
\subsection{Estimating high-dimensional VARs}
I follow \textcite{demirer_estimating_2018}'s methodology, first basing my variance decomposition on an N-variable vector autoregression, VAR(d)  
\begin{equation}
    log(\sigma^2_t) = \sum^d_{\ell=1}\phi_\ell log(\sigma^2_{t-\ell})+\varepsilon_t
    \label{var eq}
\end{equation}
where $\varepsilon_t \sim {(0,\Sigma)}$. $\ell$ is the lag order of the autoregressive terms, $d$ is the number of lags, $\phi_\ell$ is the $N \times N$ coefficient matrix for lag $\ell$, and $\varepsilon_t$ is the disturbance vector whose covariance matrix is $\Sigma$. Finally, $\sigma^2_t$ and $\sigma^2_{t-\ell}$ are $N \times 1$ vectors of stock return volatilities calculated above, and the logarithm of all volatility terms ($\sigma^2$) is taken to normalize their distribution and remove the skew. This VAR is intentionally nonstructural, as, like \textcite{diebold_financial_2015}, I do not seek to define exactly how connectedness arises.\footnote{Note that I use the VAR with the rolling window specification rather than another version of the VAR from the literature (e.g., the time-varying parameter VAR, TVP-VAR), because it best fits this use case. The leading alternative candidate, TVP-VAR, uses a Kalman filter, and its results are thus sensitive to priors and forgetting factors (\cite{antonakakis_refined_2020}). In particular, in \textcite{antonakakis_refined_2020}, the TVP-VAR is initialized with the VAR estimate of the first 60 periods, which can bias the results depending on whether this is a calm or turbulent period: a calm pre-period could bias the model to interpret shocks as outliers, while a turbulent pre-period may cause the model to overreact to noise. In addition, hyperparameter selection can also bias the results. In light of these issues, the 100-day rolling window is preferred in this context given the focus is meaningful before and after windows and a model that performs well in the presence of a large number of shocks. }

To estimate the VAR in such high dimensions, I utilize the elastic net, in line with \textcite{demirer_estimating_2018}.\footnote{Note also that while other tools like the adaptive elastic net may be an improvement over the elastic net, they cannot be calculated for the rolling window estimation due to data limitations: the window size of 100 is just enough for the elastic net calculations given the number of developers (97 Chinese firms). Given the number of firms in the sample, there are more regressors than observations when lags are included, and it is thus not possible to use the adaptive elastic net to give dynamic updates without a window of size equal to or greater than $N\times\ell$. This much larger window (in this case, equal to or greater than 291) inherently loses most of the sensitivity that the narrow 100-day one offers, so I utilize the elastic net with 100-day rolling window estimations for the bulk of the analysis. Appendix \ref{Appendix: Static Window Adaptive Elastic Net Results} offers an example of a network generated under the adaptive elastic net with a larger window size.} Performing simultaneous selection and shrinkage like the lasso, the elastic net includes both ridge and lasso penalties, with a regularization parameter $\lambda$ and an adjustable parameter $\alpha$ that balances the lasso and ridge penalties. Importantly, the elastic net can also select groups of correlated variables since the penalty term is strictly convex for all $\alpha \textit{ }\epsilon \textit{ }(0,1)$ and $\lambda>0$, a particularly valuable feature in this context given that real estate firms' volatilities are often highly correlated (\cite{zou_regularization_2005}; \cite{hastie_statistical_2015}). $\alpha$ is regarded as a ``higher-level" tuning parameter that is often set on subjective grounds (\cite{hastie_statistical_2015}). I conduct a sensitivity analysis to investigate how sensitive the results are to different $\alpha$ values in the elastic net estimations. The main specification, whose results are shown in Section \ref{sec: results}, uses $\alpha$ = 0.5, in line with \textcite{demirer_estimating_2018}.\footnote{It is also important to note that while I seek sparsity in the approximating model, I do not necessarily want to impose sparsity in the estimated real estate firm network. In line with \textcite{demirer_estimating_2018}, this shrinkage and selection is thus performed on the approximating VAR rather than the variance decomposition network directly. While performing shrinkage and selection on the VAR, the variance decomposition matrix that is used to calculate connectedness measures is a ``nonlinear transformation of the VAR coefficients and is therefore generally not sparse" (Demirer et al. 2018, p. 5). In the Chinese real estate case, the network remains fully connected.} 

\subsection{Moving from the VAR to the variance decomposition matrix}

Following \textcite{demirer_estimating_2018}, I estimate an elastic net-penalized regression with ten-fold cross-validation, extracting the coefficients from the model which minimize the cross-validation error and iterating through the columns to generate the $\phi_\ell$ matrices referenced in (\ref{var eq}). The impulse response function is then calculated, iterating through the lag periods and horizon $H$. The impulse response can be viewed as the effect of a hypothetical $N \times 1$ vector of shocks impacting the market at time $t$ compared with a baseline profile at time $t + H$, given the market's history.

Importantly, these impulse responses and variance decompositions allow estimates of connectedness at different time horizons $H$, with shorter time horizons like 1 or 2 days ($H = 1$ and $H=2$, respectively) picking up immediate market reactions while longer horizons like 30 days ($H=30$) reflect more fundamental dependencies or economic relationships. This analysis uses a horizon $H$ value of 10 days to balance longer- and shorter-term volatility comovements; a lag period of 3 is used because the VAR model often selects coefficients for several firms on the third lagged day.\footnote{When a longer lag period (e.g., 5 days) was tested, the model selected very limited numbers of coefficients from the additional days.}

Aggregating this information, the generalized forecast error variance decompositions ($\theta_{ij}(H)$), which provide firm $j$'s contribution to firm $i$'s $H$-step-ahead generalized forecast error variance, can be calculated: the numerator effectively sums up and then squares the shocks from variable $j$ on firm $i$ over all horizons up to $H-1$. This represents the cumulative effects of shocks in variable $j$ on the forecast error variance of variable $i$ up to horizon $H$. The denominator reflects the total forecast error variance, summing the contribution to the forecast error variance from all shocks affecting variable $i$; then, it is normalized by the standard deviation of the disturbance of the $j$-th equation. Subsequently, following \textcite{demirer_estimating_2018}, each $\theta_{ij}$ of the generalized variance decomposition matrix is normalized by the row sum to obtain
\begin{equation}
    d_{ij}^H = \frac{\theta_{ij}(H)}{\sum_{j = 1}^{H-1} \theta_{ij}(H)}. 
\end{equation}
This normalization means that $\sum{_{j = 1}^N} d_{ij}^H = 1$ and $\sum{_{i, j = 1}^N} d_{ij}^H = N$, allowing for the resulting $d_{ij}^H$ to be comparable across different time horizons $H$.\footnote{As will be described in Section \ref{Connectedness Measures}, these $d_{ij}^H$ values are measures of pairwise directional connectedness.} These $d_{ij}^H$ values form the matrix $D^H$, which is the core of \textcite{demirer_estimating_2018} and \textcite{diebold_network_2014}'s connectedness measures (Section \ref{Connectedness Measures}).

\subsection{Connectedness measures} \label{Connectedness Measures}

\textcite{demirer_estimating_2018}'s connectedness estimation method makes it possible to decompose how much of firm $i$'s future uncertainty at a specified horizon $H$ is due to shocks arising not from entity $i$ itself, but from each other entity $j$ in the sample. As described above, it relies upon $d_{ij}^H$, the fraction of $i$'s $H$-step forecast error variance due to shocks in firm $j$, where $D^H$ = $[d_{ij}^H]$. This full set of variance decompositions is the core of \textcite{diebold_network_2014}'s connectedness table, which is in effect an augmented variance decomposition matrix, as shown in Table \ref{tab:connectedness_gen}. 

\begin{table}[H]
\centering
\caption{Connectedness Table}
\renewcommand{\arraystretch}{1.5}
\begin{tabular}{*{5}{c}c}
\hline
 & $x_1$ & $x_2$ & $\cdots$ & $x_N$ & From others \\
\hline
$x_1$ & $d_{11}^H $ & $d_{12}^H$ & $\cdots$ & $d_{1N}^H$ & $\sum_{j=1}^{N}d_{1j}^H, j\neq 1$ \\
$x_2$ & $d_{21}^H$ & $d_{22}^H$ & $\cdots$ & $d_{2N}^H$ & $\sum_{j=1}^{N}d_{2j}^H, j\neq 2$\\
$\vdots$ & $\vdots$&$\vdots$ & $\ddots$& $\vdots$ & $\vdots$\\
$x_N$ & $d_{N1}^H$ & $d_{N2}^H$ & $\cdots$ & $d_{NN}^H$ & $\sum_{j=1}^{N}d_{Nj}^H, j\neq N$\\
\hline
To others & $\sum^{N}_{i=1} d^H_{i1}$ & $\sum^{N}_{i=1} d^H_{i2}$ & $\cdots$ & $\sum^{N}_{i=1} d^H_{iN}$  & $\frac{1}{N}\sum^{N}_{i,j=1} d^H_{ij}$ \\
& $i\neq 1$ & $i\neq 2$ &  & $i\neq N$ & $i\neq j$ \\
\hline
\end{tabular}
\label{tab:connectedness_gen}
\end{table}
The upper left block of Table \ref{tab:connectedness_gen} is the standard $N \times N$ variance decomposition matrix $D$ composed of the connectedness between each firm $i$ and $j$ for time horizon $H$, with $D^H$ = $[d_{ij}^H]$. This $D^H$ is augmented with a column of row sums in column $N+1$ and a row of column sums in row $N+1$, both for $i \neq j$. The value in cell $[N+1, N+1]$ is the grand average, again for $i \neq j$. The off-diagonal entries of the $N \times N$ variance decomposition matrix reflect the pairwise directional connectedness from firm $j$ to firm $i$. That is, 
\begin{equation}
    C^H_{i \leftarrow j } = d_{ij}^H.
    \label{pairwise connectedness formula}
\end{equation}
This means, for example, that $d_{21}^H$ is the pairwise directional connectedness from firm 1 to firm 2 while $d_{12}^H$ is the pairwise directional connectedness from firm 2 to firm 1. These values will not necessarily be equal as firms do not often impact each other symmetrically. For instance, if firm 1 is a massive price maker in the market but firm 2 is a small and relatively isolated firm, shocks to firm 1 may also have significant effects on firm 2, but conversely, shocks to firm 2 will not likely affect the powerful firm 1 in the same way. Thus, generally, $C^H_{i \leftarrow j } \neq C^H_{j \leftarrow i }$. There are consequently $N^2 - N$ separate pairwise directional connectedness measures.\footnote{$N$ is subtracted as we exclude the diagonal elements (which effectively represent self-spillover).} \textcite{diebold_network_2014} thus defines net pairwise directional connectedness as $C^H_{ij} = C^H_{j \leftarrow i } - C^H_{i \leftarrow j }$, and there are $\frac{N^2 - N}{2}$ net pairwise directional connectedness measures. 

When the diagonal entry is excluded ($d_{ij}^H$, where $i = j$), aggregating across each row yields the share of the $H$-step forecast error variance of firm $j$ that comes from shocks arising in all other firms. In line with \textcite{diebold_network_2014}, this  total directional connectedness \textit{from} others to $j$ (``from" connectedness) can be represented mathematically as 
\begin{equation}
    C^H_{j \leftarrow \bullet} = \sum^N _{i = 1, i \neq j} d^H_{ji}.
\end{equation}
The same reasoning is applied to off-diagonal column sums: for each column, excluding $d_{ij}^H$ when $i = j$ and summing the rest of the column yields the share of the $H$-step forecast error variance that firm $j$, when shocked, gives to all other firms. This is referred to as ``to" connectedness. 

This Diebold-Yilmaz approach to network connectedness is appealing because it bridges the VAR variance-decomposition and the network literature, essentially positing that ``a variance decomposition is a network" (\cite{diebold_past_2023}). The VAR can capture the dynamic interactions among multiple lagged variables without imposing strong a priori restrictions or requiring a structural model. At the same time, it assumes linear relationships among variables, and the subsequent impulse response function and generalized forecast error variance decomposition are sensitive to the shock covariance matrix derived from the VAR, which \textcite{demirer_estimating_2018} do not choose to regularize.

\subsection{Visualizing the network} \label{visualizing the network}

For the results shown in this paper, network layout is determined by Gephi's ForceAtlas2, created by \textcite{jacomy_forceatlas2_2014}, which calculates the ``net forces" acting on each node by summing the node's attraction to and repulsion from each other node it connects to.\footnote{Note that Gephi is a popular software used for visualizing networks.} For clarity, assume that all variables take on new meanings (as defined) unless explicitly specified.

The formulas for each are rooted in  \textcite{eades_heuristic_1984} and his application of physics principles to networks; the attraction formula is an altered version of Hooke's law, which reflects the compression behavior of a spring ($F = -k \times x)$, where $F$ is the spring force, $k$ is the spring constant, and $x$ is the spring compression. In Gephi, the attractive force $F_{a}$ is the product of the edge weight $w(e)^\delta$ between each pair of nodes, multiplied by the geometric distance between them, as shown below:
\begin{align}
    F_{a} = w(e)^\delta d(n_1, n_2),
    \label{attraction formula}
\end{align}
where $\delta$ is the binary variable ``Edge Weight Influence," set to either 0 or 1.

Repulsion is based on Coulomb's law, which calculates forces between electrically charged particles ($F = k\frac{n_1n_2}{d(n_1, n_2)^2}$). Here, $F$ is the resulting force; $n_1$ and $n_2$ are the point charges of particles 1 and 2, respectively; $d(n_1, n_2)$ is the distance between the two particles $n_1$ and $n_2$; and $k$ is the Coulomb's law constant. In Gephi, repulsion $F_r$ is calculated by a slightly modified version of Coulomb's law where node degree takes the place of point charges, the constant becomes scalable rather than fixed, and distance is no longer squared. The key development from previous node repulsion calculations was the addition of ``+1" to the degree, as Jacomy et al. wanted to ensure that nodes with degrees of zero still have repulsive force. The formula is thus 
\begin{align}
    F_{r} = \frac{S(deg(n_1)+1) (deg(n_2)+1)}{d(n_1, n_2)}
    \label{F_r eq}.
\end{align}
This calculation is repeated between every possible node pair in the data. Thus, $d(n_1, n_2)$ is the distance between the two nodes $n_1$ and $n_2$ in the pair, $deg(n_1)$ and $deg(n_2)$ are the degrees of each of the two nodes, and $S$ is a scalable constant that influences the repulsion level in the graph, with higher repulsion making a sparser graph.

 This combination of ``spring-like" attractive forces and ``particle-like" repulsive forces has persisted, with several authors offering slightly modified equations over the past 40 years. In Gephi, the attraction and repulsion forces can be decomposed into vector form and summed to calculate the net forces on each node ($F = F_{r} + F_{a}$); this value is then used to find the resulting displacement. The actual displacement of a node $\Delta(n)$ is then calculated by the formula: 
 \begin{align}
    \Delta(n) = s(n)*F(n)
    \label{displacement eq},
\end{align}
where $s(n)$ is the speed of node $n$. To calculate the speed, Jacomy et al. use two factors\textemdash irregular movement (``swinging") and useful movement (``effective traction"); in effect, they calculate how much the forces on a node compare between two sequential periods and (usually) use lower speeds to update a node's position if there is a large change in forces. Over several iterations calculating each node's displacement and updating its position, networks converge to a stable position. A full explanation of the formulas behind this, as well as a toy example, can be found in Appendix \ref{appendix: visualizing the network}. 

All network figures in this paper are created with the \texttt{GephiForR} package (\cite{manso_gephiforr_2024-1}), which implements the ForceAtlas2 layout algorithm in R and offers several other network visualization tools. 

\section{Results} \label{sec: results}

The figures below are color-coded by the region where the developer is primarily focused: pink is the north, light green is the south, teal is the east, bronze is the southwest, light blue is the northwest, and red-orange (of which there is only one node, CN:SOT) is the northeast. As is apparent in Figure \ref{44055}, the bulk of the nodes are eastern-focused (teal), with a significant amount focused in the north (pink) and the south (light green). Only four firms have their primary business in the southwest (bronze) and three in the northwest (light blue). These business-centricities reflect the development pattern of China, as coastal regions tended to develop first and became richer (east and south). The north is developing more at present but has the advantage of the Beijing and Tianjin municipalities being within its bounds. The northeast, southwest, and northwest are emerging regions with large areas of sparsely populated land (\cite{felice_chinese_2023}). 

Unless otherwise specified, node size is determined by ``to" connectedness, meaning that nodes with higher levels of to connectedness are larger. Node size is set as to connectedness both because to connectedness is more variable than from connectedness (as explored later) and because it offers a helpful way to understand how connectedness evolves over time. Stressed networks are indicated by nodes being drawn more tightly together, a clear core forming, and (sometimes) more regional clustering; the Appendix's Figure \ref{Covid Figure} offers a visual comparison between an unstressed and stressed network, for reference. I also clarify how sparse the $\phi_\ell$ matrices from the VAR are in Appendix \ref{appendix: var results}. 

\subsection{Three red lines} \label{three red lines section}
I first focus on the (informal) announcement of the three red lines. While the three red lines were formally announced at a meeting on 20 August 2020, there had been rumors about them in the market starting one week earlier (13 August, after close of business). I compare plots from 13 and 14 August in Figure \ref{44055} below. 

This first case provides a litmus test of network sensitivity since the rumor of an imminent policy coincides with a noticeable network contraction as the nodes draw closer together. This is important to see given that this is a rolling window estimation, meaning that the VAR is estimated on 100 days of data. Thus, the network should not (and indeed, does not) oscillate wildly between days due to the carry-over of the previous 99 days in the window, but should simultaneously allow for changes to appear sensitive to the most recent day. 

\begin{figure}[H]
  \caption{Network before and after news of the three red lines}  \centering
  \begin{subfigure}{0.5\textwidth}
    \centering
    \includegraphics[width=\linewidth]{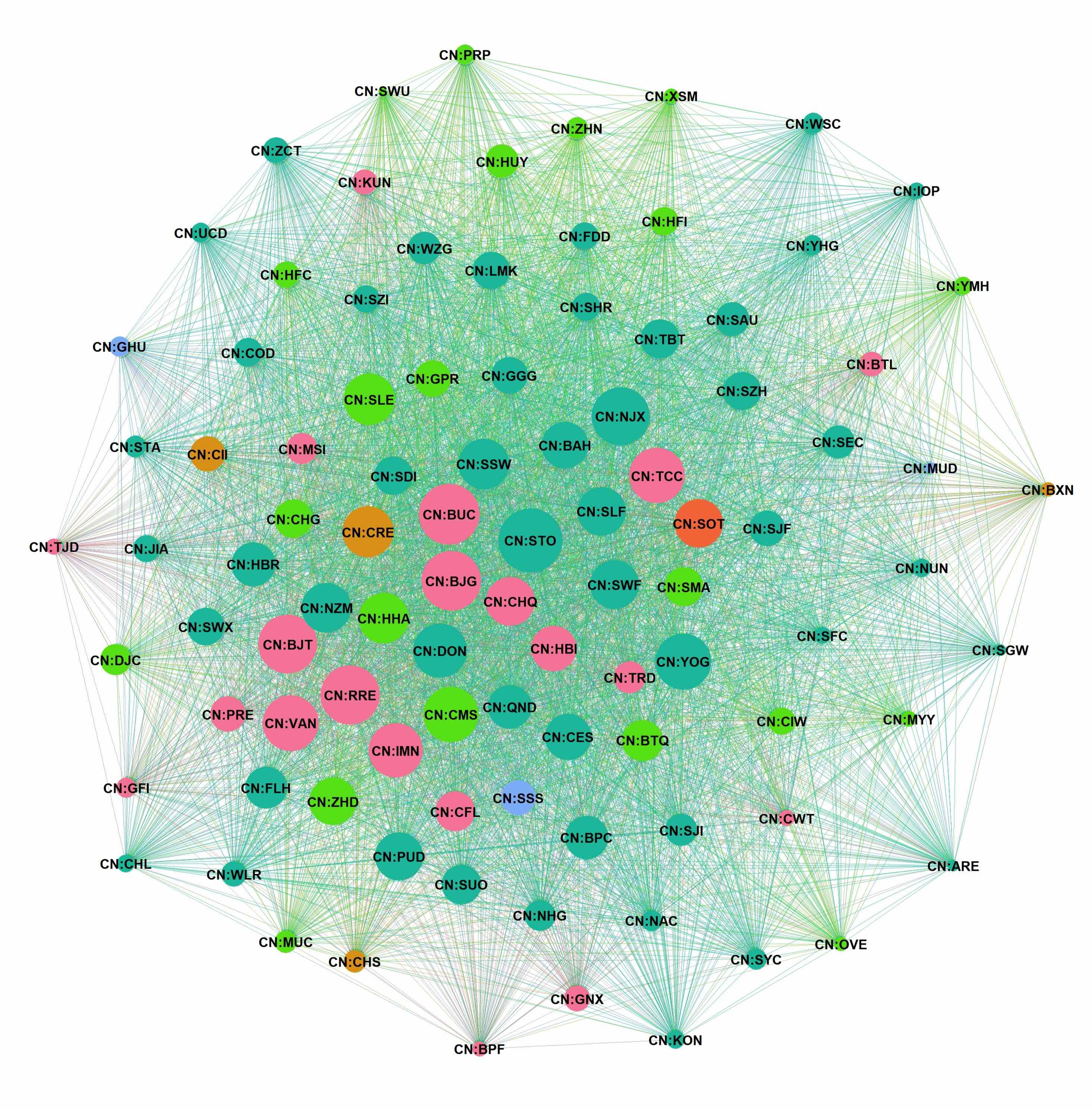}
  \end{subfigure}%
  \begin{subfigure}{0.5\textwidth}
    \centering
    \includegraphics[width=\linewidth]{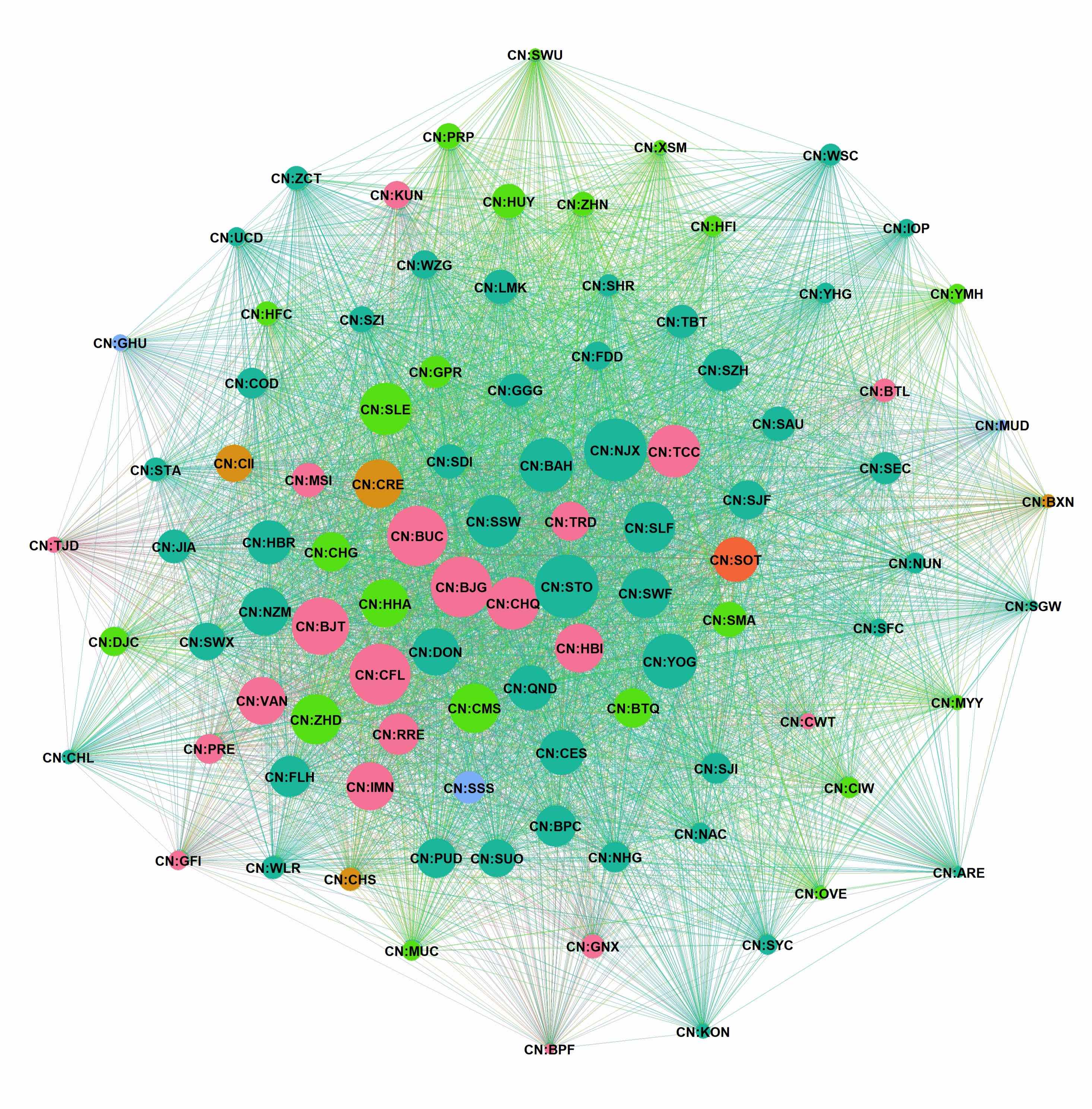}
    \end{subfigure}
  \label{44055}
  \caption*{\footnotesize Note: The left graph shows the business day before the announcement (13 August 2020), while the right graph shows the business day after the announcement (14 August 2020). Here, the colors are determined by the region where the developer is primarily focused: pink is the north, light green is the south, teal is the east, bronze is the southwest, light blue is the northwest, and red-orange (of which there is only one node, CN:SOT) is the northeast. Node size is determined by to connectedness; larger node size thus implies a higher level of to connectedness for the node.}
\end{figure}

As is apparent in the images above, the network contracts slightly with the first news of the three red lines policy. In particular, most of the nodes are drawn in more toward the core of the network\textemdash a simple visual inspection reveals that the network on 14 August appears smaller in size than that on 13 August. For the most part, nodes are not making radical moves, with many staying near their relative initial positions but pulling in slightly. This behavior suggests that the news of the three red lines indeed impacts the entire market rather than one segment disproportionately, and most nodes (80/97) experienced a decrease in ``from" connectedness. The majority of nodes (61/97) also experienced a decrease in ``to" connectedness when compared to the previous period (13 August 2020). This relatively uniform behavior likely reflects the impact of the policy change, as the estimate suggests that the majority of firms have a lower level of spillover to and from each other in this period. Simultaneously, the force calculations in ForceAtlas2 register higher attractive forces across the network (meaning some edge weights are stronger), which suggests increased interdependencies between nodes, as is often the case when a market experiences a shock. Both factors are consistent with a shock originating outside of the network, such that the connectedness of most nodes behaves similarly while higher edge weights reflect how ties between firms have strengthened, pulling nodes closer together.\footnote{It may seem counterintuitive for most nodes to experience decreases in ``to" and ``from" connectedness while the network experiences higher attractive forces. However, both the distribution and magnitude of the edge weights ($d_{ij}$), as well as the normalization, make this result possible, and I explain the phenomenon fully in Appendix \ref{appendix: visualizing the network}. In this case, when examining the edge weights of the network, roughly half of directional node pairs have stronger magnitudes of pairwise directional connectedness on 14 August than on 13 August, and their magnitude and distribution is important as well\textemdash 70\% of the node pairs have one or both edges with higher edge weights on 14 August than on 13 August, causing nearly the entire network to draw more tightly together. At the same time, to and from connectedness do not necessarily increase because they depend on the overall distribution of influence patterns in the network: when $d_{ij}$ are summed down a column or row to calculate ``to" or ``from" connectedness, respectively, the total is lower on average on 14 August than on 13 August. Certainly, this is not the case for every firm, as some experience increases in ``to" and/or ``from" connectedness, but this asymmetry, coupled with the normalization, helps lead to a decrease in ``to" and ``from" connectedness on average, as the increase in weights is not proportionate across the network.}

Here, firm interactions by region do not change significantly: no region is suddenly pulled into the core or ejected from it. Rather, as described above, nodes of all colors are pulled towards the center, and the average node's ``to" connectedness decreases. Likewise, as I seek to examine the impact of state-ownership on firms, I also generate plots of the network where the nodes are colored based on their state-ownership status. I include these state-ownership plots for this event only in the Appendix (Section \ref{appendix: three red lines additional plots}) as they are not particularly insightful: there is no evidence of differing behavior between SOEs and POEs with the announcement of the three red lines. 

\subsection{Evergrande letter to Guangdong government}

\begin{figure}[H]
  \caption{Network before, during, and after news of Evergrande's letter}
  \centering
  \begin{subfigure}{0.5\textwidth}
    \centering
    \includegraphics[width=\linewidth]{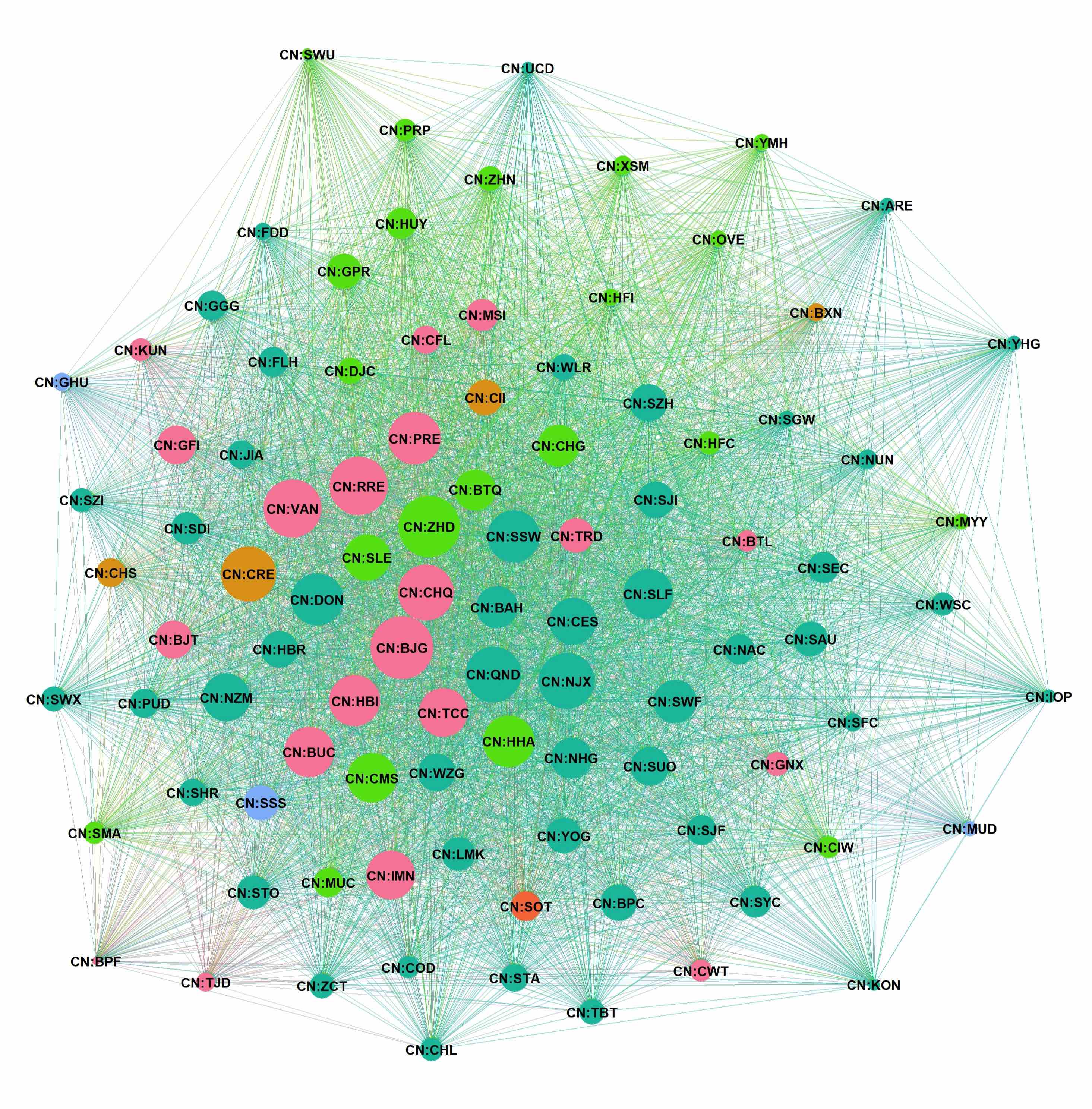}
    \subcaption{Before news circulation (21 September 2020)}
    \label{fig:before circ}
  \end{subfigure}%
  \begin{subfigure}{0.5\textwidth}
    \centering
    \includegraphics[width=\linewidth]{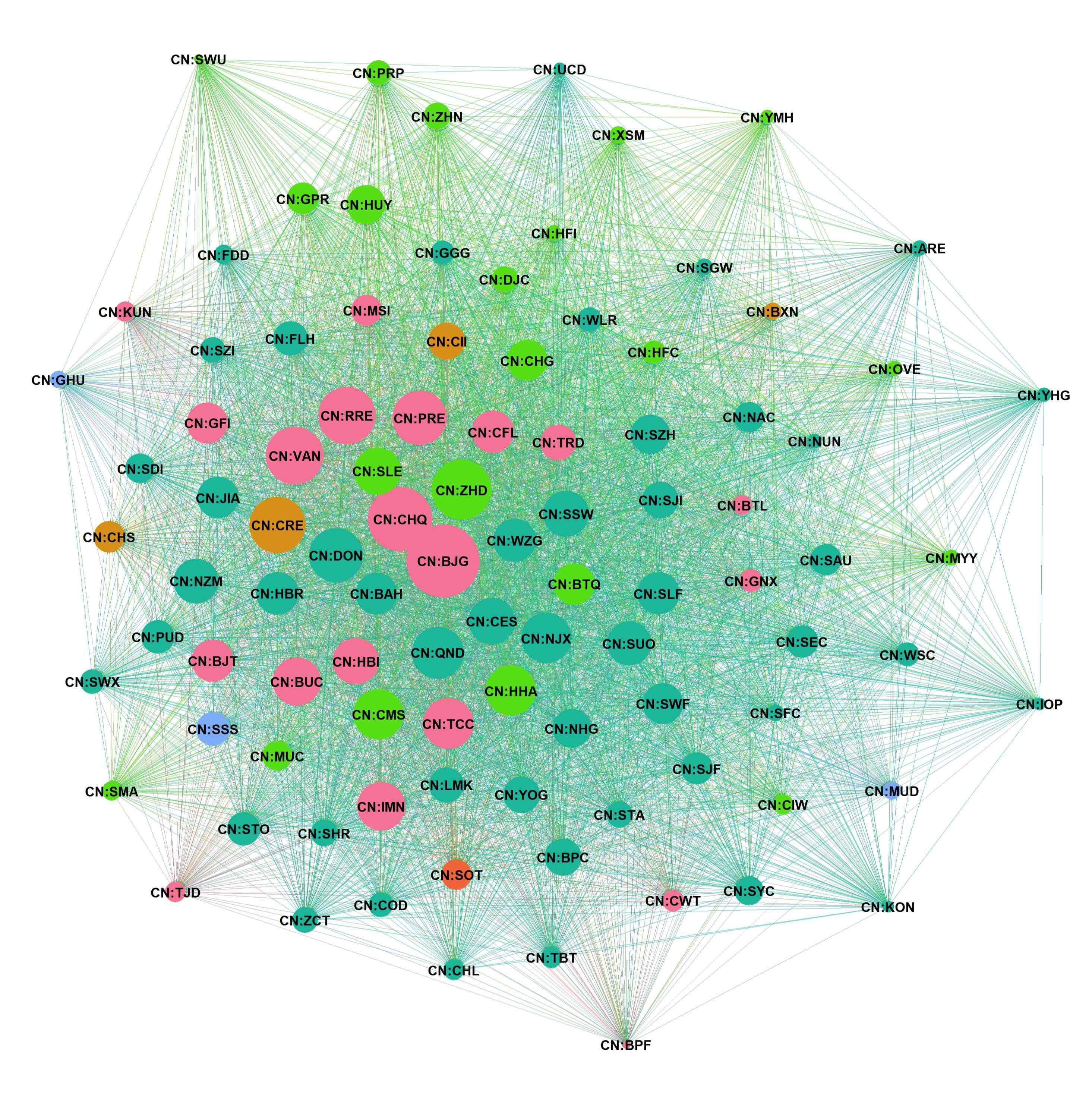}
    \subcaption{Partially circulated (23 September 2020)}
    \label{fig:circ partial}
  \end{subfigure}
  
  \begin{subfigure}{0.5\textwidth}
    \centering
    \includegraphics[width=\linewidth]{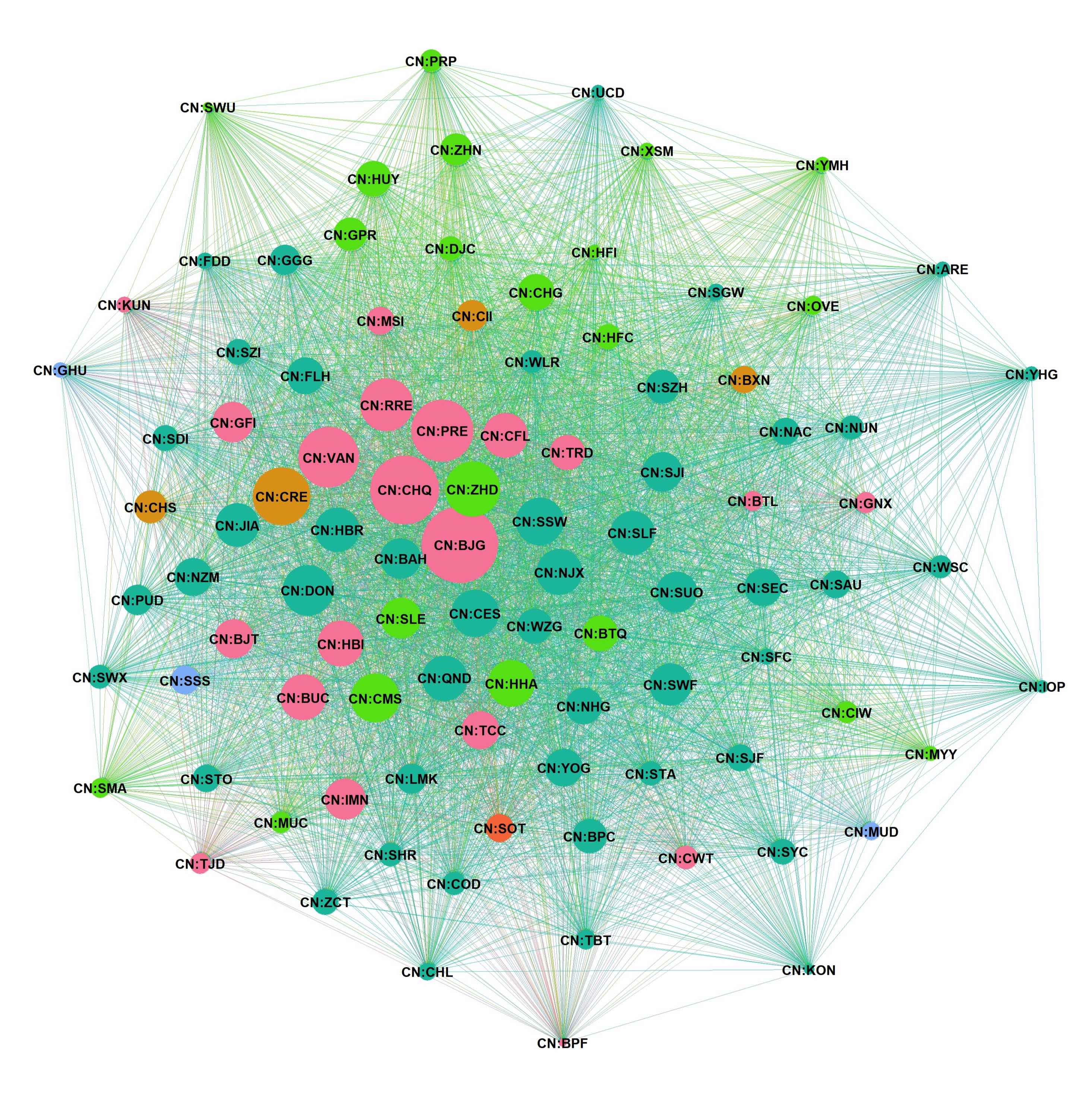}
    \subcaption{Letter goes viral (24 September 2020)}
    \label{fig:circ viral}
  \end{subfigure}%
  \begin{subfigure}{0.5\textwidth}
    \centering
    \includegraphics[width=\linewidth]{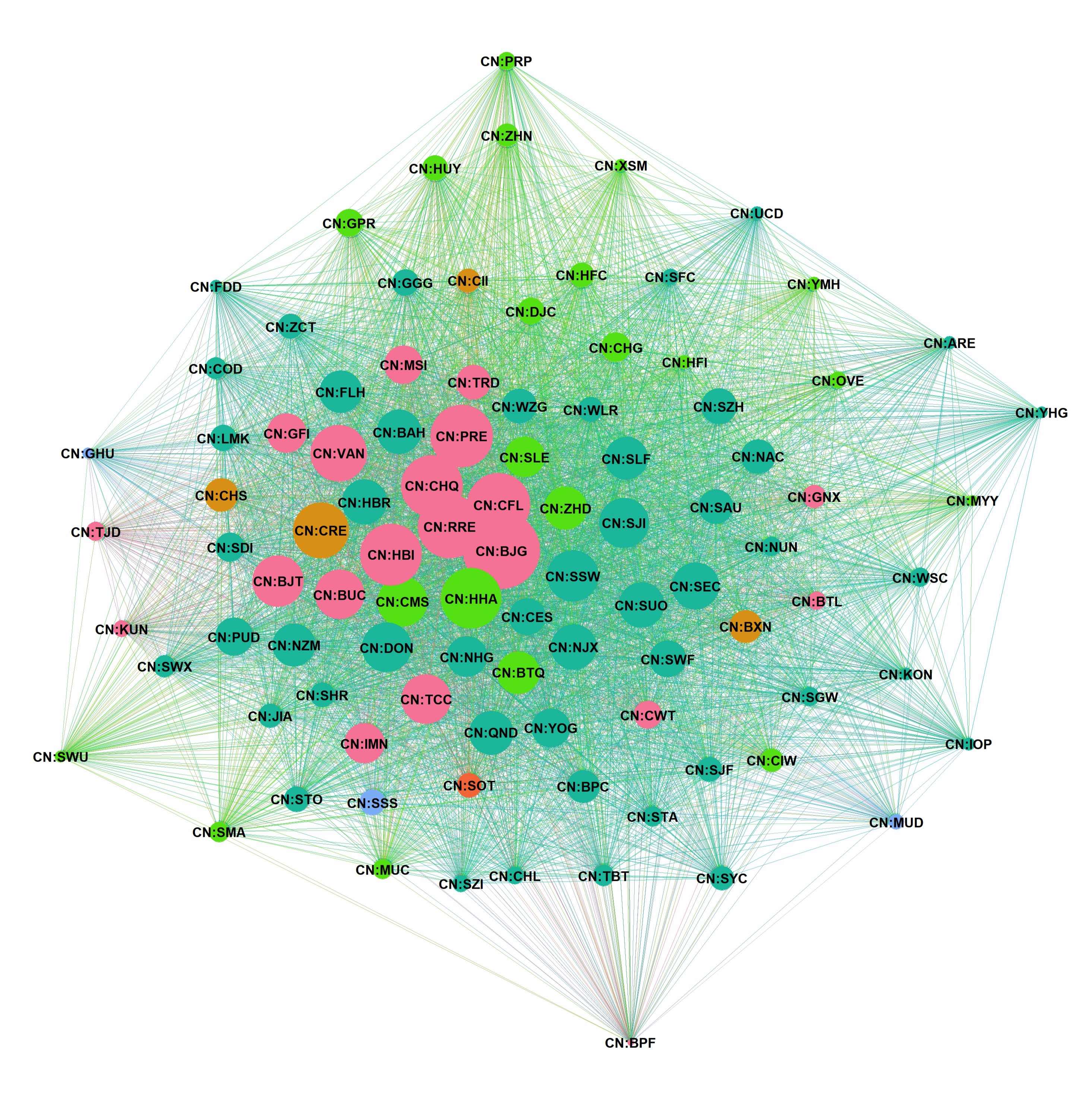}
    \subcaption{Two weeks later (9 October 2020)}
    \label{fig:two weeks later}
  \end{subfigure}
    \caption*{\footnotesize Note: The upper left graph shows the business day before the news (21 September 2020), while the upper right graph shows the network change when the news had been partially circulated (23 September 2020). The bottom left shows the day the news went viral (24 September 2020), and the bottom right shows the network roughly two weeks later (9 October 2020). As above, the colors are determined by the region where the developer is primarily focused: pink is the north, light green is the south, teal is the east, bronze is the southwest, light blue is the northwest, and red-orange (of which there is only one node, CN:SOT) is the northeast. As before, node size is determined by to connectedness; larger node size thus implies a higher level of to connectedness for the node.}
  \label{Guangdong Gov Letter Network}
\end{figure}

I next examine the first significant shock to the market from a real estate developer, which occurred when the letter from Evergrande to the Guangdong government circulated on Chinese social media and then in the news. A letter dated 24 August 2020\textemdash just four days after the government's meeting with developers where the three red lines were officially announced\textemdash began to circulate online roughly one month later on 22 September, becoming viral on 24 September. Declaring that Evergrande's capital was significantly reduced and that its cash flow had been disrupted, the letter warned of a possible default (\cite{chinaevergrandegroupco_china_2020}). Within a few hours of the letter going viral, Evergrande swiftly denounced it, but the damage was done (\cite{zhou_-_2020}; \cite{chinaevergrandegroupco_china_2020}). Figure \ref{Guangdong Gov Letter Network} shows the network at four times: just before the news broke out (21 September 2020), when it had circulated partially (23 September 2020), when it went fully viral (24 September 2020), and approximately two weeks afterward (9 October 2020). 

The network in Figure \ref{fig:before circ} already shows some clustering, with nodes in the center of the network grouping loosely by region and having relatively high ``to" connectedness levels. When the news has been partially circulated (Figure \ref{fig:circ partial}), nodes closer to the core have an increase in to connectedness, while those at the top and right sides of the network remain pushed out with lower levels of to connectedness. Particularly, teal nodes (eastern-centric firms) in the core tend to shrink slightly while their pink counterparts (north-centric firms) in the core grow, meaning that their to connectedness is increasing. This behavior continues in Figure \ref{fig:circ viral} as the news goes viral; the to connectedness of several nodes in the core increases significantly as the core nodes pull closer together. At the same time, several nodes at the periphery (particularly on the top and right) remain immune to the attractive forces pulling the other nodes in. Noticeably, to connectedness (and thus node size) increases less on average for firms with a predominantly eastern focus (teal nodes) and southern focus (light green nodes) than those from the other regions. In particular, firms with a northern focus (pink nodes) and southwestern focus (bronze nodes) seem to experience the largest increases in to connectedness, suggesting that market spectators expect some dimension of regional spillover. This higher level of spillover to northern- and southwestern-centered companies thus likely reflects that investors perceive these regions to be less stable and comparatively less insulated from a real estate shock.\footnote{Indeed, the raw stock movements confirm that the increased to connectedness of these firms is not a substitution effect wherein investors view these firms as more stable (which would drive higher prices) but that of a knock-on effect that causes investors to divest from these stocks, which they view as less stable.}

This result is especially interesting because there are no comparative increases in ``to" connectedness for southern-focused developers: one would expect that since Evergrande's headquarters and the majority of its properties are there, the developers most exposed to an Evergrande cash crunch would likely be those it interacts with most (i.e., also southern developers). It is then particularly fascinating not to see this effect immediately, as its absence suggests that investors initially anticipate effects at the regional level rather than the developer level. This behavior ties back to entrenched ideas about China's path of development and the property market in China, deriving from housing as a commodity, as discussed in Section \ref{background}. As housing is highly speculative, demand is then linked to speculators' expectations about the broader market\textemdash they expect that demand will dip most in less developed regions like the north, southwest, and northwest, and in turn, their own demand for property in those regions dips, creating a self-fulfilling prophecy. 

\begin{figure}[H]
  \caption{Network before, during, and after news of Evergrande's letter: Color by state ownership}
  \centering
  \begin{subfigure}{0.5\textwidth}
    \centering
    \includegraphics[width=\linewidth]{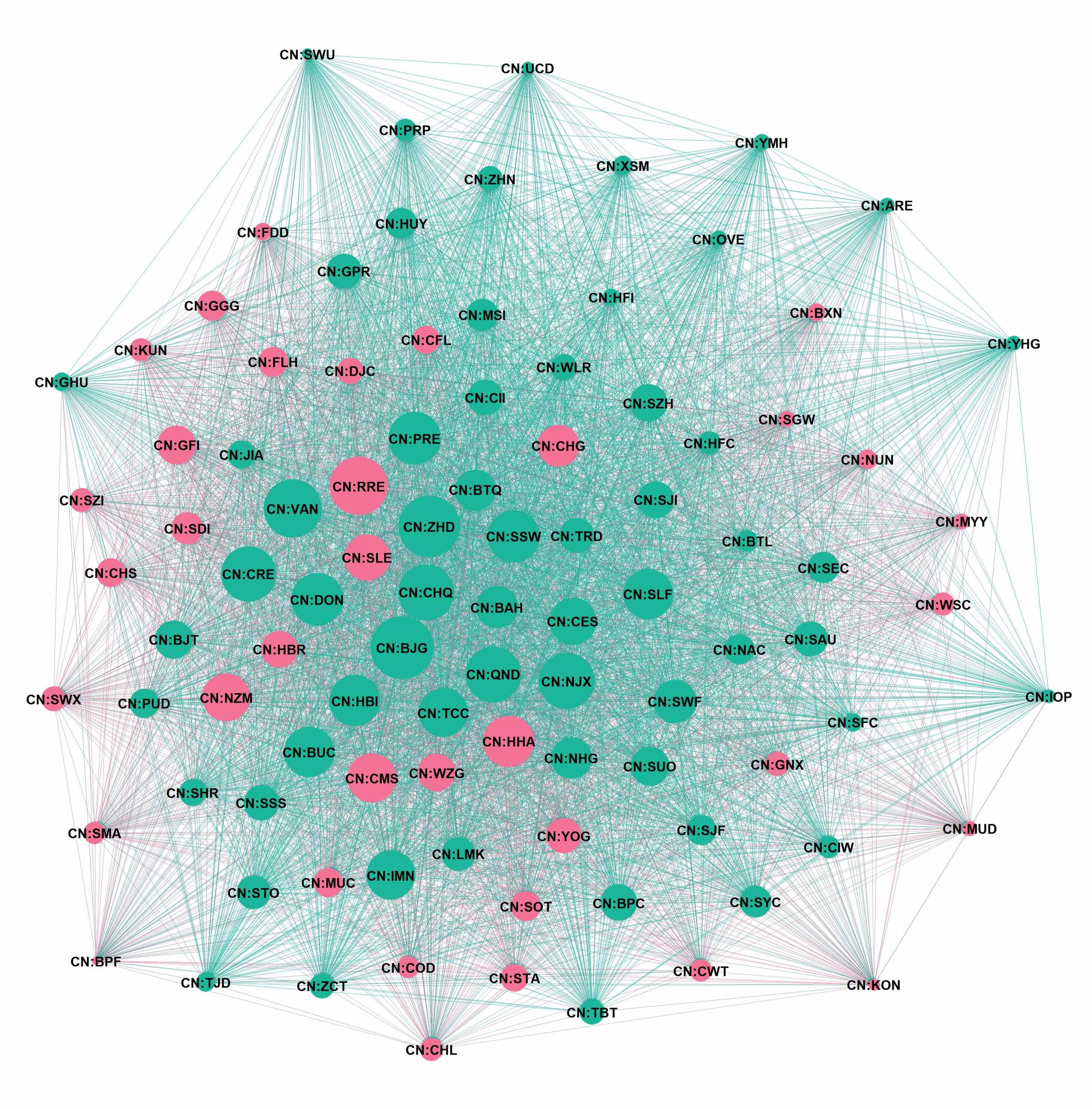}
    \subcaption{Before news circulation (21 September 2020)}
    \label{fig:SO before circ}
  \end{subfigure}%
  \begin{subfigure}{0.5\textwidth}
    \centering
    \includegraphics[width=\linewidth]{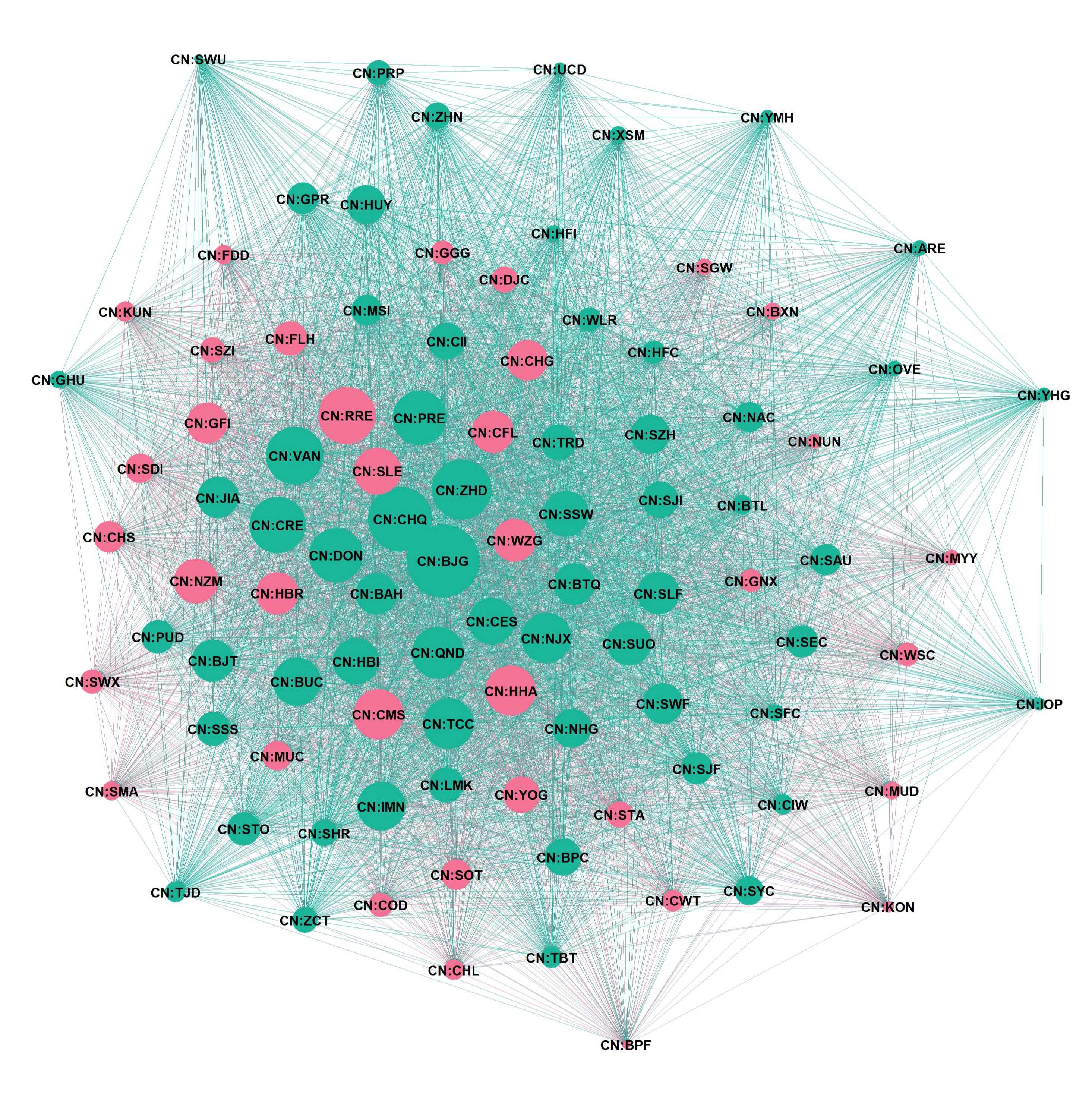}
    \subcaption{Partially circulated (23 September 2020)}
    \label{fig:SO circ partial}
  \end{subfigure}
  
  \begin{subfigure}{0.5\textwidth}
    \centering
    \includegraphics[width=\linewidth]{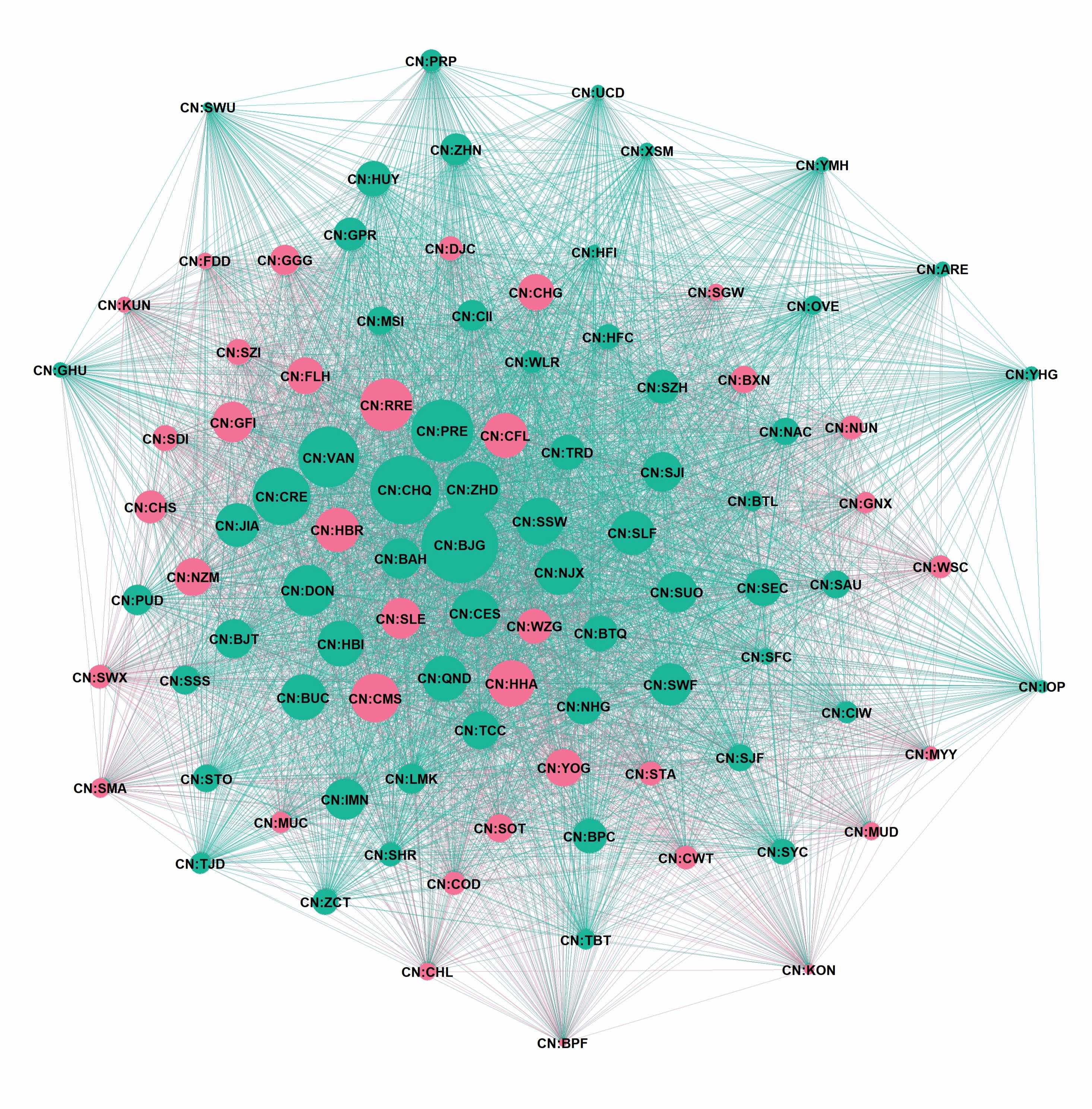}
    \subcaption{Letter goes viral (24 September 2020)}
    \label{fig:SO circ viral}
  \end{subfigure}%
  \begin{subfigure}{0.5\textwidth}
    \centering
    \includegraphics[width=\linewidth]{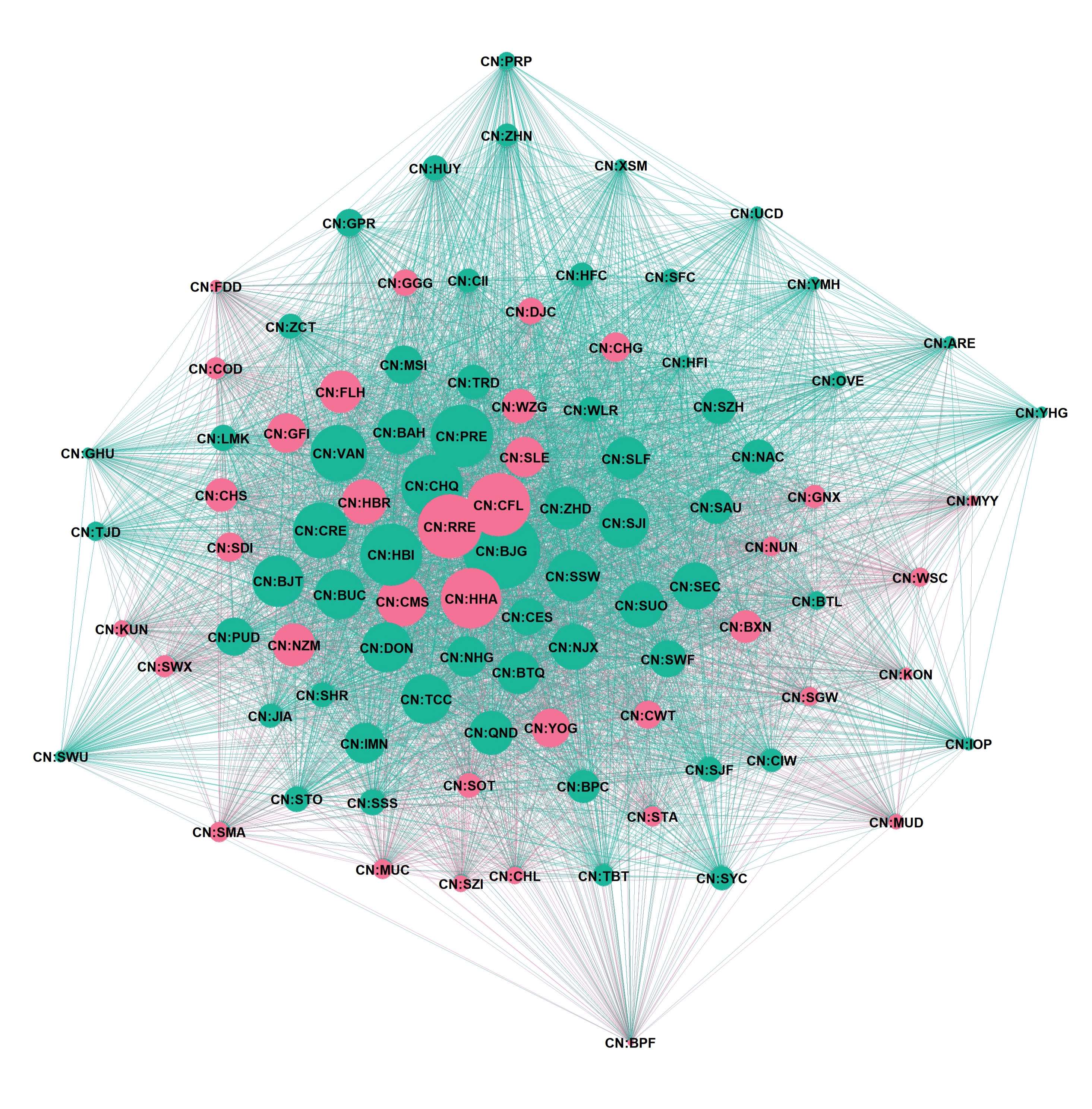}
    \subcaption{Two weeks later (9 October 2020)}
    \label{fig:SO two weeks later}
  \end{subfigure}
    \caption*{\footnotesize Note: The upper left graph shows the business day before the news (21 September 2020), while the upper right graph shows the network change when the news had been partially circulated (23 September 2020). The bottom left shows the day the news went viral (24 September 2020), and the bottom right shows the network roughly two weeks later (9 October 2020). In this series, the colors are determined by state ownership status: pink nodes are not state-owned while teal reflects state ownership. As before, node size is determined by to connectedness; larger node size thus implies a higher level of to connectedness for the node.}
  \label{SO Guangdong Gov Letter Network}
\end{figure}

Simultaneously, the continually distanced nodes on the top and right peripheries, apparent in Figures \ref{fig:circ partial} and \ref{fig:circ viral}, suggests that several nodes remain less connected with the core and are thus continually pushed outwards as they experience high repulsive forces from the other nodes. This implies a degree of segmentation in the market such that investors believe some real estate companies are particularly well insulated from Evergrande and the shock its default would generate, beyond the regional dimension. Investigating the profiles of companies on the periphery that remain pushed out reveals that they are more diversified than the developers in the core. In addition, state ownership also appears to play a role in pushing certain nodes further out than others. In particular, a plot color-coded for state ownership, shown in Figure \ref{SO Guangdong Gov Letter Network}, illustrates that in the initial shock, non-state-owned companies (pink nodes) on the periphery of the network tend to be pulled in, while their state-owned counterparts (teal nodes) tend to remain pushed out (Figures \ref{fig:SO circ partial} and \ref{fig:SO circ viral}). Indeed, most of the nodes that remain close to their initial positions farthest out on the periphery are state-owned companies rather than privately held ones; this behavior suggests that investors also hold an underlying belief about state ownership's role in firm stability\textemdash that state-owned firms are more stable than their privately owned counterparts. 

Two weeks after the shock, the network has converged toward a new base state: the core of the network has drawn inward, and most of the nodes on the periphery have drawn in as well. Instead, the network seems to have contracted more tightly than it was originally, with several firms (particularly northern ones) having higher levels of to connectedness; indeed, the to connectedness is so high in the core of the network that the nodes are almost overlapping.\footnote{This apparent overlapping is due to high levels of to connectedness; the positions prescribed by ForceAtlas2 have separation between the nodes, and the nodes do not inherently overlap, only doing so here because node size is to connectedness. } The regional dimension still seems to play a role, as the to connectedness has increased for northern-centric firms more on average than any other region. Likewise, firms in the southwest (bronze nodes) end up pulled much closer to the core of the network than they were originally.

Some southern-centric (light green nodes) and eastern-centric (teal nodes) firms have also been pulled towards the center. However, for both these southern- and eastern-centric firms, the magnitude of to connectedness is not as significant as that of the northern firms. This behavior does suggest that investors now have a more nuanced view of what specific firms will be most impacted, but there nevertheless seems to still be a regional slant towards northern firms, the bulk of which still have higher levels of to connectedness than before the news was released. 

From the state-ownership perspective (Figure \ref{fig:SO two weeks later}), the nodes furthest out on the periphery are mostly state-owned companies (teal nodes), and several private companies (pink nodes) have penetrated the core of the network, which was previously composed of mostly state-owned companies. This behavior again reflects how investors view private ownership as less stable in times of high risk, when the market is dominated by state-owned companies. Certainly, though, being state-owned does not make a firm wholly protected from any financial shock: state intervention takes time and is not perfectly efficient, meaning one would not expect all state-owned firms to simultaneously exit the core and have very low levels of ``to" connectedness: these firms are still market participants, and their profit will dip in the event of a market contraction, regardless of the contraction's origin8.

\subsection{Kaisa suspension}
While these two events have a relatively clear-cut impact on the network given that it was not in distress before, later events are more complex as the market became constantly stressed. After the previously discussed Evergrande letter, the market spiraled through various cycles and contractions as more developers faced financial trouble\textemdash and it is these events, in which new developers have their first major loss of public confidence, that allow us to understand spillover patterns that emerge when the network is shocked.

I thus analyze the suspension of Kaisa on 5 November 2021, which occurred a day after one of its affiliates missed a payment to onshore investors (\cite{galbraith_kaisa_2021}). Kaisa became the first Chinese developer to default on its dollar bonds in 2015 but had largely recovered since then; however, combining the bad market with rating downgrades, the developer was under pressure and struggling (\cite{jim_exclusive_2021}). The shares were suspended pending the release of ``inside information" with the resumption of trading ultimately occurring 17 days later (\cite{zhu_kaisa_2021}). Figure \ref{Region Kaisa} below illustrates the network around this change, with Figure \ref{fig: Kaisa before} depicting the network on 3 November 2021, the day before the missed payment to onshore investors and two days before the suspension. Figure \ref{fig: Kaisa after} illustrates the network the day the suspension is announced and begins (5 November 2021).

\begin{figure}[H]
  \caption{Network before and after Kaisa suspension: Color by region}
  \centering
  \begin{subfigure}{0.5\textwidth}
    \centering
    \includegraphics[width=\linewidth]{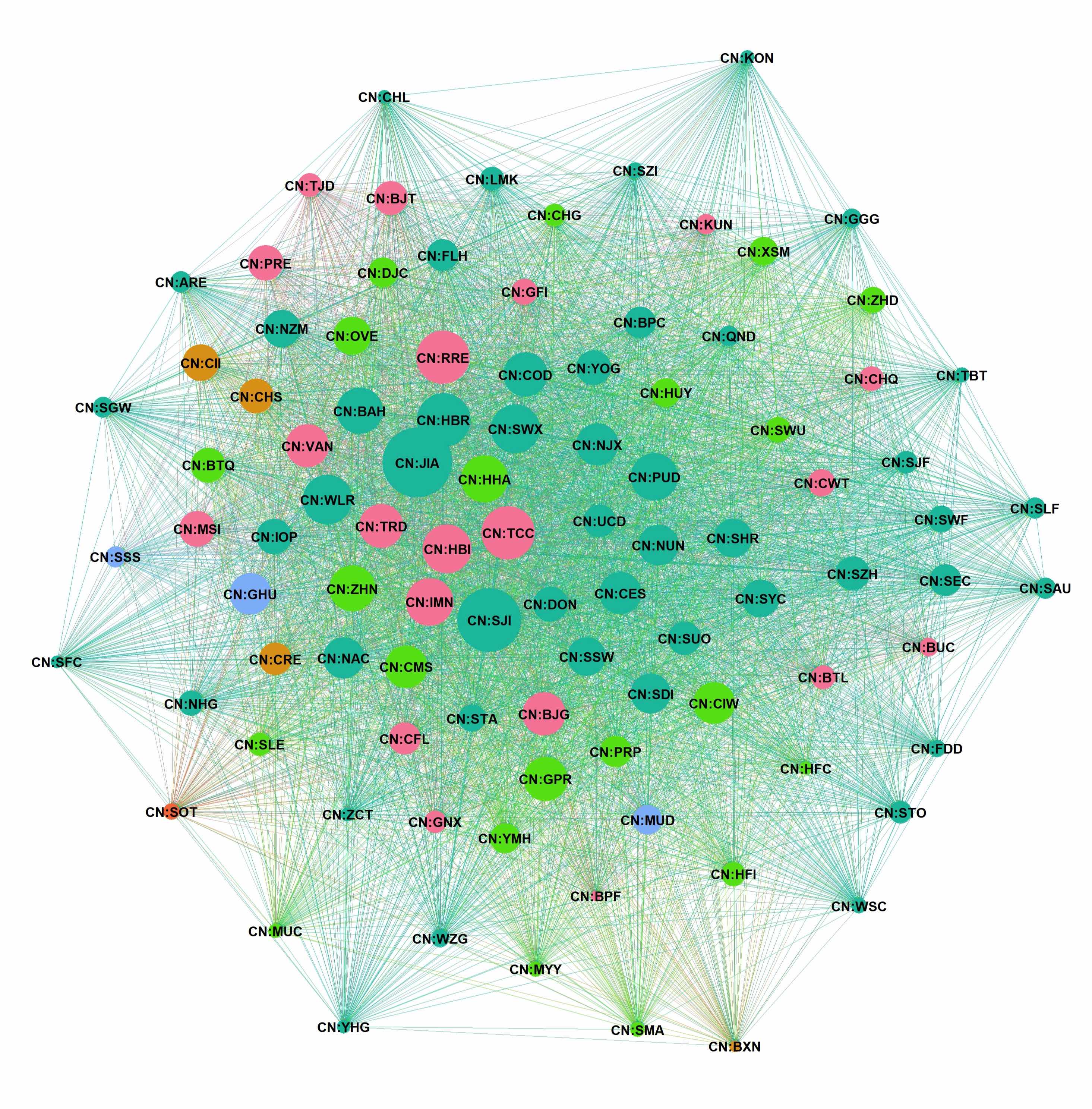}
    \subcaption{Before suspension (3 November 2021)}
    \label{fig: Kaisa before}
  \end{subfigure}%
  \begin{subfigure}{0.5\textwidth}
    \centering
    \includegraphics[width=\linewidth]{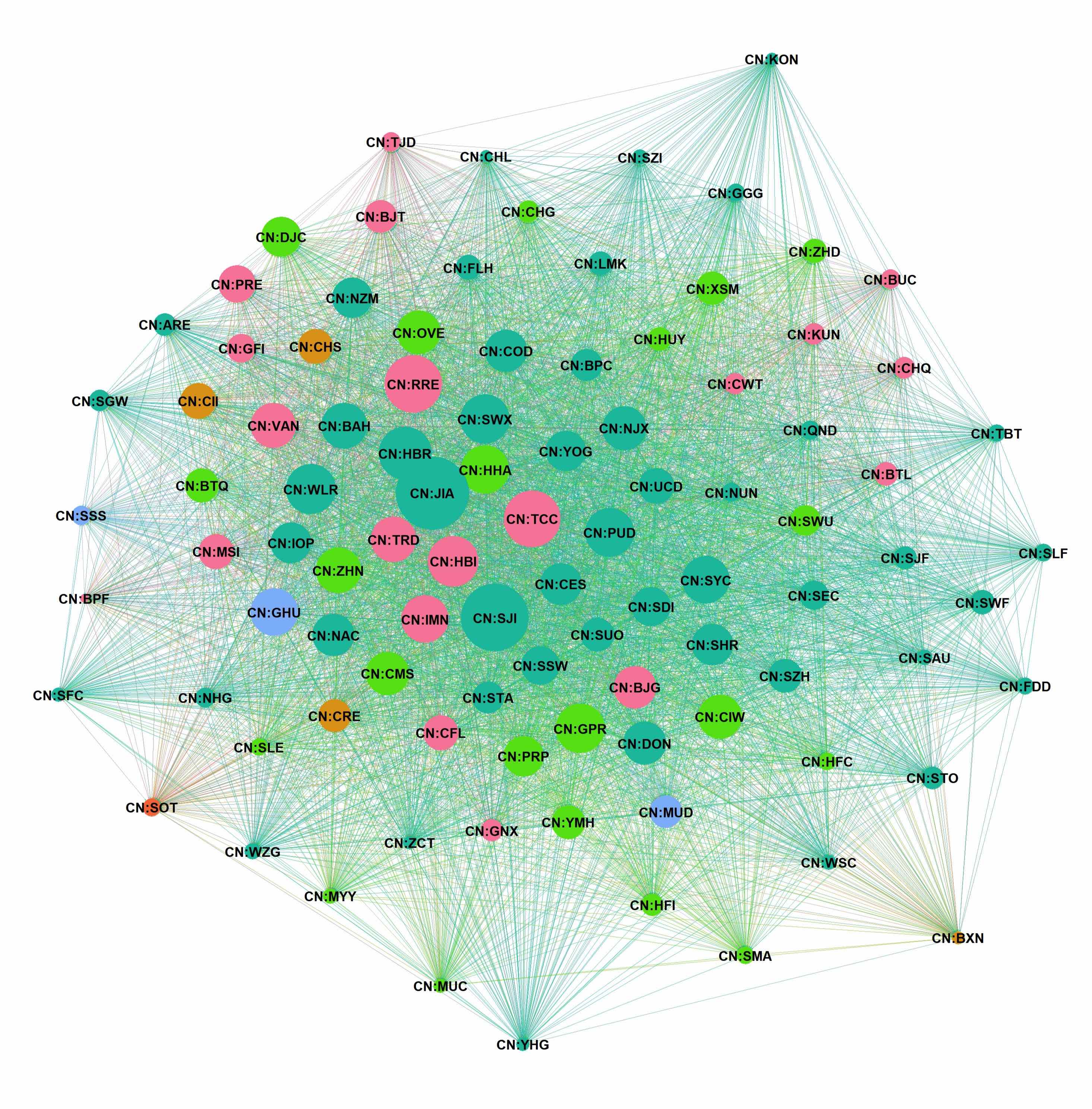}
    \subcaption{After suspension (5 November 2021)}
    \label{fig: Kaisa after}
  \end{subfigure}
  \label{Region Kaisa}
      \caption*{\footnotesize Note: The left graph shows the business day before the missed payment to onshore investors, two days before the suspension (3 November 2021). The right graph shows the network change on the day of the suspension (5 November 2021), as the stock suspended at 9 am\textemdash the time the exchange opens for the day. The colors are determined by the region where the developer is primarily focused: pink is the north, green is the south, teal is the east, bronze is the southwest, light blue is the northwest, and red-orange (of which there is only one node, CN:SOT) is the northeast. Node size is determined by ``to" connectedness.}

\end{figure}

The core of the network contracts, with several nodes pulling in towards the core while those on the periphery remain on the edge of the network. There are a few interesting observations here: first, unlike previous periods, the companies at the core of the network with the highest levels of connectedness are now eastern-centric companies (teal nodes) rather than northern- or southern-centric ones (pink or light green nodes, respectively). Certainly, there are still pink and light green nodes in the core and many that are drawn in from being more towards the periphery between Figures \ref{fig: Kaisa before} and \ref{fig: Kaisa after}. Yet, the bulk of nodes that experience an increase in to connectedness in Figure \ref{fig: Kaisa after} are eastern-centric. Contrarily, the nodes on the periphery, even those that are teal, experience a decrease in to connectedness on average and thus shrink; 20/31 periphery nodes shrink, with 12/17 of teal nodes on the periphery experiencing decreases in to connectedness. ``From" connectedness is largely the same across both periods; excluding the seven firms who have larger increases in from connectedness as they are pulled in from the periphery, most firms experience a negligible from connectedness increase of 0.51 between 3 and 5 November.

In earlier periods (such as when Evergrande's letter to the Guangdong government was exposed), eastern-centric companies were largely grouped to one side of the network and comprised most of the periphery. Their new position\textemdash pulled more towards the center of the network and largely surrounded by their southern and northern peers, rather than vice-versa\textemdash suggests investors have an updated belief about regional spillover: the areas that were initially regarded as comparatively more stable are now much more at risk, manifesting in the new node position apparent in Figures \ref{fig: Kaisa before} and \ref{fig: Kaisa after}.

This behavior is interesting given Kaisa has the bulk of its properties in the south, east, and southwest of China: certainly, the southwestern- and southern-centric companies (bronze and light green, respectively) are drawn in slightly, but again, neither comprise the bulk of the core. The shock of the suspension is thus felt through increasing to connectedness mainly in eastern-centric developers as well as increasing pairwise connections between these developers and those around them, as reflected by the contraction of the layout and the increase in attractive force between nearby core nodes. 

\begin{figure}[H]
  \caption{Network before and after Kaisa suspension: Color by SOE status}
  \centering
  \begin{subfigure}{0.5\textwidth}
    \centering
    \includegraphics[width=\linewidth]{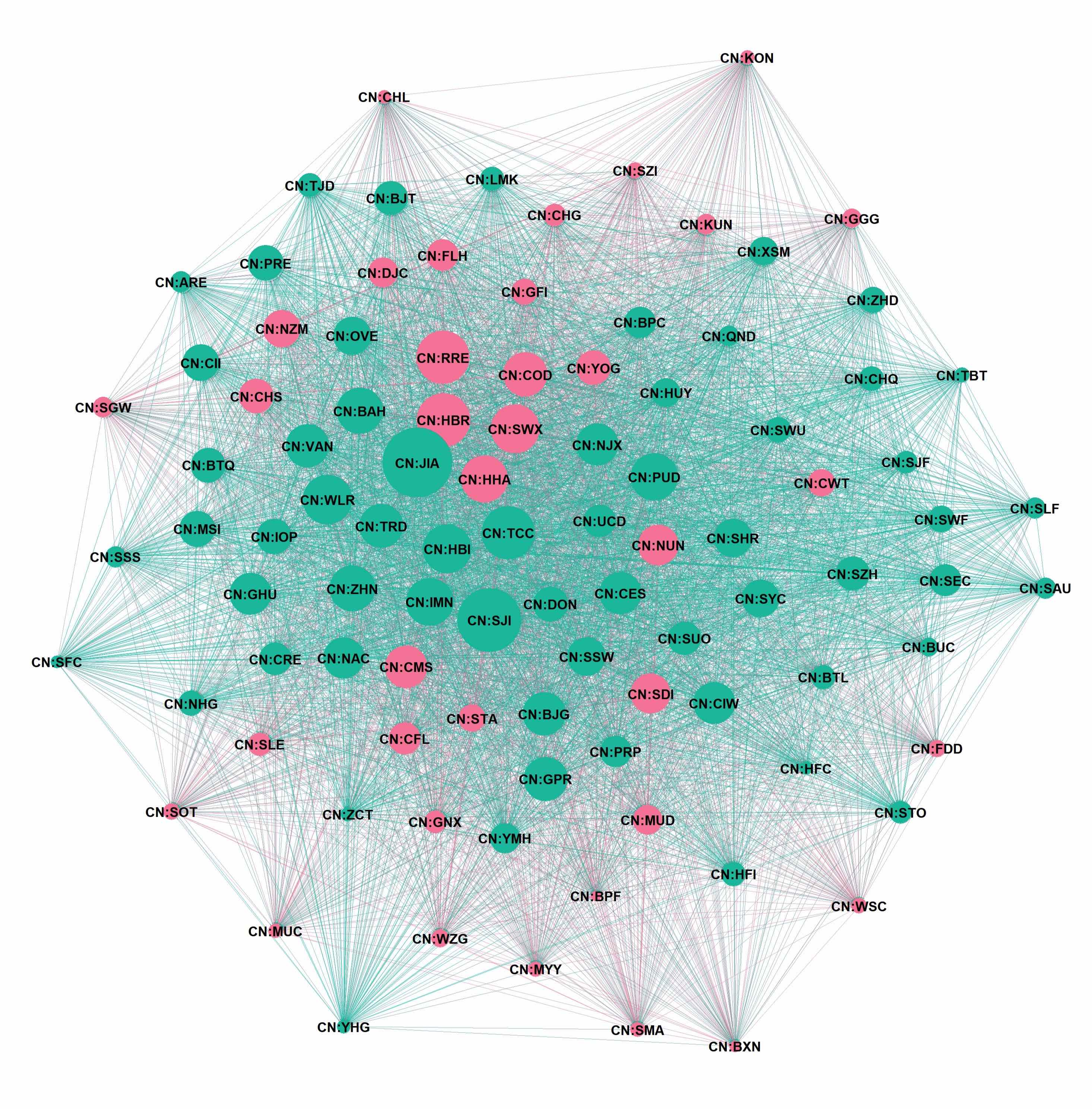}
    \subcaption{Before suspension (3 November 2021)}
    \label{fig: SO Kaisa before}
  \end{subfigure}%
  \begin{subfigure}{0.5\textwidth}
    \centering
    \includegraphics[width=\linewidth]{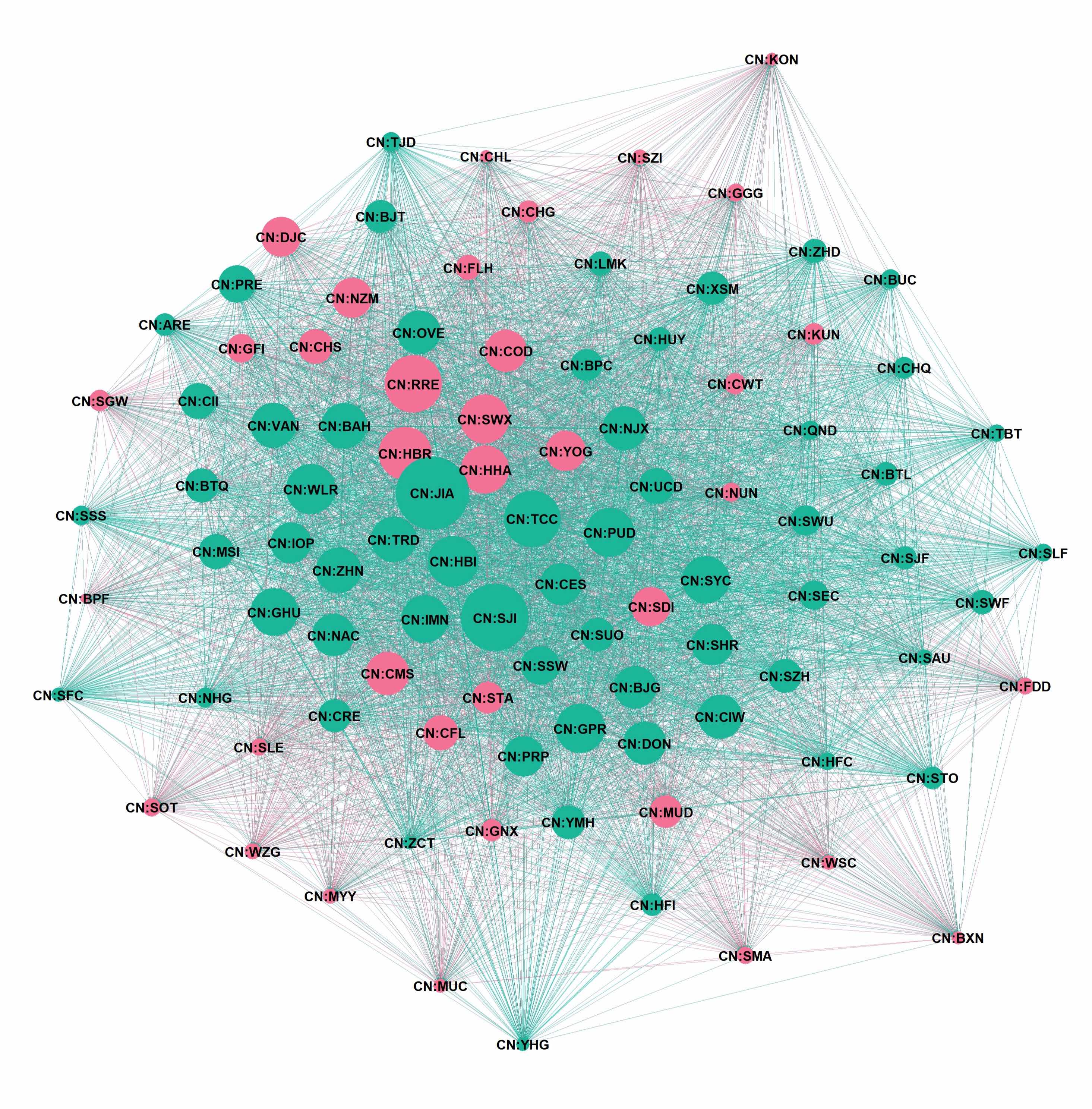}
    \subcaption{After suspension (5 November 2021)}
    \label{fig: SO Kaisa after}
  \end{subfigure}
  \label{SO Kaisa}
  \caption*{\footnotesize Note: The left graph shows the business day before the missed payment to onshore investors, two days before the suspension (3 November 2021). The right graph shows the network change on the day of the suspension (5 November 2021), since the stock suspended at 9 am\textemdash the time the exchange opens for the day. In this series, node colors are determined by the state ownership status: pink nodes are not state-owned while teal reflects state ownership. Node size is determined by ``to" connectedness.}
\end{figure}

In terms of state ownership (Figure \ref{SO Kaisa}), several private firms remain pulled into the core, as in Figure \ref{fig:two weeks later}. While most of the periphery is composed of private firms in Figure \ref{fig: SO Kaisa before}, these firms are largely pulled towards the core in \ref{fig: SO Kaisa after}, such that even though they mostly remain on the periphery after the suspension, their pairwise connectivity with the core nodes increases, resulting in closer positions. Noticeably, most of the private firms that are pulled towards the core (such as CN:HHA, CN:SWX, CN:HBR, and CN:COD) experience very small or no increases in to connectedness; comparatively, those that are state-owned in the core, like CN:JIA, CN:SJI, and CN:TCC, and even state-owned firms on the edges of the core, like CN:PRP, CN:DON, and CN:BJG, experience increasing to connectedness, as reflected by a larger node size.\footnote{The full company names of the nodes mentioned are as follows: the private firms pulled towards the core are CN:HHA (Hubei Fuxing Science and Technology Co.), CN:SWX (Shanghai Shimao `A'), CN:HBR (Hangzhou Binjiang Real Estate Group Co.), and CN:COD (Dima Holdings `A'). The state-owned firms in the core are CN:JIA (Greenland Holdings `A'), CN:SJI (Everbright Jiabao `A'), and CN:TCC  (Tianjin Tianbao Infrastructure `A'). The state-owned firms on the edges of the core are CN:PRP (Shenzhen Properties \& Resources Development Group Ltd.), CN:DON (Bright Real Estate), and CN:BJG (Beijing North Star `A').} Underpinning this behavior is an interesting substitution effect towards private firms: returning to the raw stock return data reveals that the majority of privately owned firms on the periphery (10/17) experience increases in closing price for their stocks on 5 November, compared to 3 November. Meanwhile, most state-owned companies on the periphery (11/16) experience price decreases, like the nodes in the core; only a select few state-owned companies, mostly on the right edge of the periphery, experience price increases. Appendix \ref{appendix: Kaisa further figures} contains plots showing periphery nodes with this color coding. 

This behavior suggests investors implicitly consider state ownership in their risk calculations: unlike in earlier periods, state-owned firms appear riskier than their private counterparts, and with the shock to Kaisa, investors seem to divert funds away from state-owned firms and towards privately held ones. This is a fascinating effect that runs counter to standard expectations; it is likely because at this point, state-backed developers were asked to purchase or take on some projects of struggling private developers, whether through buying the property outright or increasing their stake to be the majority shareholder in a development (\cite{chow_evergrande_2024}). These deals often were designed to give the struggling private developer a substantial cash injection and/or a share of the future revenue of the property, meaning the deal was often not profitable for the SOE buyer (\cite{jim_china_2021}). 

Thus, these plots seem to reflect investor sentiment that with policy changes in 2021, state-owned firms are no longer the safest bet\textemdash and may even be riskier because of the suboptimal bailout investments they are made to take on. The private firms on the periphery are mostly those with diversified investments and operating sectors, factors which likely catalyze the substitution effect apparent above. 

\subsection{Country Garden's profit warning} \label{COGA body}

As the crisis dragged on, a new phase emerged in mid-2023: developers who had previously avoided financial trouble were now getting dragged into the fray, chief among them Country Garden. In early August, Country Garden\textemdash at this point the largest private property developer in China, with four times as many pending developments as Evergrande\textemdash began to exhibit signs of financial trouble (\cite{liu_how_2023}; \cite{choi_country_2023}; \cite{jim_country_2023}). I analyze two parts of this event: the scrapping of a share sale (1 August 2023) and the issuing of a profit warning (10 August 2023).

The aborted US\$300 million share sale happened at the last minute on 1 August 2023 as Country Garden shared that it had not reached a ```final agreement' for the deal to go ahead" (\cite{jim_country_2023-1}). Then, after a relatively calm couple of days, rumors began to quietly circulate again: Country Garden had apparently missed a payment that was due on 8 August, and then, after close of business on 10 August, it issued a profit warning.\footnote{I do not include the networks before and after the 8 August news in the body of the text, largely because, as with the case of the profit warning on 10 August, there is little change in the network. They are included in the Appendix, Section \ref{Appendix: additional plots}, for reference.} The company revealed that it expected to record a net loss between US\$6.24-7.63 billion for the first 6 months of 2023 (unlike its net profit of US\$265 million for the first 6 months of 2022) (\cite{countrygardenholdingscompanyltd_profit_2023}). 

\begin{figure}[H]
  \caption{Network before and after Country Garden news: Color by region}
  \centering
  \begin{subfigure}{0.5\textwidth}
    \centering
    \includegraphics[width=\linewidth]{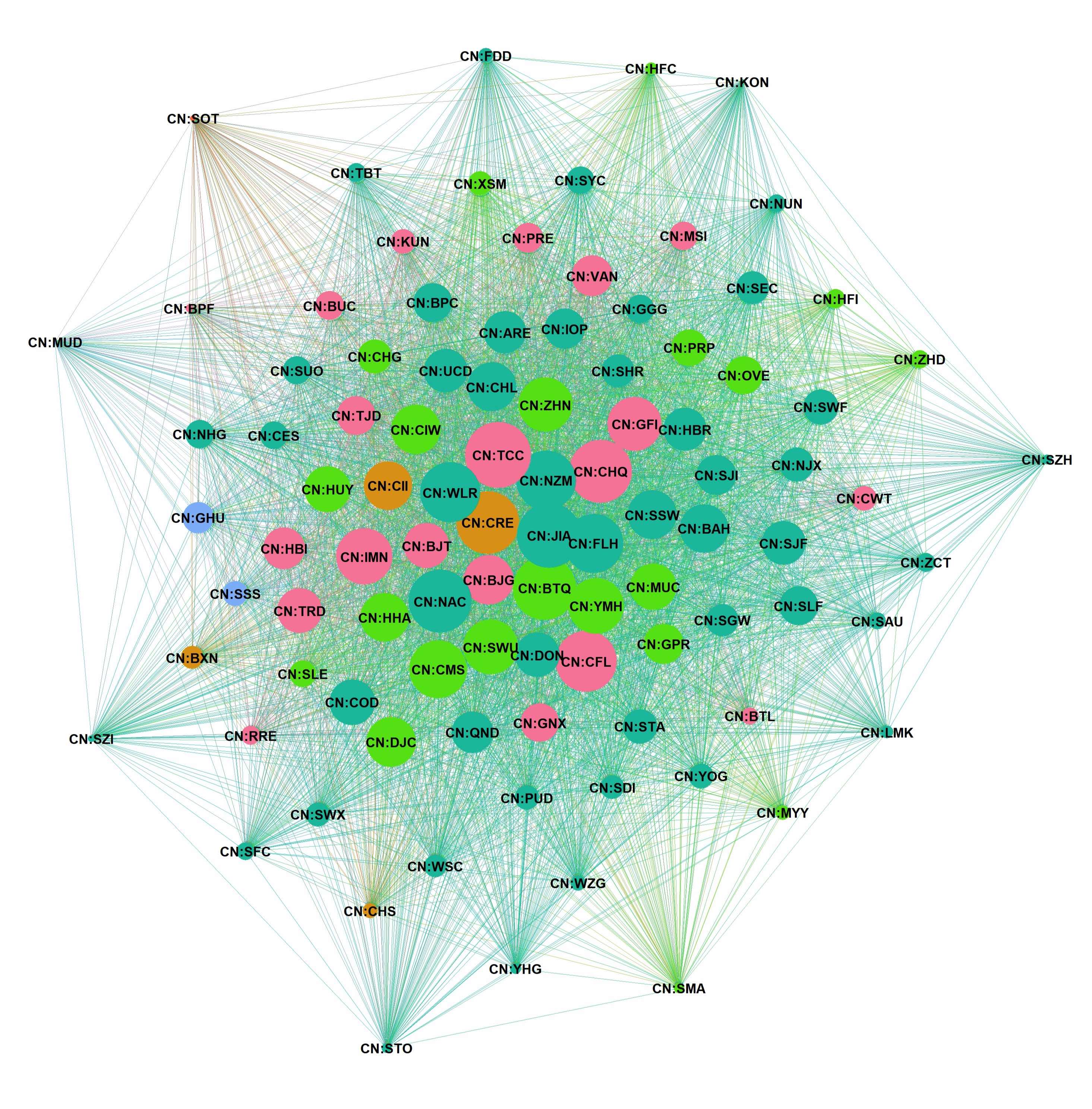}
    \subcaption{Before any news circulates (31 July 2023)}
    \label{fig: COGA before share sale}
  \end{subfigure}%
  \begin{subfigure}{0.5\textwidth}
    \centering
    \includegraphics[width=\linewidth]{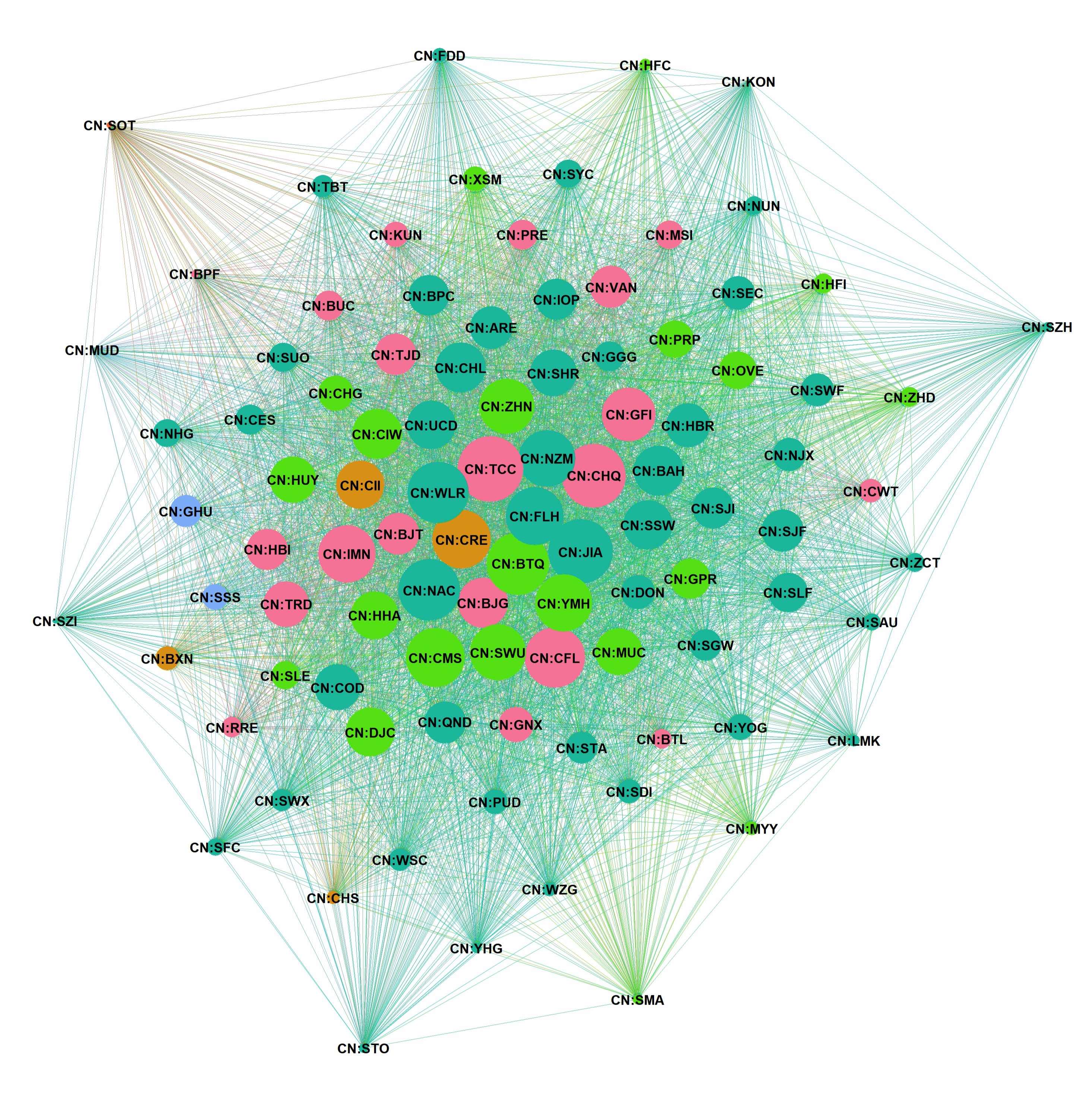}
    \subcaption{News of aborted share sale breaks (1 August 2023)}
    \label{fig: COGA after share sale}
  \end{subfigure}
  
  \begin{subfigure}{0.5\textwidth}
    \centering
    \includegraphics[width=\linewidth]{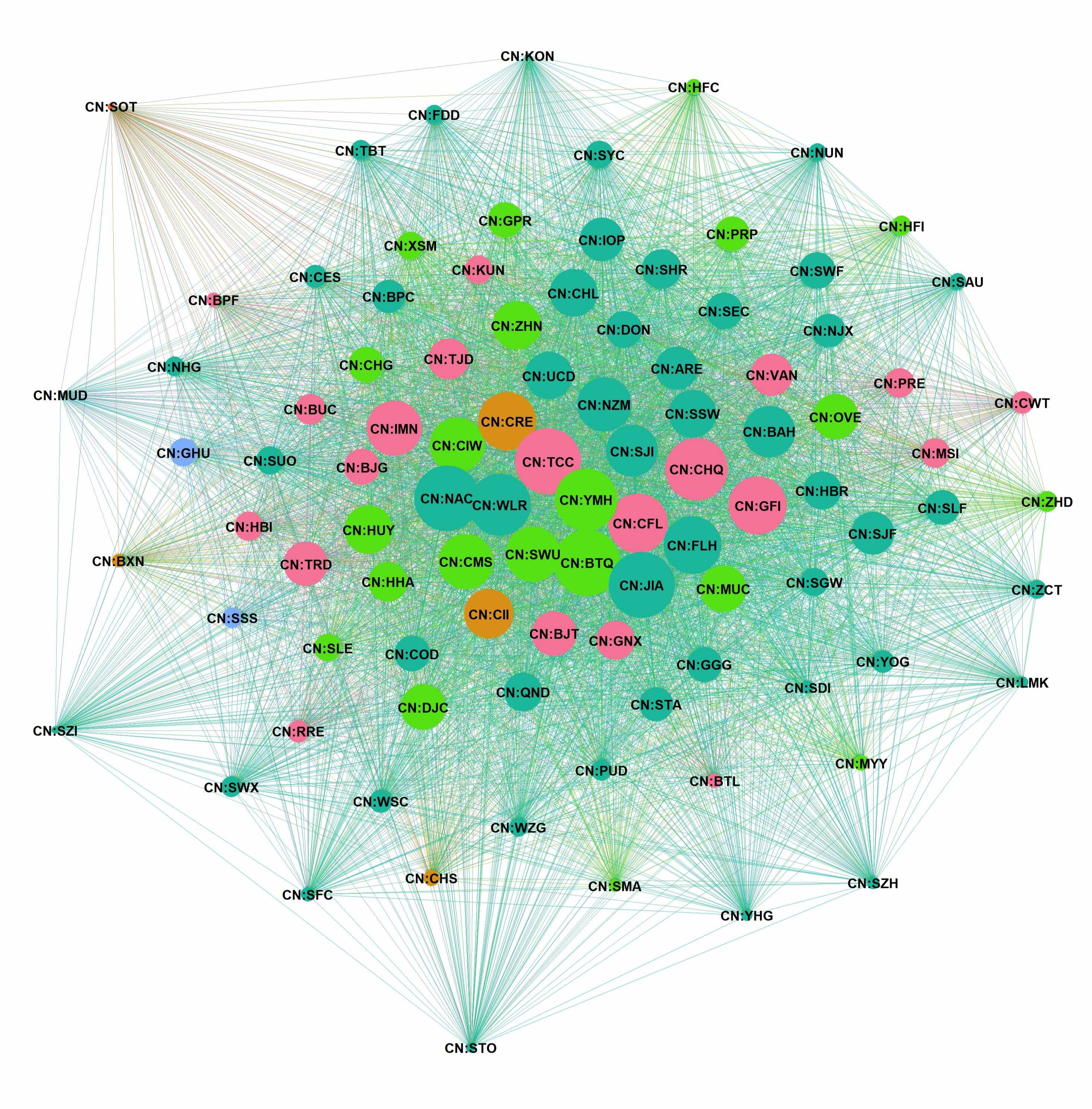}
    \subcaption{Day before profit warning (10 August 2023)}
    \label{fig: COGA before profit warning}
  \end{subfigure}%
  \begin{subfigure}{0.5\textwidth}
    \centering
    \includegraphics[width=\linewidth]{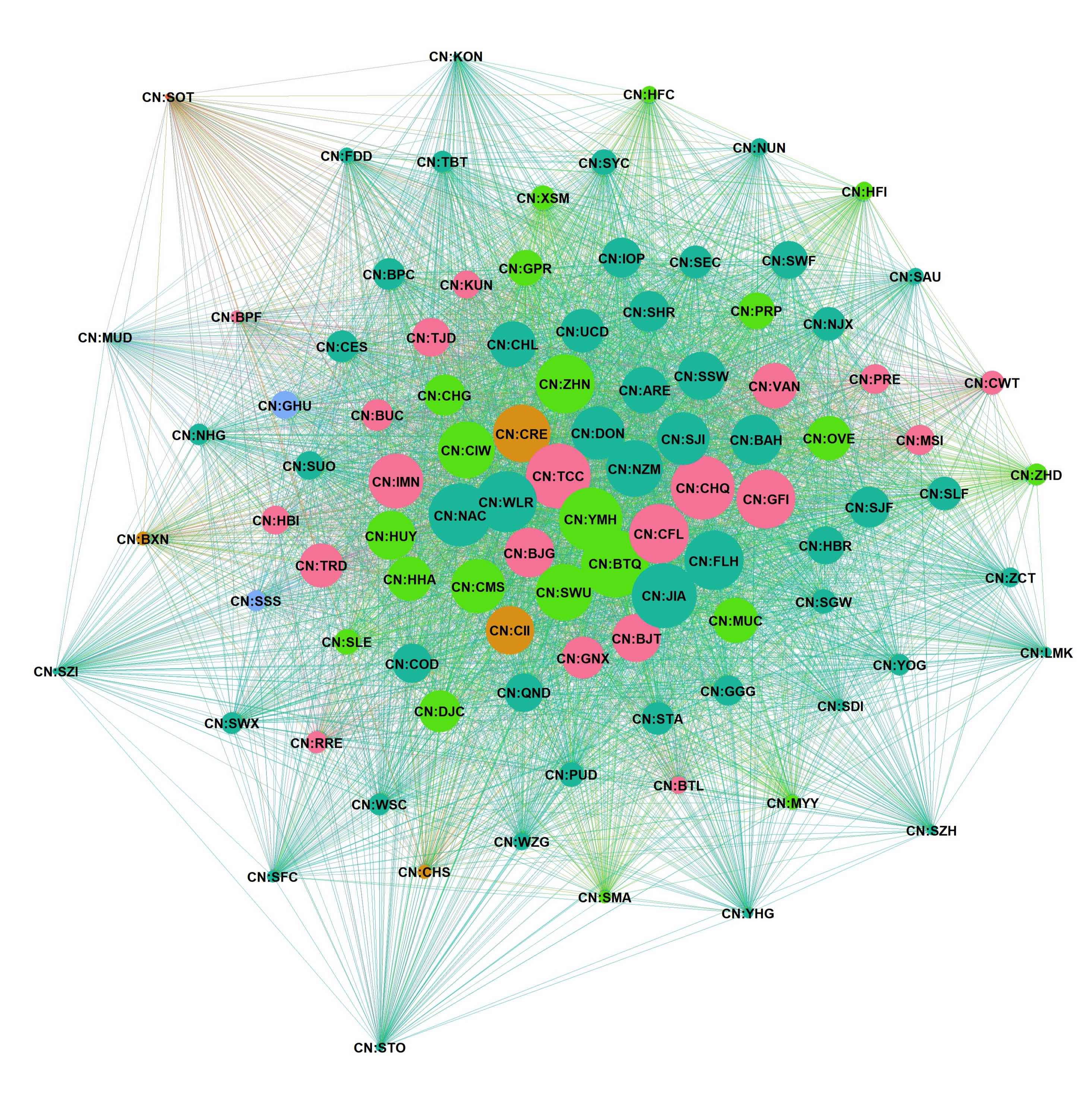}
    \subcaption{Day after profit warning (11 August 2023)}
    \label{fig: COGA after Profit warning}
  \end{subfigure}
    \caption*{\footnotesize Note: The upper left graph shows the business day before any Country Garden-related news circulates (31 July 2023), while the upper right graph shows the network change when the news of the aborted share sale breaks (1 August 2023). The bottom left shows the business day before the profit warning was issued (10 August 2023), and the bottom right shows the network the day after the profit warning was issued (11 August 2023). Node colors are determined by a developer's region of primary focus: pink is the north, green is the south, teal is the east, bronze is the southwest, light blue is the northwest, and red-orange is the northeast. Node size is determined by to connectedness.}
  \label{COGA}
\end{figure}

In Figure \ref{COGA}, I thus include 4 dates: the business day before any Country Garden-related news circulates (31 July 2023), the day the news of the aborted share sale breaks (1 August 2023), the business day before the profit warning was issued (10 August 2023), and the day after the profit warning was released (11 August 2023).\footnote{Note that 10 August is the ``before" period for the profit warning because Country Garden only issued it after close of business; the aborted share sale news broke early in the morning, so 1 August 2023 is the appropriate ``after" period for the share sale news.} 

As is apparent in Figure \ref{COGA}, the network has a larger reaction to the news of the aborted share sale than to the profit warning\textemdash an interesting result given that aborting the share sale does not in itself imply financial struggle. Indeed, Country Garden's rhetoric around the canceled share sale took great pains to emphasize that it was not the buyers who backed out, but Country Garden itself who prevented the deal (\cite{jim_country_2023-1}). The above plots indicate that market participants were not entirely convinced that the firm was free from trouble as it claimed. 

The effect on the network comparing Figure \ref{fig: COGA before share sale} and \ref{fig: COGA after share sale}, then, is more obvious than that between \ref{fig: COGA before profit warning} and \ref{fig: COGA after Profit warning} but, at the same time, is nowhere near the larger-scale changes seen in earlier plots like Figure \ref{Guangdong Gov Letter Network}. Indeed, with the news breaking out on 1 August, only 56/97 firms experience increases in to connectedness, and for all but two firms, this increase is very minor. Compared to other events such as Evergrande's letter to the Guangdong government and Kaisa's suspension, this is the first time that a southwestern-centric firm (bronze node) has been in the very core of the network. Indeed, two of these nodes are in the core in Figure \ref{fig: COGA before share sale} and get pushed out slightly between 31 July and 11 August. The core is thus extremely diverse; southern-focused, northern-focused, eastern-focused, and now southwestern-focused companies all occupy key positions in the core. While southern companies (light green nodes) seem to get pushed slightly away from the network's core with the news of the aborted share sale, nodes at the center are drawn closer as the core tightens, and some of the periphery nodes become drawn in as well, particularly on the bottom and the left of Figure \ref{fig: COGA after share sale}. Overall, though, the network is remarkably stable.

Now looking at the announcement of the profit warning, the stability of the network between periods is even more apparent between Figures \ref{fig: COGA before profit warning} and \ref{fig: COGA after Profit warning}; apart from some slight movement around the core, most nodes remain relatively close to their initial positions\textemdash even the periphery has minimal movement. In terms of to connectedness, only 44 firms experience an increase in to connectedness between 10 and 11 August, and the magnitudes of increase are relatively small across the board.

Both of these events were accompanied by sharp drops in Country Garden's share price, per the raw stock data, with the stock dropping to a ``record low" after the 10 August profit warning\textemdash making the network's stability that much more interesting (\cite{lim_country_2023}). Indeed, analyzing these two closely related events in tandem reveals an interesting picture: more so than in other periods, there is less concern about regional spillover and more emphasis on specific firm attributes. The firms who do experience increases in to connectedness and maintain positions in the network's core are a specific subset with no central geographic focus or SOE vs. POE status: they instead seem to be those with the most perceived exposure to Country Garden, like China Vanke (CN:VAN) and Dima Holdings (CN:COD). 

Indeed, one of the most critical things to observe here is that node movement and to connectedness changes are much more limited than in the previous events. This is likely because investors have been forced to become more reticent about their investments as the real estate crisis has dragged on: gone are the immediate, region-based, knee-jerk reactions to the three red lines (2020) or the emergency shock suspensions of developers like Kaisa (2021). Now, a full two years later, investors have examined company financials, performed their due diligence, and mapped expected exposure patterns. The result is that there is less sensitivity to ``shocking" news because the majority of market participants have already ``priced in" the information: at the first hint of concern, price-makers quickly divest from exposed developers, such that when major events occur\textemdash like the aborted share sale or the profit warning announcing an expected loss of over US\$6 billion\textemdash the network does not experience a quick and drastic change like before. This is even the case for major, market-making events such as the liquidation of Evergrande in January 2024 and the suspension of Country Garden in March 2024; I include plots of these in the Appendix (Section \ref{Appendix: additional plots}).

\begin{figure}[H]
  \caption{Network before and after Country Garden news: Color by state-ownership}
  \centering
  \begin{subfigure}{0.5\textwidth}
    \centering
    \includegraphics[width=\linewidth]{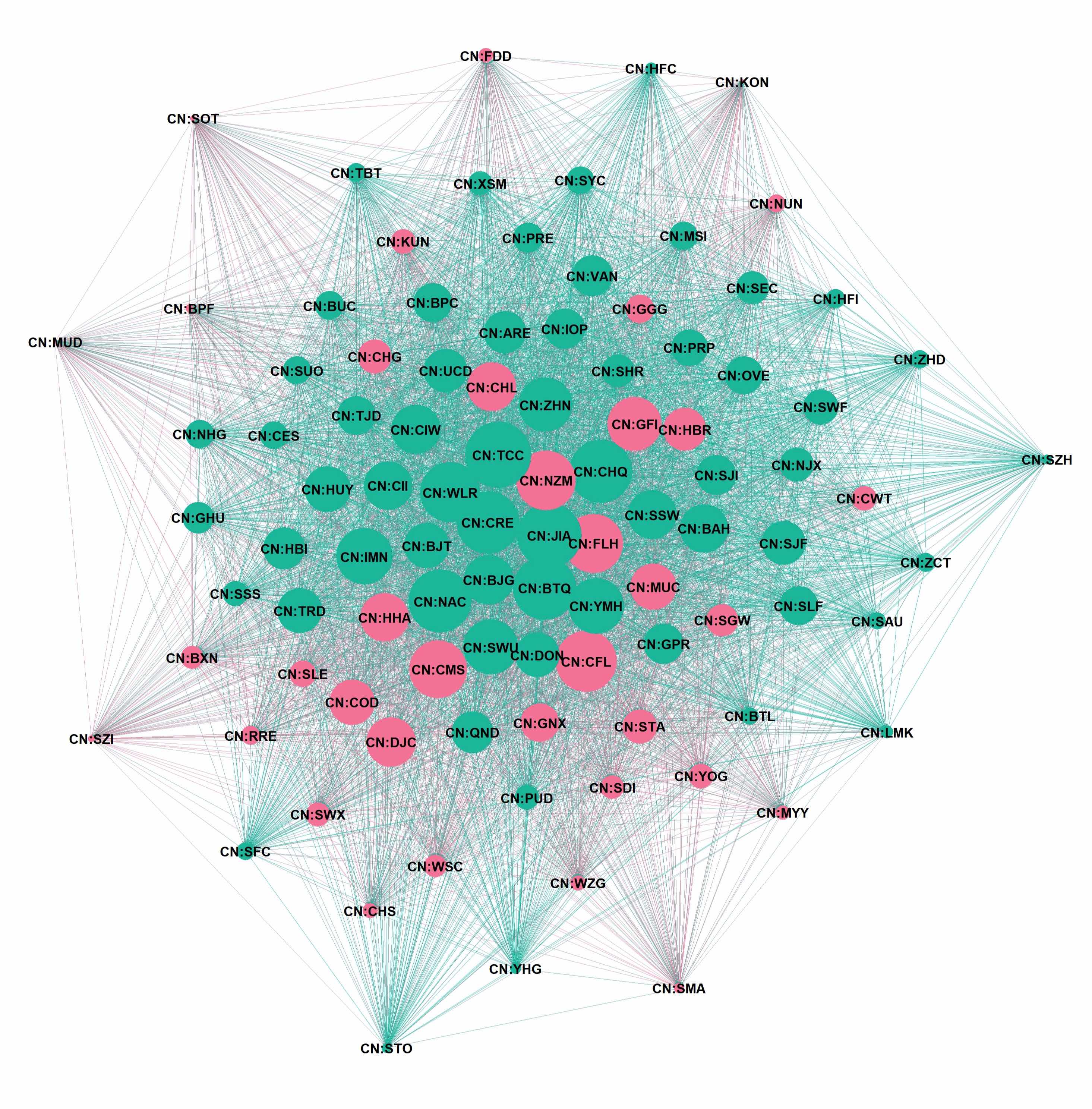}
    \subcaption{Before any news circulates (31 July 2023)}
    \label{fig: COGA SO before circ from}
  \end{subfigure}%
  \begin{subfigure}{0.5\textwidth}
    \centering
    \includegraphics[width=\linewidth]{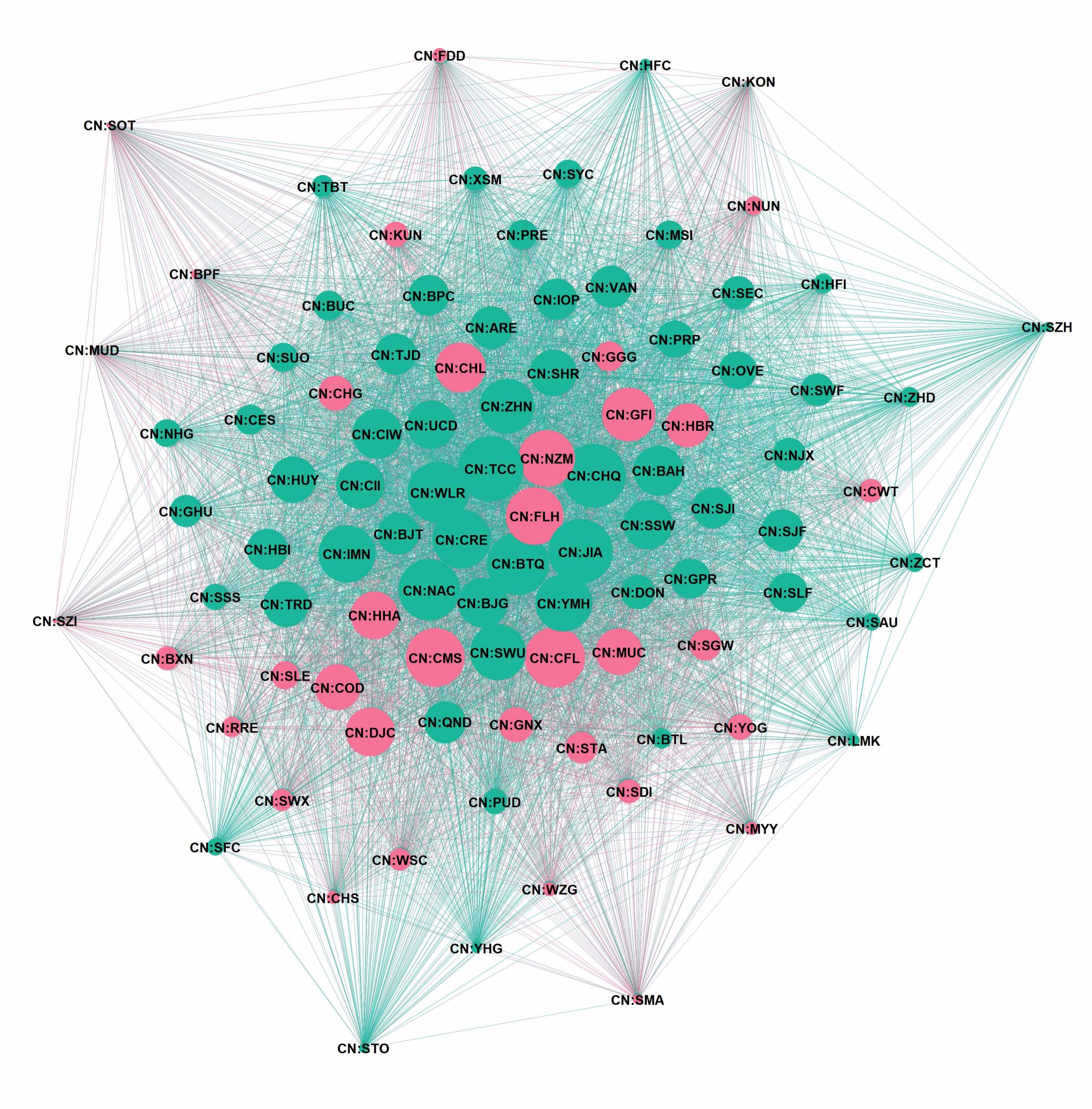}
    \subcaption{News of aborted share sale breaks (1 August 2023)}
    \label{fig: SO COGA share sale}
  \end{subfigure}
  
  \begin{subfigure}{0.5\textwidth}
    \centering
    \includegraphics[width=\linewidth]{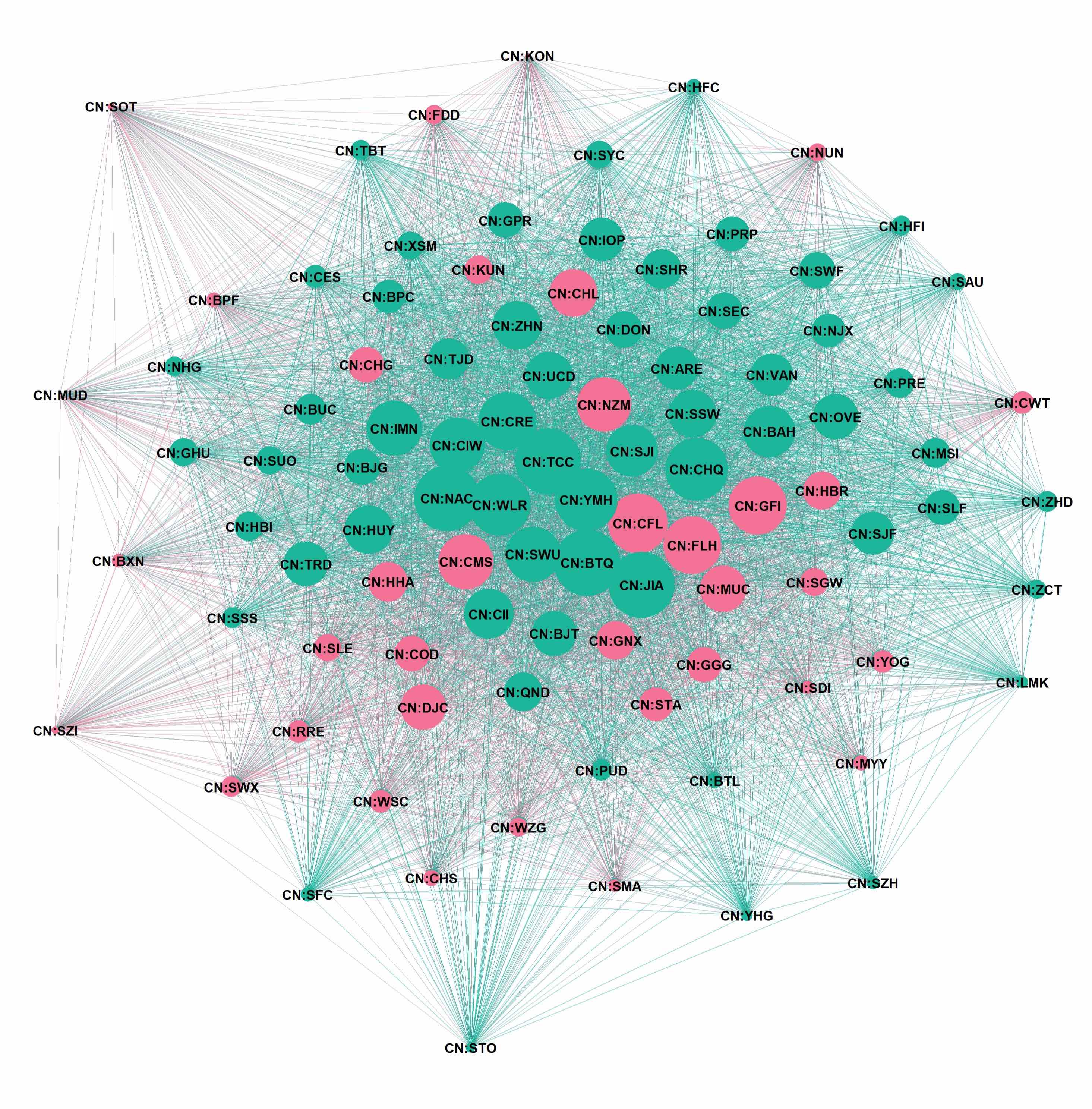}
    \subcaption{Day before profit warning (10 August 2023)}
    \label{fig: SO COGA before profit warning}
  \end{subfigure}%
  \begin{subfigure}{0.5\textwidth}
    \centering
    \includegraphics[width=\linewidth]{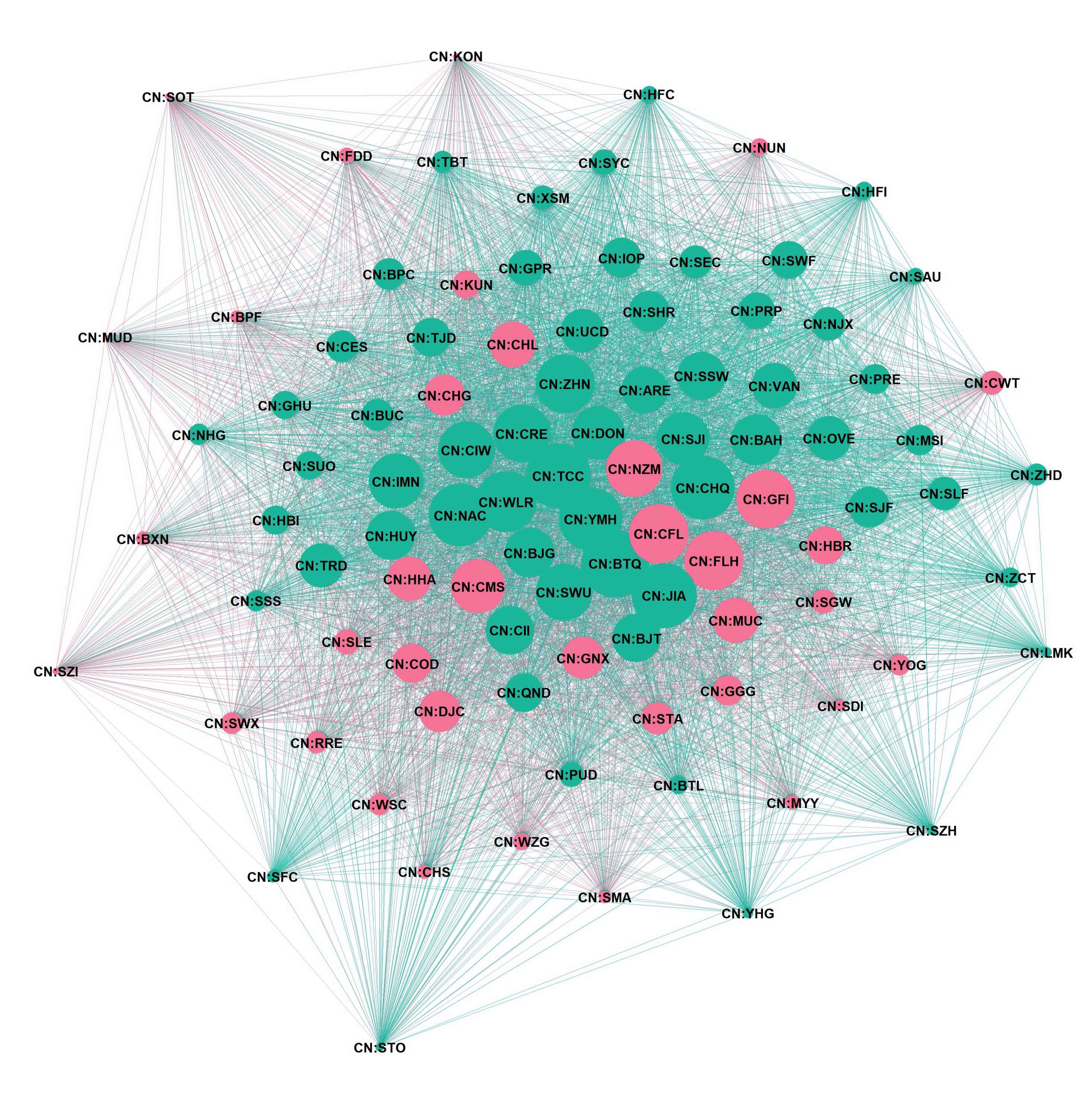}
    \subcaption{Day after profit warning (11 August 2023)}
    \label{fig:SO COGA Profit warning}
  \end{subfigure}
      \caption*{\footnotesize Note: The upper left graph shows the business day before any Country Garden-related news circulates (31 July 2023), while the upper right graph shows the network change when the news of the aborted share sale breaks (1 August 2023). The bottom left shows the business day before the profit warning was issued (10 August 2023), and the bottom right shows the network the day after the profit warning was circulated (11 August 2023). Colors are determined by the state ownership status: pink nodes are not state-owned while teal are state-owned. As before, node size is determined by to connectedness; larger node size thus implies a higher level of to connectedness.}
  \label{SO COGA}
\end{figure}

Examining state ownership status also provides insight into a fascinating dynamic occurring in Figures \ref{fig: COGA SO before circ from} and \ref{fig: SO COGA share sale}. On the surface, nearly all of the state-owned enterprises have initial positions more towards the core than the periphery. Indeed, in Figure \ref{fig: COGA SO before circ from}, there are few SOEs at the outskirts of the network compared to POEs. As time passes, several POEs tend to be pulled more towards the core of the network\textemdash particularly those on the bottom of the network, as is most apparent in Figure \ref{fig:SO COGA Profit warning}.

On the surface, this behavior suggests that private firms are seen as more exposed than state-owned ones, but the effect is not uniform. Instead, returning to the raw price data offers further insight. In fact, the substitution effect of the Kaisa suspension has reversed: in that case, the to connectedness measures and network layout suggested that private firms were seen as more stable than their state-owned counterparts, likely due to investor expectations about state-owned firms being compelled to enter into disadvantageous agreements with failing private developers. Now, there is prominent evidence of the opposite happening: when the first signs of trouble are apparent at Country Garden, there is a shift to state-owned developers. Of the companies on the periphery\textemdash whose position further away from the core suggests that they are more insulated from real estate shocks\textemdash the bulk of those who experience a price increase (7/10) are state-owned. 

In fact, this seems to be the prevailing sentiment among market participants, as 2023 land market data suggest a prominent shift toward state-owned developers: the top six Chinese developers of 2023 in terms of home sales all had state-backing, and many of the most prominent private developers slid down the rankings (\cite{jim_chinas_2024}). Some predictions even expect real estate troubles will last for the next ten years, positing that private developers will continually struggle to make debt payments as sales remain sluggish (\cite{ao_chinas_2024}). In the face of these recently emerging expectations, this substitution towards state-owned developers is largely expected.

Certainly, many more developers experience decreases in share value than increases between 31 July and 1 August, but the pattern is prominent among those who do experience increases in share value. The behavior again reflects how firms of a certain regional focus or state-ownership status are not homogeneous in the eyes of investors: the balance sheet trumps all, and discerning investors will work to uncover where exposure lies. 

\section{Conclusion} \label{sec: conclusion}
Offering evidence of an external shock to the network with the announcement of the three red lines, the 100-day rolling window estimation allows insight into the more nuanced behavior of speculators and investors as the real estate crisis progressed. Further, it confirms there is a basic level of connectedness that comes from being listed on a Chinese exchange and having real estate exposure. 

In one of the first major events of the real estate crisis\textemdash the release of Evergrande's letter to the Guangdong government\textemdash investors immediately expected stronger contractions in less developed regions. There is also evidence that state-owned firms are seen as slightly more insulated from shocks than their privately owned counterparts. As usual, the most diversified companies took up positions on the periphery, while those most exposed to real estate tended to be drawn towards the core. 

One year into the crisis, network behavior during the suspension of share trading for Kaisa offers updated insight into investors' beliefs: unlike before, eastern-centric firms are drawn most heavily into the core of the network, capturing how areas that were regarded as relatively stable earlier in the crisis are now increasingly at risk\textemdash it is not just the less developed regions of China feeling pressure on real estate, but the richest and most developed as well. The network also reflects a change in sentiment surrounding the relative stability of state-owned enterprises, exhibiting a fascinating substitution effect towards private developers. This behavior is likely reflective of the recent failure of prominent SOEs, as well as knowledge about how SOEs were forced to bail out private developers in disadvantageous deals. 

Then, fast-forwarding to nearly two years later, by the time of Country Garden's profit warning in August 2023, the network captures just how much investor sentiment has changed. Unlike in earlier periods, there is much less homogeneity across regions and state-ownership status, likely reflecting how developers have adapted to investing amidst the crisis. At this point, likely because they have delved into the financial statements of companies, conducted their due diligence, and mapped exposure patterns, companies behave in more nuanced ways: this means that when a ``shocking" event occurs, the network change is minimal because investors have already priced in their beliefs. Nevertheless, there is evidence of a substitution effect towards state-owned enterprises\textemdash the reverse of that seen under the Kaisa case\textemdash and this seems to reflect the market sentiment that state-owned developers are now ``safer" bets for investment.  

Taken together, these cases offer insight into how investor behavior and spillover patterns have changed as the real estate crisis evolved. No longer expecting spillover by region or state-owned status, investors actively conduct dynamic, nuanced market research. Moreover, with the VAR estimation appearing to be relatively consistent and robust across sequential rolling window specifications, the most significant implication of these findings and methodology is the insight they can offer active market participants\textemdash those investors themselves who have become reticent. If the relevant data could be pulled daily and the estimation recalculated for each real-time window, investors could have an up-to-date picture of firm connectedness and general market sentiment; it is certainly possible to determine firms ``less exposed" to the real estate crisis based on the network results and connectedness measures, as well as those most at risk of being net receivers of spillover. 

Future extensions of this paper will delve into more complex modifications to increase the robustness of estimation and circumvent network size limitations, but this specification applying \textcite{demirer_estimating_2018} in a new context still highlights the potential gains from analyzing connectedness\textemdash and just how much it can capture market sentiment. 

\pagebreak
\appendix
\section*{Appendix}
\global\long\def\thesection{A}%
\global\long\def\theequation{A.\arabic{equation}}%
\global\long\def\thetable{A\arabic{table}}%
\global\long\def\thefigure{A\arabic{figure}}%
\setcounter{table}{0}\setcounter{equation}{0}\setcounter{figure}{0}
\subsection{Dated timeline} \label{Appendix: Dated Timeline}
\begin{itemize}
    \item 31 August 2021- Evergrande warns it may default on its debt if it cannot raise cash in an interim earnings statement (\cite{hale_heavily_2021}).
    \item 22 September 2021- Developer Sunac bought back US\$34 million of its bonds and denied requesting assistance from the State. Despite claiming to meet two out of the three red lines at the beginning of the month, an internal ``draft" letter from Sunac surfaced online which revealed that recent government regulations controlling property prices had left some properties unable to break even, sending the stock tumbling (\cite{li_sunac_2021}). 
    \item 4 October 2021- Evergrande suspends trading of shares in Hong Kong (ultimately resuming trading on 21 October), citing a ``possible general offer," but did not initially make any announcements about the offer. It was reported that rival Hopson Development was set to buy a 51 percent stake in Evergrande's property services subsidiary for US\$5 billion (\cite{jim_timeline_2024}; \cite{westbrook_evergrande_2021}).
    \item 5 October 2021- Developer Fantasia missed a payment on a US\$206 million bond that matured the day before, triggering a default. While a relatively small developer with a market value of only \$415 million, Fantasia had reported ``no liquidity issues" just weeks prior (\cite{westbrook_evergrande_2021}; \cite{heng_chinese_2021}).
    \item 7 October 2021- The Shanghai-based developer Sinic Holdings warned in a filing to the Stock Exchange of Hong Kong that it was unlikely to be able to repay a US\$250 million bond due on 18 October, after missing earlier domestic payments in September. At the time, it had US\$694 million of outstanding bonds (\cite{huang_chinese_2021}). 
    \item 8 October 2021- US law firm Kirkland and Ellis and investment bank Moelis, who were hired by international bondholders leading up to Evergrande's failed interest payment, informed bondholders that they expected Evergrande's default to be imminent and that the company had failed to engage with them meaningfully (\cite{hale_advisers_2021}). 
    \item 11 October 2021- Beijing-based Developer Modern Land attempted to extend the maturity of a US\$250 million bond due later in the month, noting it may not be able to repay the bond in full (\cite{laforga_modern_2021}). 
    \item 15 October 2021- The Chinese government offered a rare (reportedly, the first) comment on the Evergrande situation, attributing it to poor management, imprudent business practices, and a blind diversification strategy. They stated that financial contagion was controllable and that financial institutions had limited exposure to Evergrande (\cite{galbraith_evergrande_2021}).
    \item 16 October 2021- The Chinese government was considering implementing a nationwide real estate tax in the belief it would bring down prices. However, there was strong pushback, with many officials arguing that it could cause a precipitous drop in consumer spending and harm the economy. The proposal was reportedly moving forward in a much more limited scope than originally planned, with President Xi Jinping writing in the 16 October ``issue of the party’s top theoretical journal, \textit{Qiushi}, `We should actively and steadily promote the legislation and reform of real-estate tax, and do a good job in the pilot work'" (\cite{wei_tackling_2021}).
    \item 19 October 2021-  Sinic defaulted on US\$246 million worth of bonds (\cite{hale_chinese_2021-2}). The same day, official figures showed real estate output in China was down 1.6\% in the third quarter year-on-year, the first time it has been negative since before the COVID-19 pandemic (\cite{liu_chinas_2021}).
    \item \sloppy 20 October 2021- Hopson Development announced that the possible deal with Evergrande\textemdash wherein Hopson would buy the 51\% stake in Evergrande's property service division\textemdash fell through. Evergrande applied to the Stock Exchange of Hong Kong to reopen trading on its shares (resuming 21 October). Except for a stake in a regional bank, as of this date, ``there has been no material progress on [the] sale of assets of the group" according to Evergrande. Shares dropped 13.6\% in response. On this same day, new data from the National Bureau of Statistics of China indicated that home prices had dropped month-on-month for the first time since April 2015, dropping in more than half of the 70 cities surveyed (\cite{hale_china_2021}). 
    \item 23 October 2021- A five-year trial of the proposed property tax was authorized for select regions with high real estate prices, likely Shenzhen, Hangzhou, and Hainan (\cite{noauthor_china_2021}).
    \item 25 October 2021- Modern Land defaults on the bond it had previously asked for an extension on, due to a cash crunch and credit downgrade (\cite{laforga_modern_2021}). 
    \item 5 November 2021- Kaisa suspends its stock due to cash flow concerns (\cite{zhu_kaisa_2021}). 
    \item 10 November 2021- After missing bond payments on nearly US\$280 million offshore bonds on 23 and 29 September and 11 October, Evergrande avoided default on the three bonds and made an interest payment within the 30-day grace period (\cite{noauthor_evergrande_2021}; \cite{dong_evergrande_2021}).
    \item 19 November 2021- It was announced that Evergrande would be removed from Hong Kong's Hang Seng China Enterprises Index. The index does not usually give reasons for delisting (and did not in Evergrande's case), but it is typically due to poor company performance (\cite{john_evergrande_2021}). 
    \item 24 November 2021- Kaisa resumes trading (\cite{zhu_kaisa_2021}). 
    \item 25 November 2021- Evergrande chairman and founder Xu Jiayin sells 9\% of his stake in the company for US\$344 million (\cite{jim_evergrande_2021}).  
    \item 7 December 2021- Evergrande officially defaulted for the first time, failing to make US\$82.5 million in interest payments that were due last month before the end of the 30-day grace period; the default could cause up to US\$19 billion in cross-defaults (\cite{jim_explainer_2021}; \cite{choong_wilkins_evergrande_2021}).   
    \item 8 December 2021- Trading in shares of Kaisa Group\textemdash the second-largest offshore debt holder among Chinese developers\textemdash was suspended after an anonymous source said that Kaisa would likely not meet a deadline for a US\$400 million offshore debt payment (\cite{roantree_trading_2021}). 
    \item 9 December 2021- Fitch downgrades Evergrande and Kaisa from ``C" to ``RD" (restricted default). Shares fell to their lowest level since Evergrande was first listed on the market. Evergrande, meanwhile, promised to negotiate a restructuring plan with overseas creditors (\cite{noauthor_fitch_2021}; \cite{noauthor_fitch_2021-1}; \cite{webb_china_2021}).
    \item 14 December 2021- News circulates of the developer Shimao's recent asset sales and cancelled apartment deals, sending bond prices tumbling. Shimao was one of China's top-ten developers during 2020 and was investment-grade rated until November (\cite{jones_shimao_2021}).
    \item 15 December 2021- Guangzhou R\&F seeks to extend the maturity of a US\$725 million offshore bond due in January by six months. The trouble came on the heels of rating downgrades catalyzed by the tightening market (\cite{galbraith_developer_2021}). 
    \item 17 December 2021- Credit rating agency S\&P declares Evergrande in selective default for its failed coupon payments (\cite{kingsbury_evergrande_2021}).
    \item 27 December 2021- Chinese officials announced that the Danzhou city government ordered Evergrande to demolish 39 buildings in one of its uncompleted projects. The project (Ocean Flower Island) was one of Evergrande's flagship developments, and the action thoroughly undermined investor confidence, with many considering it a signal of declining State support (\cite{yu_insight_2022}).
    \item 3 January 2022- Trading of Evergrande's shares was halted for the second half of the day pending an announcement ``containing inside information” (\cite{stevenson_china_2022}). 
    \item 4 January 2022- Evergrande, as its shares were set to resume trading, announced that the demolition order would not affect the rest of its project at the Ocean Flower Island. Shares rose almost 10\% as trading resumed; Evergrande noted that its 2021 sales were US\$70 billion, down 39\% from 2020 (\cite{sweney_evergrande_2022}). 
    \item 6 January 2022- Several large developers receive a notice from the government that ``the merger-and-acquisition loan could be excluded from their declared debt level," and they could further increase their debt level by 5 percent" (\cite{zhang_high_2024}; \cite{huld_explainer_2022}). 
    \item 11 January 2022- Shimao files a letter of clarification to the Stock Exchange of Hong Kong after news circulated that it defaulted on a trust loan payment of US\$101 million, causing S\&P to downgrade its credit rating to B-. The letter stated that it was not selling its flagship Shimao International Plaza in Shanghai and denied having missed any payment. It, however, was planning fire asset sales to make future payments (\cite{shimaogroupholdingsltd_voluntary_2022}). Only three months before, it was one of the few Chinese developers to meet all three red lines (\cite{farrer_chinese_2022}). Shimao also recently announced that it missed its lowered 2021 sales guidance by 7\%, with December’s sales tumbling 68\% year-over-year and by 25\% sequentially, according to data from the China Real Estate Information Corporation (\cite{chui_property_2022}). 
    \item 8 February 2022- China eased a year-long restriction ``on loans for the real-estate sector to fund public rental housing" (\cite{liu_chinas_2022}; \cite{zhu_china_2022-2}). This was just one of the policies it implemented to avoid a ``collapse in credit" and the resulting financial shock; others included an interest rate cut and a promise to ``open its monetary policy tool box wider" (\cite{liu_chinas_2021}). 
    \item 16 February 2022- Several Evergrande subsidiaries reported that assets worth over US\$150 million (including bank deposits and real estate) had been frozen due to Chinese court orders (\cite{jim_china_2022}).
    \item 18 February 2022- Local media report that banks in several cities across China have cut mortgage down-payments for some homebuyers in an effort to support the sector (\cite{winters_how_2022}). 
    \item 15 March 2022- Evergrande's stock sinks to a new all-time low of US\$0.15 per share (HK\$1.16) (\cite{noauthor_china_2024}). That same day, China's third largest developer Sunac is downgraded to a ``B-" credit rating from ``BB-" by Fitch Ratings due to the company's limited ability to access capital markets and falling contracted sales, coupled with hefty debt payments (nearly US\$4 billion) for 2022 (\cite{noauthor_fitch_2022}).
    \item 16 March 2022- The government announces its decision not to expand the property tax trial conducted in selected cities across China between late 2021 and early 2022 (\cite{winters_how_2022}).
    \item 22 March 2022- Evergrande suspends trading of its shares, citing both its inability to produce audited yearly results before the Stock Exchange of Hong Kong's 31 March deadline and the recently launched investigation into its property services unit. This investigation led banks to confiscate nearly US\$2 billion in deposits as the money was pledged as security for third party guarantees ``without the property services unit's knowledge" (\cite{farrer_evergrande_2022}; \cite{jim_timeline_2024}). 
    \item 1 April 2022- Developers China Aoyuan, Kaisa, Fantasia, Modern Land, and Sunac suspend their shares after failing to produce yearly audited results by the Stock Exchange of Hong Kong's 31 March deadline (\cite{yiu_six_2022}).
    \item 6 April 2022- News breaks that over 60 municipal governments across China have relaxed restrictions on home buying in the first quarter (\cite{winters_how_2022}). This is one of the behind-the-scenes steps the government has been taking in the spring of this year to quietly try to increase housing demand and ease the loan stress of developers (\cite{winters_how_2022}). 
    \item 11 May 2022- Sunac defaults on a US\$29.5 million coupon on a dollar bond (\cite{huang_major_2022}). 
    \item 24 June 2022- One of Evergrande's top investors, Top Shine Global, files a liquidation suit to the Hong Kong court on the basis of Evergrande's inability to repay debts of US\$110 million (\cite{laforga_evergrande_2022}). 
    \item 3 July 2022- Shimao defaults on its first debt payment, failing to repay a US\$1 billion bond with no grace period (\cite{he_chinas_2022}). 
    \item 11 July 2022- This is roughly the beginning of the popular mortgage boycott, wherein Chinese homebuyers started boycotting mortgage payments on properties that were not yet constructed or had poor construction. On 11 July, the number of properties boycotted was only 28, growing to 58 on 12 July, and then more than 100 in 50 cities on 13 July (\cite{frost_china_2022}). By the end of July, the number grew to 301 projects (\cite{choong_wilkins_xi_2022}). Data from the website ``WeNeedHome" (later moved to GitHub) showed that homebuyers were boycotting payments for 343 projects in over 100 cities in mid-September 2022 (\cite{noauthor_weneedhome_2022}). 
    \item 27 July 2022- The Chinese State Council approved a plan involving the People's Bank of China issuing roughly US\$148.2 billion in low-interest loans to State commercial banks to help ``refinance stalled real estate projects" (\cite{leng_chinas_2022}).
    \item 31 July 2022- Evergrande promised to have a preliminary debt restructuring plan by the end of July but failed. The deadline was pushed back to ``within 2022"  (\cite{zhu_china_2022-1}). It was also recently forced to sell shares in Shengjing Bank, which it had largely owned and used as a source of cheap loans over the years (\cite{chow_evergrande_2024}; \cite{zhu_china_2022-1}). 
    \item 6 September 2022- Developer Jiayuan International Holdings is served with a winding-up petition for failing to pay a US\$14.5 million debt (\cite{jiayuaninternationalgroupltd_announcement_2022}).
    \item 14 November 2022- Beijing stepped up its financial interventions in the crisis, issuing a 16-point plan to financial institutions with instructions to support struggling property developers. The plan detailed that lenders should support SOEs and POEs equally, prioritizing developers that focus on their core business and have sound governance\textemdash in other words, the plan was not a blanket bailout (\cite{noauthor_goldman_2022}). Indeed, leading investment firms took this as a signal that Evergrande and the other most over-leveraged firms would not be able to access the funds, which was reportedly the case (\cite{chow_evergrande_2024}; \cite{hale_restructure_2023}).
    \item 20 December 2022- Evergrande announces that it has resumed work on 631 pre-sold and previously undelivered projects (\cite{sridharan_china_2022}). 
    \item 6 January 2023- News circulated that the government would relax the three red lines, easing caps on developer borrowing and pushing back the grace period for complying with the policy's debt targets. Many viewed the action as an attempt to ``restore market confidence" (\cite{cai_china_2023}).
    \item 17 January 2023- PwC resigns as Evergrande's auditor, noting that it did not receive sufficient information on key matters from 2021 (particularly relating to an investigation into the company's electric vehicle unit) (\cite{leng_pwc_2023}). 
    \item 22 March 2023- Evergrande announced plans for the restructuring of its offshore debt, worth US\$22.7 billion, and offers holders could swap their debt into new bonds and equity-linked instruments backed by the group's listed subsidiaries (Evergrande Property Services Group and Evergrande New Energy Vehicle Group) (\cite{jim_china_2023}). 
    \item 3 April 2023- Evergrande and key bondholders reached an agreement, signing a restructuring deal that was said to pave the way for later agreements with other bondholders (\cite{huang_evergrande_2023}).
    \item 2 May 2023- Jiayuan International Group is liquidated by order of the Hong Kong court (\cite{ma_china_2023-1}). 
    \item 17 July 2023- Evergrande reports huge net losses: in 2021 they totaled US\$66.36 billion, and in 2022, they were US\$14.76 billion. In 2020, the group comparatively generated US\$1.13 billion in profits. The firm said the losses were caused by diminished land returns, the ``write-down of properties, loss on financial assets and finance costs" (\cite{jim_evergrandes_2023}).
    \item 10 August 2023- The largest private property developer in China at the time, Country Garden, warned of a large net loss for the first six months of 2023 due to ``impairment on property projects and declining profit margins" (\cite{choi_country_2023}). It expected a loss upwards of US\$6.25 billion, compared to a net profit in the same period in 2022. The company's shares fell up to 14.4\% to a record low on the day following the profit warning, closing below HK\$1 for the first time since it was listed. The actual loss for the first half year announced on 30 August 2023 was US\$6.7 billion (\cite{jim_country_2023}).
    \item 18 August 2023- Evergrande filed for Chapter 15 bankruptcy protection in a New York court to protect its assets in the U.S. while structuring a deal (\cite{jim_evergrande_2023}).
    \item 28 August 2023- Evergrande resumes trading after a 17-month suspension, tumbling by 79\% (equivalently, dropping in value by US\$2.2 billion) to a share price of US\$0.04 (\cite{jim_evergrande_2023-1}).
    \item 14 August 2023-  state-backed developer Sino-Ocean announced that it had missed almost US\$21 million in interest payments (\cite{monahan_chinese_2023}). On the same day, Country Garden sought to extend the maturity of a bond for the first time and suspended trading of 11 onshore notes it had issued (\cite{cai_country_2023}).
    \item 30 August 2023- Country Garden warned it was on the ``brink" of default in a filing to the Stock Exchange of Hong Kong, as it had ``failed to grasp and react to the risks of the ongoing real estate slump, most notably in smaller cities that are home to most of its developments" (\cite{kuo_chinas_2023}; \cite{countrygardenholdingscompanyltd_2023630_2023}).
    \item 28 September 2023- Evergrande's debt restructuring agreement with bondholders fell through after its chairman and founder Xu Jiayin was placed under ```mandatory measures' on suspicion of involvement in `illegal crimes'" (\cite{hale_evergrande_2023}). The company then suspended its stock again in the wake of Xu and other top Evergrande officials' arrests, attempting to renegotiate an agreement with its offshore bondholders to avoid liquidation (\cite{ward_evergrande_2023}). 
    \item 10 October 2023- Country Garden announced that it had failed to make a US\$60.04 million payment on debt by its due date. The company shared that it did not expect to be able to fulfill all its offshore payment obligations in the coming days, including its dollar-denominated bonds. The company's shares slumped over 10\% following the statement (\cite{murdoch_country_2023}). 
    \item 25 October 2023- Country Garden defaults on its dollar bonds for the first time, missing the grace period to pay US\$15.4 million in interest on a dollar bond that was due (\cite{ma_china_2023}).
    \item 30 October 2023- Evergrande was given another chance to negotiate with its offshore bondholders as the Hong Kong High Court adjourned its wind-up hearing to 4 December 2023 (\cite{jim_china_2023-1}).
    \item 4 December 2023- Although the previous extension was reported to be the last opportunity for Evergrande to reach an agreement with its debtors, Evergrande was given another extension to negotiate until 29 January 2024 (\cite{jim_china_2023-1}; \cite{jolly_chinas_2023}). 
    \item 5 January 2024- Chinese shadow banking giant Zhongzhi Enterprise Group Co. filed for bankruptcy; the firm had very strong ties to the real estate sector, with reportedly over half its assets being linked to real estate. Its downfall ``marks one of China's biggest-ever bankruptcies" and illustrates the stress the real estate downturn brings to the broader Chinese financial system (\cite{liu_troubled_2024}).
    \item 8 January 2024- Evergrande reveals the vice chairman of its electric vehicle subsidiary has been detained and is under ``criminal investigation" (\cite{jim_china_2024-1}).
    \item 16 January 2024- Ping An Bank\textemdash one of the major lenders in China\textemdash placed 41 real estate developers on a list as those eligible for funding support (\cite{cai_chinas_2024}). Over half were state-backed. These companies likely overlap heavily with the companies whose projects are on the ``whitelist"\textemdash State-designated properties eligible for funding (\cite{zhu_china_2024}).
    \item 29 January 2024- Unable to reach an agreement with its bondholders, Evergrande was liquidated by a Hong Kong court, permanently halting trading of its shares (\cite{jim_china_2024}). 
    \item 28 February 2024- A creditor (Ever Credit) filed a winding-up petition against Country Garden in an attempt to force a quicker restructuring. Country Garden had failed to repay a loan worth about US\$204 million  (\cite{hale_chinese_2024}). 
    \item 11 March 2024- China has asked banks to ``enhance financing support for state-backed [developer] China Vanke and called on creditors to consider [a] private debt maturity extension" in an unusual direct intervention from Beijing to aid struggling developers (\cite{jim_exclusive_2024}). Investors, meanwhile, had been dumping Vanke bonds and shares in the weeks prior to this announcement given concerns about the company's liquidity (\cite{jim_exclusive_2024}). 
    \item 11 March 2024- Moody's was the first ratings agency to withdraw Vanke's BAA3 rating, moving it to BA1 and marking it as ``review for downgrade" (\cite{liu_chinese_2024}). Fitch, meanwhile, only downgraded Vanke from ``BB+" to ``BB-" on 23 May (\cite{noauthor_fitch_2024}). 
    \item 13 March 2024- Country Garden missed its first yuan-denominated bond payment, on a bond of 96 million yuan (US\$13 million). The bond, however, had a 30 trading-day grace period that was set to expire in May (\cite{zhu_country_2024-1}). 
    \item 19 March 2024- Evergrande stated that the China Securities Regulatory Commission had found that the company overstated its revenue by US\$29.7 billion in 2019 and by US\$48.6 billion in 2020 since it recognized sales in advance. This amounts to overstating yearly sales by 63\% and 87\%, respectively (\cite{yu_evergrande_2024}). 
    \item 28 March 2024- Country Garden announces its expected suspension on 2 April after saying it would not publish its annual report by the Stock Exchange of Hong Kong's 29 March deadline (\cite{zhu_country_2024}). 
    \item 8 April 2024- State-owned bank China Construction Bank (Asia) filed a liquidation suit against developer Shimao over unpaid debts of US\$201.75 million. This was the first time a domestic bank, rather than overseas-based creditors, had been the party to instigate legal processes against a major Chinese developer (\cite{murdoch_chinas_2024}).
    \item 11 May 2024- Country Garden repays the domestic bonds it had previously defaulted on, five days before the end of the grace period (\cite{cash_chinas_2024}).
    \item 23 May 2024- Fitch downgrades Vanke from ``BB+" to ``BB-," citing weaker than expected sales year-to-date, despite strong government support (\cite{noauthor_fitch_2024}). 
\end{itemize}

\pagebreak
\subsection{Data cleaning procedures} \label{Appendix: Data Cleaning Procedures}
\indent I remove depository receipts. In the mainland Chinese and Hong Kong markets, there are three types of shares, ``A", ``B", and ``H." ``A" shares are shares of companies incorporated in the People's Republic of China and traded on the Shanghai Stock Exchange (SSE) or the Shenzhen Stock Exchange (SZSE) (\cite{richardson_guide_2016}). ``B" shares are secondary stocks of a listed company designed to attract foreign investment; they are traded on the SSE in US dollars (USD) or on the SZSE in Hong Kong dollars (HKD). Originally, only offshore investors or domestic investors in the secondary market with foreign currency accounts were able to invest in ``B" shares (\cite{richardson_guide_2016}). However, Stock Connect, which links the HKSE with the SSE and the HKSE with the SZSE, was launched in 2016 and effectively allows offshore investors to purchase ``A" shares through qualified brokers/intermediaries. ``B" shares have, in effect, lost much of their original function, causing many companies to shift their ``B"-share stock to the HKSE to become an ``H"-share stock denominated in HKD. For those that remain on the SSE or SZSE, ``B" shares often have lower market valuation and poorer liquidity than their ``A"-share counterparts, making them unattractive investment options for foreign investors (\cite{ma_huaxin_nodate}). 

In my analysis, I thus remove all ``B" share listings, keeping only ``A" share listings for each company. For instance, in the case of Shanghai Waigaoqiao Free Trade Zone group, which has two listed stocks on the SSE (CH600648 and CH900912) with one in USD and one in CNY, the USD denominated stock is excluded from analysis. Similarly, for those companies with active ``H" and ``A" share listings, I select the stock with the highest volume: in all cases, volumes in the primary currency were significantly larger (often at the ratio of 10:1). This means, for instance, that real estate companies with their corporate address in Hong Kong had higher volumes for their ``H" share listing than their ``A" share listing, while those headquartered in mainland China had higher ``A" share volumes than ``H" share volumes. In each case, I kept the primary. There were other unique cases; for instance, several companies have multiple branches that manage different facets of a group's property business. For example, Poly Property Group (HK0119) is the parent company of Poly Property Services (HK6049) and Poly Developments and Holdings Group (CH600048), with each performing distinct functions. In these cases, to avoid cutting distinct but connected companies from the data, I keep all entities included, but mark the parent/child relationship. I only give the parent/child relationship marker to those with truly dependent relationships (i.e., the ``child" is a legal subsidiary of the ``parent" or the majority owner of the ``parent" is also the majority owner of the ``child"). These filtering steps ensure that no company has multiple exchange listings for the same enterprise but at the same time allow for relationships between meaningfully separate companies to be marked and preserved for analysis. 

I also remove listed closed-end funds and bank-managed Real Estate Investment Trusts (REITs) of other listed real estate companies to ensure that the companies included reflect sovereign entities independently maximizing profit. I remove infrastructure public REITs, which are REITs established by the China Securities Regulatory Commission and the National Development and Reform Commission to focus on a single specific infrastructure project in a given area, such as an expressway or an industrial park (\cite{noauthor_first_2021}).

Ultimately, all stocks included in the data are either in Hong Kong dollars for those on the SEHK or in Chinese yuan (CNY) for those listed on the SSE or SZSE. I do not convert either to a common currency, given investors often make implicit decisions about currency value when buying stocks. 

I also drop all dates that are either a Hong Kong or mainland Chinese public holiday. This is by necessity, as the exchanges in mainland China (SSE and SZSE) are connected with Hong Kong's (SEHK) via the aforementioned Stock Connect, a ``mutual stock market access mechanism" allowing ``international and mainland Chinese investors to trade securities in each other's markets through the trading and clearing facilities of their home exchange" (\cite{noauthor_shanghai-hong_nodate}). There are two groups of Stock Connects: Northbound, wherein investors from Hong Kong and abroad invest in the SZSE and SSE, and Southbound, wherein investors from mainland China invest in the SEHK. While China and Hong Kong have shared cultural heritage and have many of the same holidays, the duration of the festivities is often different, both because Hong Kong has the autonomy to set its own public holiday schedule and because reforms of the public holiday system in mainland China have changed the duration of holidays (\cite{bao_trading_2023}). This means, for instance, that while National Day is just a 1-day holiday in Hong Kong, it is a 7-day holiday in mainland China. While the Hong Kong market would be open for the other trading days that are not public holidays, investors from mainland China would be unable to invest due to the closure of the Stock Connect. Prior to 2023, investors could buy and sell shares through Stock Connect on trading day T only when both the Hong Kong and mainland China markets are open for trading and banking services are available ``in both markets on the corresponding money settlement days" (T+1) (\cite{noauthor_hkex_2022}). These trading rules are designed to prevent day trading, as one cannot sell a share bought on day T before the money settlement day (T+1) (\emph{Stock Connect Northbound Trading Service}). 

It is important to note, though, that the Northbound and Southbound trading services are separate and behave differently: for instance, if there is a public holiday in Hong Kong but not mainland China on day T+1, on day T, the Northbound Stock Connect (wherein Hong Kong and international investors buy or sell shares in mainland China) would be closed, as the originator exchange (Hong Kong) would not be open on the money settlement day T+1 to facilitate the trade. The Southbound Stock Connect would be open on day T as the SZSE and SSE are both open on the money settlement day T+1. On the day of the holiday (T+1), both the Northbound and Southbound Stock Connects would be closed (\cite{noauthor_shanghai-hong_nodate}). In 2023, however, the Stock Connect policy was updated such that on days before a public holiday, money settlement is carried out on the evening of day T; this means that both Stock Connects would be open the day before a public holiday in either country (\cite{noauthor_stock_2023}).\footnote{Alternatively, if the money settlement is not made in the evening of the day T before the public holiday, the money is settled before 12:30 pm on the first trading day after the public holiday.} This kind of behavior makes for very complex dynamics and no static holiday schedule year-to-year. Thus, in my data handling, I drop trading days where the SEHK, SSE, or SZSE are closed (implying the Northbound and Southbound Connects would be closed) and any trades will only reflect the behavior of a small cross-section of market participants.\footnote{Importantly, volatility estimates for companies with their exchange closed would need to be imputed based on information from other days, presenting a statistical challenge that would introduce a large amount of bias. I thus refrain from this imputation and drop holidays as described.} I do include pre-holiday days where at least one of the Northbound or Southbound Stock Connects are open. This strikes an important balance between excluding days where only part of the market participates and including days wherein nearly all market participants are active and executing their pre-holiday trades. 

In general, I drop the observed holidays for the New Year (early January), Chinese New Year (late January or early February), the Ching Ming Festival (April), Labour Day (early May), the Tuen Ng Festival (June), Hong Kong Special Administrative Region (HKSAR) Establishment Day (early July), the Mid-Autumn Festival (September), National Day (October), and Christmas (Late December) (\cite{noauthor_trading_2024}). There are also a small number of days where an exchange had an unplanned closure, such as due to a typhoon (see \textcite{zhang_hong_2023} and \textcite{li_hong_2023}, for instance). I omit these dates as well given the Stock Connect would also be closed. 

I also briefly note that Stock Connect has a daily quota for Northbound trades (those on Chinese exchanges from Hong Kong) and Southbound trades (trades on the Hong Kong exchange from China). As of 4 July 2022, the Northbound daily quota is set at 52 billion CNY for each of Shanghai and Shenzhen, and the Southbound daily quota is set at 42 billion CNY for each of Shanghai and Shenzhen (\cite{noauthor_stock_2023}). Importantly, this is a buy-only quota; there is no limit on volume of sales, though they must, of course, be conducted when the exchange (and Stock Connect, if applicable) is open. 

After removing holidays and dropping ``B" shares, duplicates across ``A" and ``H" listings, and REITs among other unique cases, I remove some listings with very low frequency of trading. On the SEHK, SSE, and SZSE, there were several companies among the original real estate list with relatively low trading frequency, such that the stock went many days on end without changes in opening, closing, high, and low price. These were often stocks with values less than 1 USD. Unlike stocks that were suspended, these companies' stocks were available for trading but did not experience enough trades for the price to change. As I ultimately use the logarithm of daily high, low, opening, and closing price information to calculate volatility and subsequently take the logarithm of these volatilities in my VAR estimation, I cannot have days with volatility $\sigma^2$ = 0. Rather than imputing volatility or carrying down the last non-zero $\sigma^2$ value for several weeks in a row, I drop those companies with several periods of extended low frequency. For those with smaller gaps (from skipping 1-4 days to a rare 8- or 9- day gap), I carry down the most recent non-zero $\sigma^2$ value, as is standard in the literature and financial analysis; this method has further been proven to offer similar results to more complex generalized linear mixed model (GLMM) imputations (\cite{overall_last-observation-carried-forward_2009}). Ultimately, I strive to find a balance between keeping companies that have the majority of days with non-zero volatilities and dropping those with significant gaps due to no trading: I expect that some of the companies with some $\sigma^2$ = 0 values are often seen as more stable than their peers and/or less sensitive to the price movements of their peers, and dropping the most stable or insulated companies would likely increase the connectedness of the network, biasing the results. It is also the case that some of the companies have consistent daily returns in the beginning of the date range but have a sharp decrease in value and ultimately have days with $\sigma^2$ = 0 towards the end of the date range. With these drops in value and demand likely reflecting financial trouble, scandal, or mismanagement, these cases are likewise important to include in the sample to capture network connectedness. I ultimately find this methodology allows me to drop companies with massive gaps but at the same time keep ones with occasional gaps in the sample, reflecting the true Chinese real estate network as much as possible. 

It is also important to note that mainland Chinese exchanges have minimum trading volume requirements for ``A" share stocks. The Shanghai exchange minimum, for instance, is 3.75 million yuan over 90 consecutive trading days; stocks trading below 1 yuan for 10 consecutive days are at risk of being delisted by the exchange (\cite{noauthor_rules_2023}). This is helpful as well because it ensures that there is a minimum level of transaction volume that inherently limits the amount of imputation needed. 

\pagebreak

\subsection{Static window adaptive elastic net results} \label{Appendix: Static Window Adaptive Elastic Net Results}

\begin{figure}[H]
  \caption{Static estimation before and after news of the three red lines: Color by region}
  \centering
  \begin{subfigure}{0.5\textwidth}
    \centering
    \includegraphics[width=\linewidth]{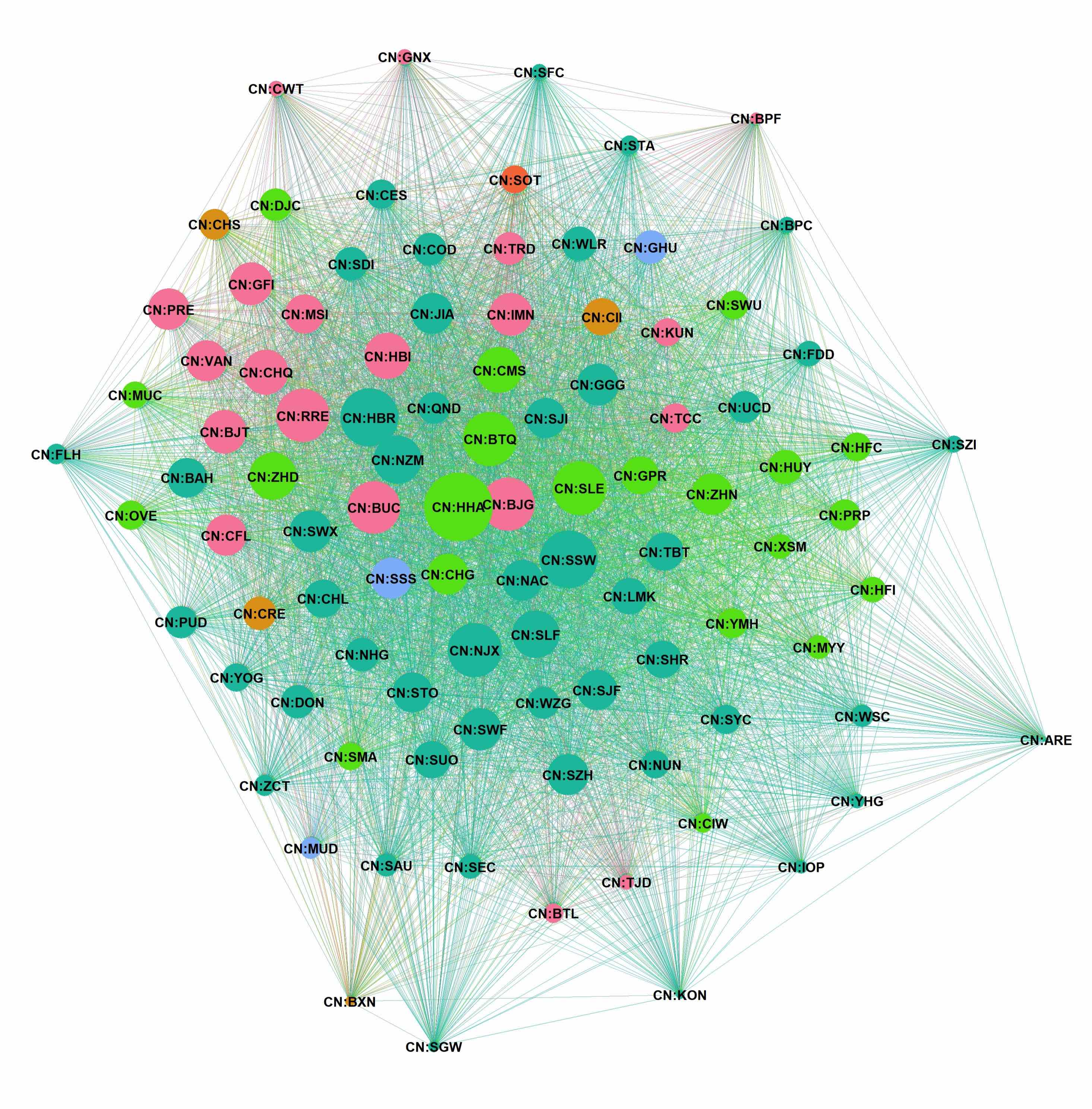}
    \subcaption{2 January 2019 to 13 August 2020}
    \label{fig: static before}
  \end{subfigure}%
  \begin{subfigure}{0.5\textwidth}
    \centering
    \includegraphics[width=\linewidth]{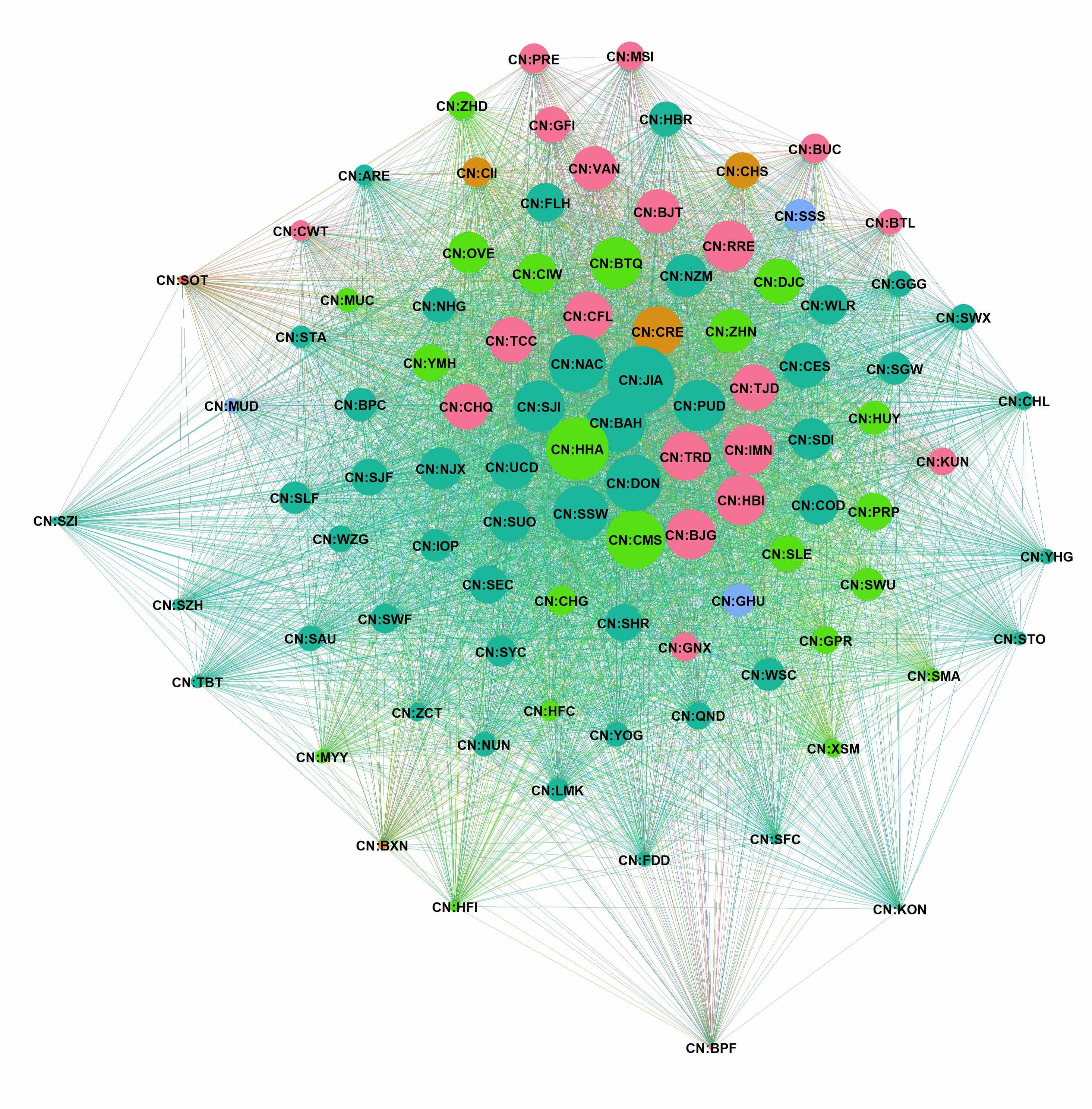}
    \subcaption{14 August 2020 to 19 April 2024}
    \label{fig: static after}
    \end{subfigure}
  \label{static fig}
  \caption*{\footnotesize Note: The left graph shows the static network estimation from 2 January 2019 to 13 August 2020, while the right static network estimation for 14 August 2020 to 19 April 2024. The colors are determined by the region where the developer is primarily focused: pink is the north, light green is the south, teal is the east, bronze is the southwest, light blue is the northwest, and red-orange (of which there is only one node, CN:SOT) is the northeast. Node size is determined by to connectedness; larger node size thus implies a higher level of to connectedness for the node.}
\end{figure}

As described in Section \ref{sec: methodology}, I include the static adaptive elastic net estimation for the three red lines event. Specifically, Figure \ref{fig: static before} shows the static network estimation from 2 January 2019 to 13 August 2020, while Figure \ref{fig: static after} illustrates the static network estimation for 14 August 2020 to 19 April 2024. Given the larger window sizes, the adaptive elastic net can now be used instead of the elastic net. 

As is apparent, the regional clustering effect is much stronger, as nodes are separated largely by color. In Figure \ref{fig: static before}, the bulk of the northern-centric firms (pink nodes) are on the upper left, while the bulk of the eastern-centric firms (teal nodes) are on the bottom half of the network. In Figure \ref{fig: static after}, the northern-centric firms have moved almost exclusively to the top right, while the majority of the eastern-focused companies occupy the bottom left of the network. This behavior suggests that over longer timescales, the pairwise directional connectedness between firms within a region are generally stronger, likely reflecting strong price comovement. This is particularly true for northern- and eastern-focused firms, but less so for southern-, northwestern-, and southwestern-focused ones, who experience some regional clustering but tend to be interspersed throughout the network. Further, Figure \ref{fig: static after} certainly shows the network is more stressed in the later period (as expected), given its high clustering and the relatively larger degrees of to connectedness for nodes in the core. 

However, it is difficult to declare the effect as deriving from the three red lines given the date range is so large, and the data are completely non-overlapping, unlike that for the rolling window estimations (wherein the majority of dates carry over between sequential windows). These plots thus allow more for a comparison of the ``before" vs. ``after" periods, rather than showing specific information about an event. 

\pagebreak

\subsection{Visualizing the network} \label{appendix: visualizing the network}

\subsubsection{ForceAtlas2 algorithm}

To calculate the speed, it is first critical to understand the two key metrics that shape it: irregular movement (``swinging") and useful movement (``effective traction"). Swinging is defined as the ``divergence between the force applied to [a node] $n$ at a given step and the force applied to $n$ at the previous step" (Jacomy et al. 2014, p. 7). $swg_{(t)}(n)$\textemdash the swinging of node $n$ at time $t$\textemdash is thus calculated as the difference between the net force at time $t$ applied to node $n$ ($F_{(t)}(n)$) and the net force applied to node $n$ at time $t-1$ $(F_{(t-1)}(n))$:
\begin{align}
    swg_{(t)}(n) = |F_{(t)}(n) - F_{(t-1)}(n)|.
    \label{swinging eq}
\end{align}
As Jacomy et al. highlight, for a node moving toward its balancing position, $swg(n)$ is close to 0, but a node that experiences forces significantly different from the previous period has a high swinging value. 

Subsequently, the global swinging value $swg(G)$ can be calculated by summing the local swinging values, weighted by the degree of each node; like in equation (\ref{F_r eq}), Jacomy et al. add 1 to each node's degree to account for nodes of 0 degree. Global swinging is thus $swg(G) = \sum_n (deg(n) + 1)swg(n)$. 

On the other hand, the effective traction $tra(n)$ of a node is the amount of ``useful force" applied to that node\textemdash that is, forces that contribute to the node's convergence. It is calculated as the average of the net force applied to node $n$ in time $t$ and that applied in the previous period ($t-1$): it is therefore 
\begin{align}
    tra_{(t)}(n) = \frac{|F_{(t)}(n) + F_{(t-1)}(n)|}{2}.
    \label{traction eq}
\end{align}
This formula means that nodes who keep their course have $tra(n) = F(n)$ and those that revert to their former positions (a perfect swing) have $tra(n) = 0$. Global effective traction $tra(G)$ is the sum of local effective traction values, weighted by the degree of each node, again adding 1 to each degree in order to account for possible nodes of degree 0. It is therefore calculated as $tra(G) =  \sum_n (deg(n) + 1)tra(n)$. Global speed $s(G)$ ``keeps the global swinging $swg(G)$ under a certain ratio $\tau$ of the global effective traction $tra(G)$" (Jacomy et al. 2014, p. 9). It is thus defined as 
\begin{align}
   s(G) = \tau\frac{tra(G)}{swg(G)}. 
    \label{global speed eq}
\end{align} 
$\tau$, the tolerance to swinging, can be set by the user, but Gephi also includes an algorithm to calculate it based on the density of the network and the value in previous iterations. The speed of each node is then calculated as: 
\begin{align}
    s(n) = \frac{k_ss(G)}{(1+s(G)\sqrt{swg(n)})},
\end{align}
where $k_s$ is a constant set to 1, unless the user requires a mode (like no node overlap) which sets it to 0.1 in Gephi's implementation. As \textcite{jacomy_forceatlas2_2014} describe, the logic of having a local speed for each node $s(n)$ rather than just having a global speed $s(G)$ is rooted in providing more precision for nodes that struggle to converge. The global speed is as high as possible while being limited by the tolerance $\tau$; the local speed, on the other hand, can slow nodes down but cannot speed them up. Putting this behavior together means that the ``local speed regulates the swinging while the global speed [indirectly] regulates the convergence" (Jacomy et al. 2014, p. 9). 

Ultimately, the more a node swings, the slower it moves, while nodes with very little swinging move at rates near the global speed. Finally, with the speed $s(n)$ calculated, the displacement $\Delta(n)$ can be computed per equation (\ref{displacement eq}). Final node position $p$ in each iteration $r$ is thus $p_r(n) = p_{r-1}(n) + \Delta_r(n)$, and it is ultimately a reflection of the forces acting on the node and the speed at which each node moves, with the network gradually converging toward a stable position across iterations for each time period. In my package, the built-in value is \verb|iterations = 100|, although I do more in my own calculations, as described in \ref{Contextualizing these Formulas in the VAR}.

Importantly, one of the key modifications of my algorithm from the Gephi original is that it allows layout continuity across periods: while Gephi randomly assigns each node a position for the first iteration, my R function allows the beginning position for each node to be determined by a layout that has already been calculated. In practice, this means the previous day's layout can be passed in as the initial position, so it is possible to directly observe the movement of nodes across the network continuously from day-to-day. This behavior does not have much effect on any of the calculations described above but does have the added benefit of stabilization across periods for nodes who do not experience significant changes in $F_r$ and $F_a$\textemdash and facilitates the detection of those who do have large changes in position.

\subsubsection{Contextualizing these formulas in the VAR} \label{Contextualizing these Formulas in the VAR}

In my network, as all nodes are fully connected, these formulas simplify slightly. For repulsion, equation (\ref{F_r eq}) essentially simplifies to a constant over the distance between two nodes, as the numerator is the same for all cases (given all nodes have the same degree): $F_r = C/d(n_1, n_2)$, where $C = S(deg(n_1)+1) (deg(n_2)+1) = 10(98)(98) = 96040$. 
As I calculate the sequence of rolling windows for the entire date range under examination (2 January 2019 to 29 March 2024), for each date in the rolling window estimation, the initial position for the first iteration ($p_1$) is the final layout from the previous day. That is, going into the first iteration on day $t$, $p_{1, t}(n_1)$ = $p_{1, t-1}(n_1)$.\footnote{Note that the initial positions on the first day 2 January 2019 are randomly assigned a 1000x1000 region of the coordinate plane. All other periods have the previous day's final layout passed in as the initial position in the first iteration such that $p_{1, t}(n)$ = $p_{1, t-1}(n)$.} In each sequential iteration $r$ on a given day $t$, the position is then $p_{r, t}(n)$ = $p_{r-1, t}(n)$ + $\Delta(n)$. In each iteration $r$, the distance between two nodes $d(n_1, n_2)$ is calculated by the Euclidean distance formula ($d = \sqrt{(n_{2x} - n_{1x})^2 + (n_{2y} - n_{1y})^2}$, where the $x$ and $y$ subscripts reflect the $x$ and $y$ coordinates, respectively). 

\begin{figure}[H]
\caption{Pairwise directional connectedness as edge weight diagram}  \centering
    \centering
    \includegraphics[width=0.32\linewidth]{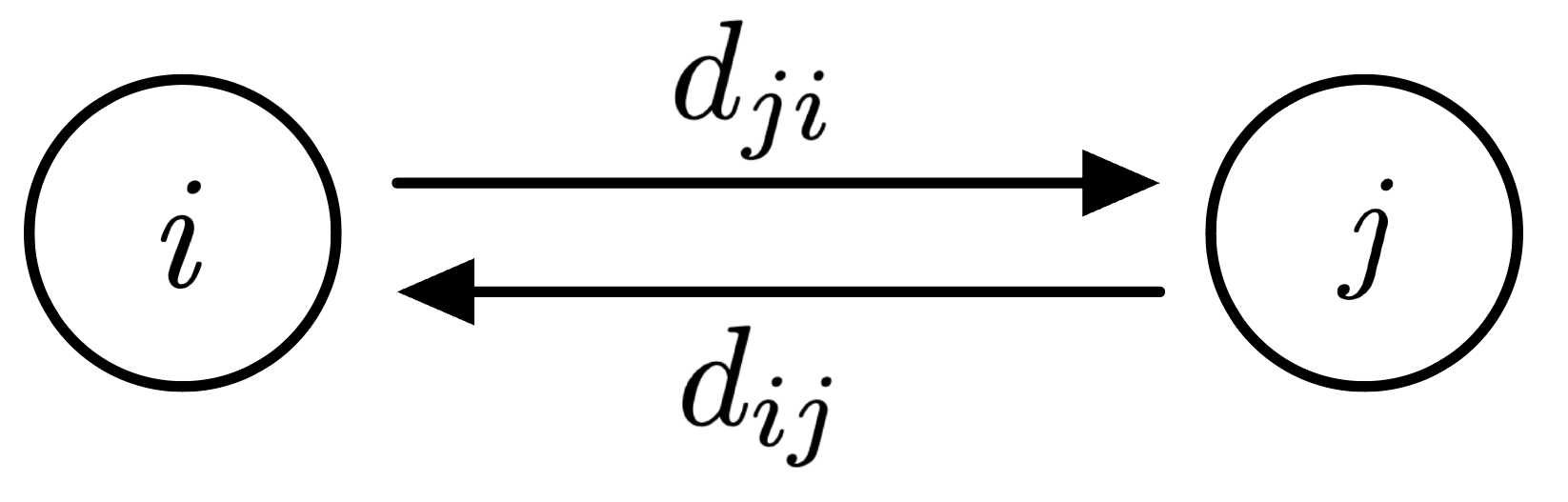}
    \label{fig: pairwise diagram}
\end{figure}

For the attraction formula, equation (\ref{attraction formula}), the distance between two nodes ($d(n_1, n_2)$) is also included, this time in the numerator. The distance is multiplied by the edge weights $w(e)^\delta$, and because I include edge weights in my attraction formula (as is standard), $\delta = 1$. In my specification, following \textcite{demirer_estimating_2018}, the edge weights are the pairwise directional connectedness measures; that is, using the terminology and notation of equation (\ref{pairwise connectedness formula}), the edge weight from firm $j$ to firm $i$ is $d_{ij}$ from matrix $D^H$. Each of these edges weights are directional, such that the edge weight from firm $j$ to firm $i$ ($d_{ij}$) is not equal to the edge weight from firm $i$ to firm $j$ ($d_{ji}$): that is, $d_{ij} \neq d_{ji}$. My edge weight matrix thus has $N^2 - N$ entries, as the diagonal entries of $D^H$ are excluded. When ForceAtlas2 is called, it keeps the directionality of the edges, meaning that it computes separate $F_a$ and $F_r$ calculations for each edge between firms $i$ and $j$ separately.\footnote{Note that the $F_r$ calculation result would not change since $d(n_j, n_i)$ would be the same for each of these two edges.} Figure \ref{fig: pairwise diagram} illustrates the two edges between nodes $i$ and $j$, as well as the edge weights between them. 

Because all non-diagonal values in the $D^H$ matrix are preserved as edge weights, it is therefore possible to have the following features at the same time: 1) the network can have higher attractive forces (meaning higher edge weights), which also causes nodes to pull closer together, while 2) ``to" and ``from" connectedness decrease on average. This behavior is highly dependent on the distribution of higher $d_{ij}$ values: going back to the connectedness table (Table \ref{tab:connectedness_gen}), ``to" connectedness for a firm $j$ is computed by summing the $j$-th column (excluding when $i =j$) while ``from" connectedness is calculated for a firm $j$ by summing the $j$-th row (excluding when $i =j$). It is therefore possible that a subset of $d_{ij}$ values in each column increase while the total column sum decreases. If the magnitude of increase for these selected $d_{ij}$ values is high, $F_a$ would be much stronger for these edges. In turn, even if $d_{ji}$ experiences a slight decrease, if the magnitude of $d_{ij}$ is greater, the two nodes will ultimately pull closer together. If this behavior repeats across the matrix, with either $d_{ij}$ or $d_{ji}$ increasing while the other remains roughly constant or drops slightly, the network can become more dense as attraction between a node pair increases, becoming tighter. At the same time, ``to" and ``from" connectedness can decrease on average because they depend on the overall distribution of influence patterns in the network: some node pairs can have lower edge weights while others increase. Then, since both to and from connectedness are the results of normalized values\textemdash the $\theta_{ij}$ of the generalized forecast error variance decomposition is normalized to then become $d_{ij}$\textemdash this normalization process can lead to a decrease in to connectedness when there is asymmetry in weights across the network. It need not be every mode pair that experiences greater attraction for the network to pull tighter, so this behavior is entirely possible, though initially counterintuitive. 

Then, as detailed in Section \ref{visualizing the network}, $F_a$ and $F_r$ for each node are summed, and displacement is subsequently calculated, along with speed $s(n)$. I include an example of how speed is calculated in the toy example below, clarifying what swinging and traction can look like in practice. For each day in my analysis, I set \verb|iterations = 600| to ensure a stable position is reached in every period, but the bulk of adjustments occur within the first 100 iterations.

\subsubsection{A simple example} \label{simple toy example}

\begin{figure}[H]
  \caption{Layout plots of a simple example: r = 0, r = 1, r = 39, and r = 80}  \centering
  \begin{subfigure}{0.25\textwidth}
    \centering
    \includegraphics[width=\linewidth]{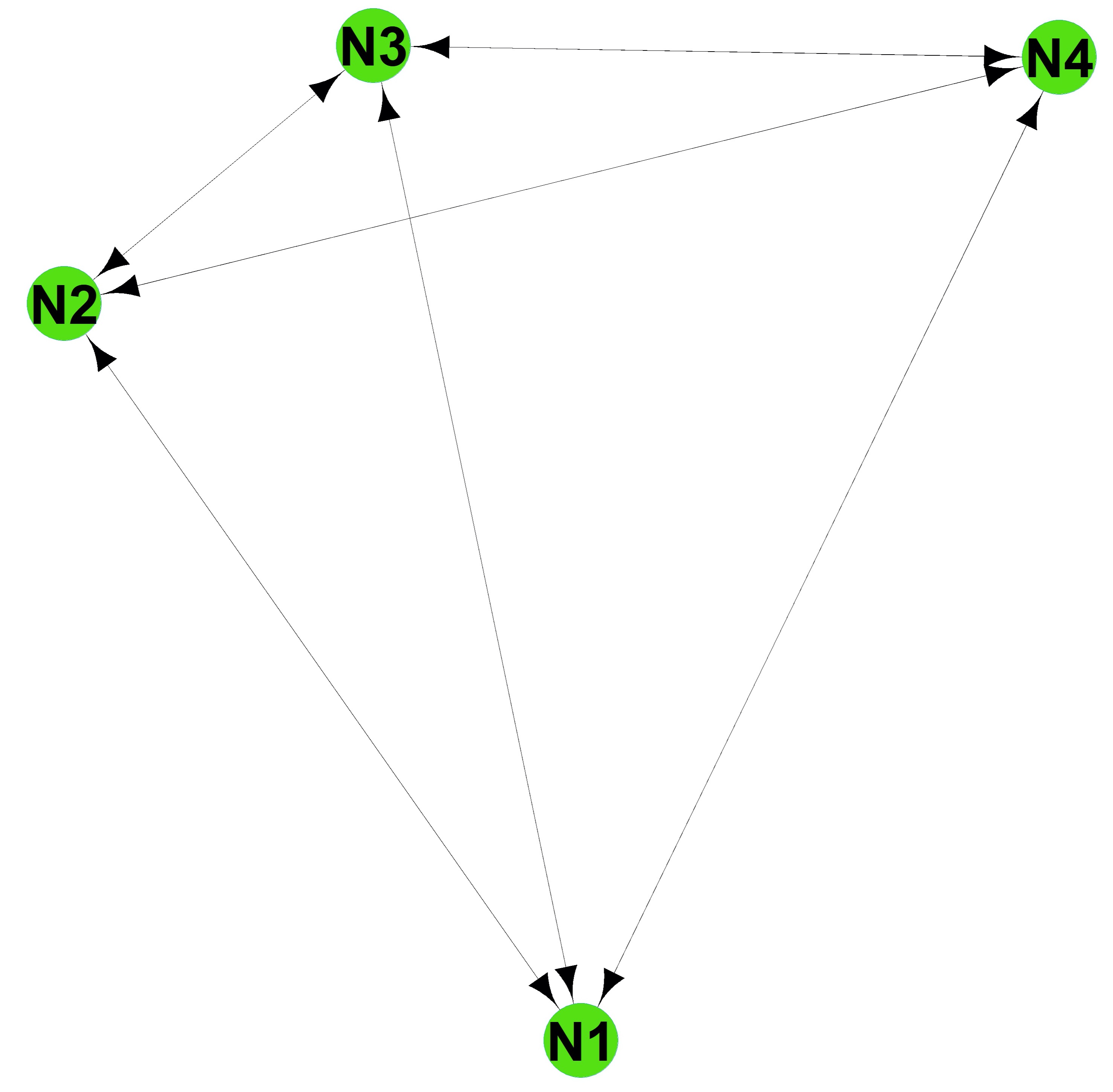}
    \subcaption{Initial Position (r = 0)}
    \label{minigraph t=0}
  \end{subfigure}%
  \begin{subfigure}{0.25\textwidth}
    \centering
    \includegraphics[width=\linewidth]{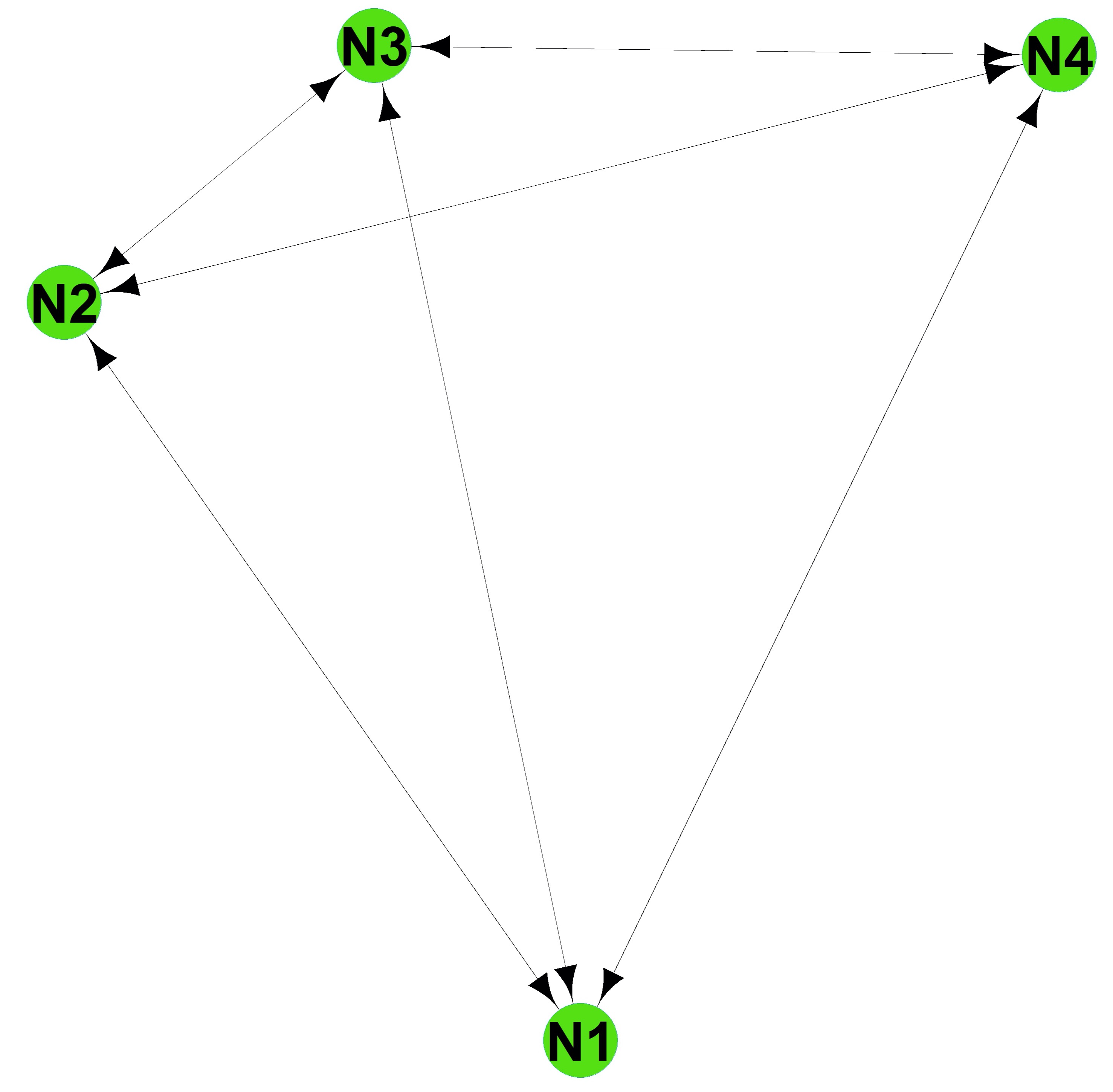}
    \subcaption{1 Iteration (r = 1)}
    \label{minigraph t=1}
    \end{subfigure}%
    \begin{subfigure}{0.25\textwidth}
    \centering
    \includegraphics[width=\linewidth]{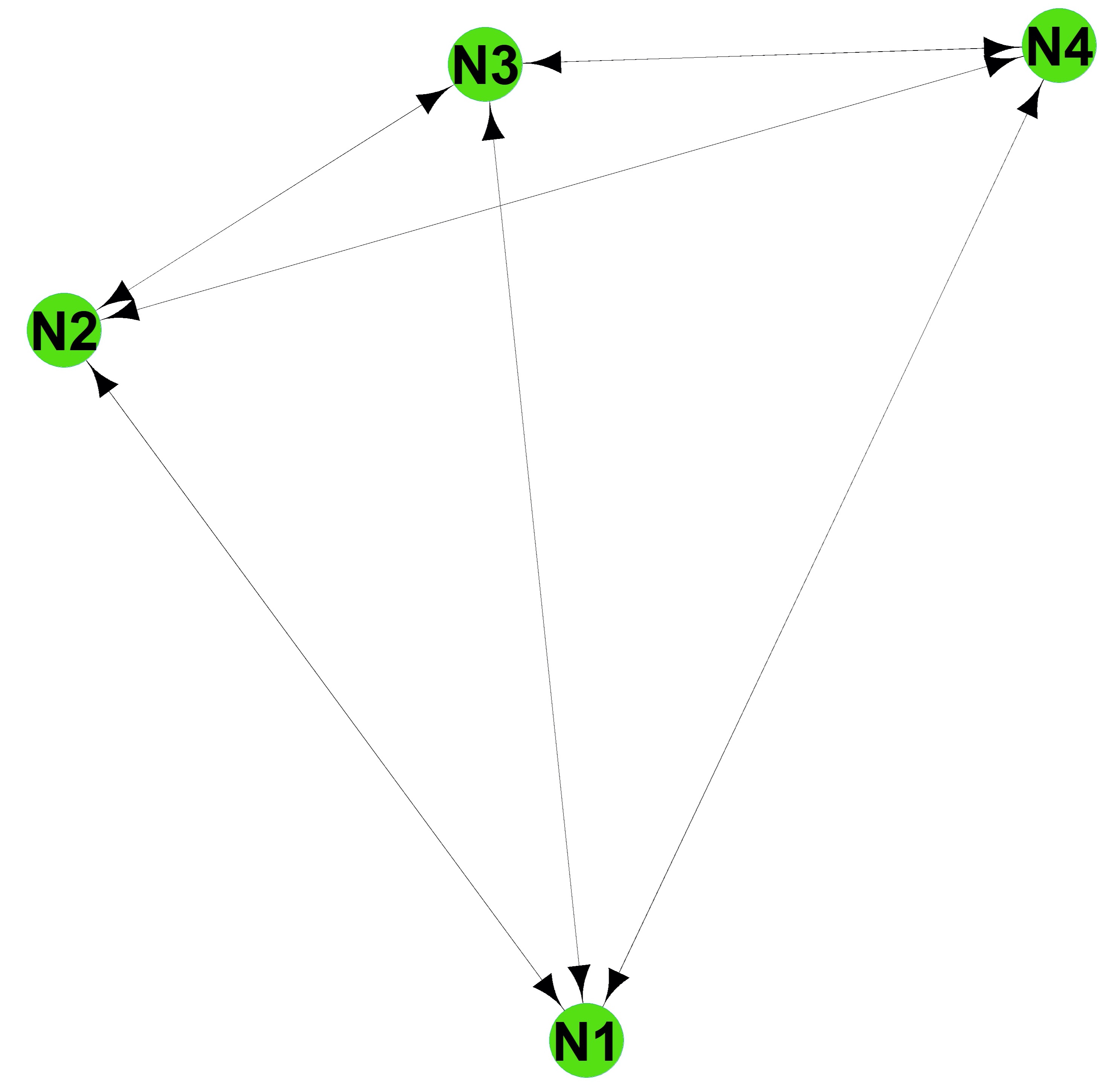}
    \subcaption{39 Iterations (r = 39)}
    \label{minigraph t=39}
    \end{subfigure}%
      \begin{subfigure}{0.25\textwidth}
    \centering
    \includegraphics[width=\linewidth]{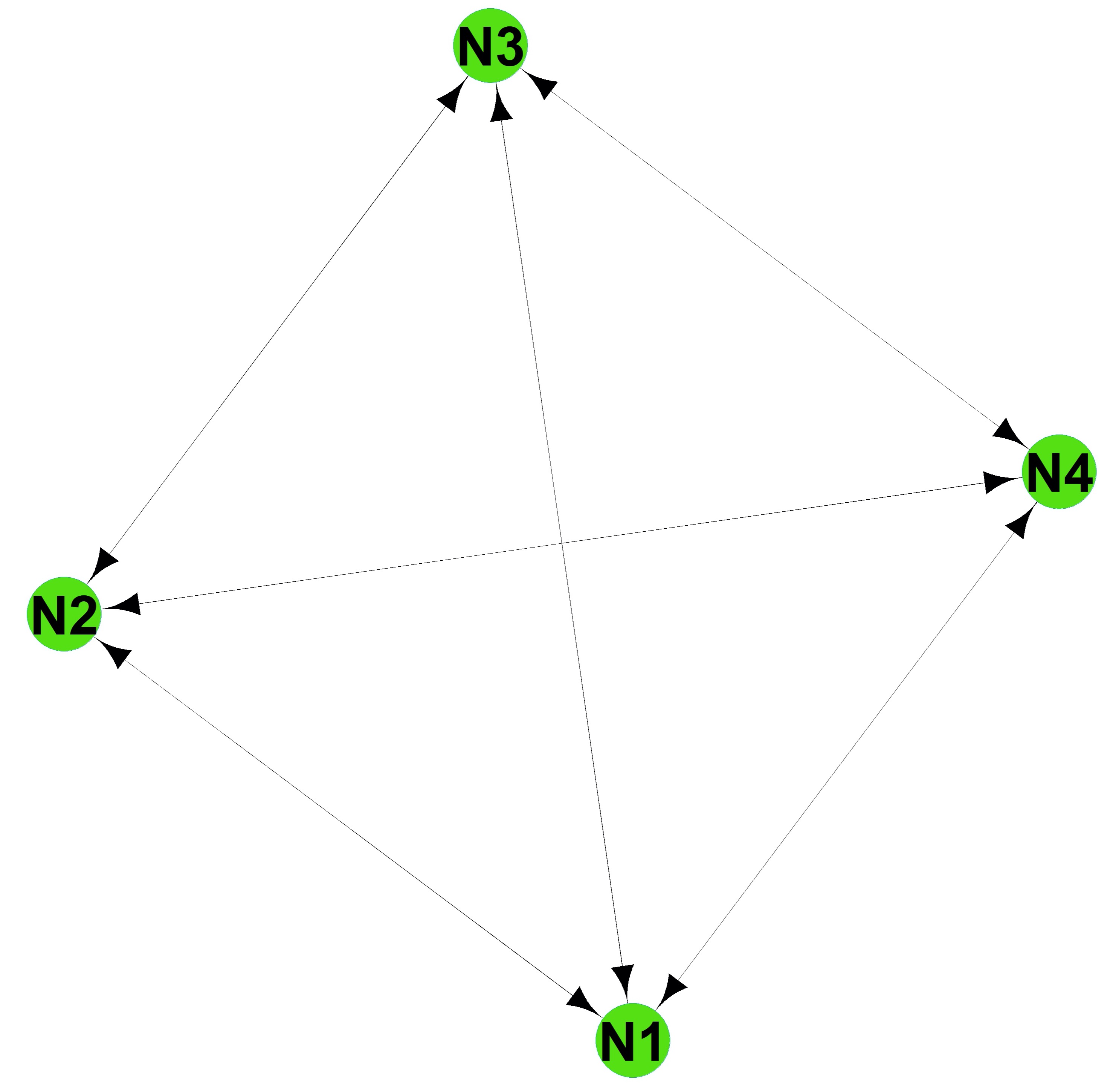}
    \subcaption{80 Iterations (r = 80)}
    \label{minigraph t=80}
    \end{subfigure}
    \caption*{\footnotesize Note: Moving from left to right, the above graphs shows the network's randomly assigned initial position (r = 0), its position after one update (r = 1), its position after 39 iterations (r = 39), and its final position after 80 iterations (r = 80). Nodes are labeled N1-N4, and all have directional connections with each other, as reflected by the arrows.}
  \label{Toy ex initial fig}
\end{figure}

As this is a technical series of formulas, I provide a simple example of four fully-connected nodes, like those in my network. Assigning random initial positions for the first period yields the layout shown in Figure \ref{minigraph t=0} on a 1000x1000 coordinate plot. I also include the result after the first iteration for comparison in Figure \ref{minigraph t=1}, as well as the result after 39 iterations (Figure \ref{minigraph t=39}) and the final layout after all 80 iterations (Figure \ref{minigraph t=80}).

A simple visual inspection of the above Figures \ref{minigraph t=0} and \ref{minigraph t=1} reveals that the position update that occurs in the first iteration (r = 1) is quite small: as is apparent in Figure \ref{minigraph t=1}, N2 and N4 move up slightly, but the change is very minimal. This is largely because the $F_{(t-1)}$ used in equations (\ref{swinging eq}) and (\ref{traction eq}) are both 0 given this is the first position update, so according to these formulas, the swinging value is double the traction. Using the dynamically calculated $\tau$ value, this means that the global speed $s(G)$ is very small at 0.005. Applying the speed formula to each node yields that the speeds for each node range from 0.0009 to 0.0016, meaning that the actual position updates are very small (ranging from a 1.37 to 3.69 unit movements, when the distance between nodes 3 and 4, for reference and scale, is around 215). 

In the next several iterations, now that $F_{(t-1)}$ is not 0, the ratio of global traction to global swinging is much closer, as highlighted by equation (\ref{global speed eq}). With $\tau$ again being dynamically calculated, $s(G)$ grows but remains relatively low, often between 0.024 and 0.05, and the average node moves at a speed of 0.01. By the 39th iteration, the layout still looks quite similar, as shown in Figure \ref{minigraph t=39}.\footnote{Note that this is easier to visualize in scaled plots, but it is still happening here.} Between update r = 1 and r = 25, the distance between N3 and N1 decreased from nearly 400 to around 45. The change in distance between N2 and N4 is even greater, moving from over 800 in r = 1 to around 130 in r = 25.

At the same time, the vectors between the nodes largely maintain the same ratios: comparing Figures \ref{minigraph t=1} and \ref{minigraph t=39}, the nodes do not appear to have moved much, with N3 being the exception. Relative to N2 and N4, N3 has drawn closer to N1, as is apparent in Figure \ref{minigraph t=39}. N3 is also slightly closer to N4 than N2, unlike before. As highlighted by equations (\ref{swinging eq}) and (\ref{traction eq}), the swinging is much higher than the traction during these iterations since the change in forces on each node is sizeable when moving such large distances across the Cartesian plane in each update. Because the swinging is high given the amount of distance covered, the average node speed throughout these updates remains quite low, ranging from 0.008 to 0.012 for most updates. 

Then, in subsequent updates, as the nodes have drawn closer together across the Cartesian plane, they begin to update their positions relative to one another. In the iterations following r = 39 (Figure \ref{minigraph t=39}), N2 and N4 shift downward while maintaining roughly the same distance from each other. Indeed, it is only at iteration 48 when global traction is greater than global swinging ($tra(G)>swg(G)$), and as the ratio approaches this point, the nodes begin to adjust their positions relative to each other more quickly after iteration r = 39. The final position is apparent in Figure \ref{minigraph t=80}. Even if more iterations are performed afterwards, the nodes do not update their positions further; they have converged towards the equilibrium. 

This behavior helps to clarify why not all node movement is effective traction: with each update step, each node experiences different magnitudes and directions of forces, even in a small network such as this one. In this case, the bulk of the first several iterations are composed of adjusting the geodesic distances between the nodes, with little updates to the positions of the nodes relative to each other\textemdash it is this behavior that causes the plots in Figure \ref{minigraph t=0} to look similar to \ref{minigraph t=1} and \ref{minigraph t=39}. It is only when nodes become close enough to have $F_t$ and $F_{t-1}$ stabilize between periods (approaching the point where $tra(G)>swg(G)$) that the nodes update their positions relative to each other. 

This effect is even more magnified when there are several hundred nodes in the network. In my much larger network of Chinese real estate firms, given the previous period's position is passed in as the initial position for each rolling window estimation, nodes begin updating their positions relative to other nodes right away. That is, unlike the simple example described here, the initial position is not randomly assigned, and there is no need to spend the first 30+ iterations moving the nodes closer together across space. There is still swinging, though, because each node is adjusting to the new forces acting on it as it moves, and with forces adjusting significantly in the early iterations, not all of that movement is smooth convergence to a node's final position.

\subsection{Region assignment and State-ownership status} \label{Appendix: Region Assignment and State-Ownership Status}

\subsubsection{Region assignment} \label{Appendix: region assignment}
As masters of diversification, Chinese real estate developers frequently have properties in multiple regions of the country, spanning provinces and tiered cities. Selecting the region where they are focused is a challenge simply because each has so much exposure and can experience spillover from any region disproportionately at times of local shock. Thus, in assigning region of focus, for each developer, I developed a list of regions where they had some properties. Nearly half (42/97) had some degree of exposure to more than 1 region, and 16 had exposure to three or more regions. For these companies, I then investigated further, looking at the developer's website and property list to see where their properties were concentrated, assigning their region of focus based off this. For a few developers that still remained hard to classify due to substantial investment in multiple regions (less than five), I examined the network graphs over time to get a sense of where the developer was concentrated, reasoning that market participants likely had the greatest knowledge of where the company's regional exposure laid and that it should be apparent in the network as well. In each case, the developer in question was surrounded by nodes of a certain regional focus; for instance, developer $i$ with exposure in the north, south, and east would appear surrounded by nodes with eastern-centric businesses, moving with them across large chunks of time. I thus follow this classification when necessary. Once a ``x-region-centric" classification is assigned, it is maintained throughout all network graphs for consistency.  

However, I recognize that because of the diversification of firms, this method of ``x-region-centric" is not an ironclad classification method. While I considered classifying by tiered cities, this proved too challenging to do, as nearly every developer had at least two tiers of properties, if not all three, and since there are only three tiers, there is less room for firm differentiation than from the 6 regions. I thus monitor regional spillover over time by generating several network graphs segmented by regional exposure, coloring the nodes by whether the firm has any exposure to the region. For example, I generate a set of networks over time that show exposure to the north; a firm with any properties in the north is colored pink while those without any northern properties are colored teal. I then generate these sets of plots for every region in China, working to see if there is any asymmetric spillover to a certain region for one of the events that I analyze, for instance, as reflected in ``to" or ``from" connectedness changing rapidly or in node movement across the network. More often than not, there is no such disproportionate regional spillover, and the node seems to move according to its ``x-region-centric" classification. This result does make sense given how the classifications were developed. It also suggests that even developers with some exposure to a region may not be shocked when a mini-crisis disproportionately impacts that region\textemdash because the firm's diversified business effectively insulates it from reactive market forces. 

\subsubsection{State-ownership status assignment} \label{appendix: state ownership status}

Teasing out state-ownership status in China can be an incredibly complex process since there are multiple levels of state ownership (i.e., at the district level, the city level, the municipality level, the province level, or at the national level). Further, when looking at companies' ownership records and shareholders, state-ownership will not always show as something obvious like ``Government of the City of Hefei" or ``Government of Jiangsu Province," rather reflecting a nondescript-sounding company name that is in fact owned by the State. To determine state-ownership, I rely upon the family tree tool of Eikon's Refinitiv software, which traces ownership through shell companies to find the ultimate owner of an enterprise: those that Eikon identifies as state-owned are classified as such, and those that Eikon did not show state-ownership of are not classified as state-owned (although I verify these latter cases manually to confirm). 

Certainly, this means there are gradients of state-ownership that are difficult to capture in a binary variable, and some edge cases exist. Vanke (CN:VAN), for instance, is one such case: it is a ``state-backed developer" where 33.4\% of the company is owned by Shenzhen Metro, which is owned by Shenzhen's State asset regulator (\cite{jim_exclusive_2024}). Following \textcite{chow_evergrande_2024}'s broader definition of SOEs\textemdash which defines a SOE to be a company whose largest shareholder is the State\textemdash I include Vanke as a SOE since the State is indeed the single largest shareholder of the company. 

\pagebreak

\subsection{Numerical results from the VAR} \label{appendix: var results}

I first aim to give a sense of how sparse the $\phi_\ell$ matrices from the VAR are. As I have three lags throughout my analysis, I investigate the non-zero values in $\phi_1$, $\phi_2$, and $\phi_3$. 

Beginning in the pre-COVID era (12 June-24 July 2019), the maximum number of non-zero coefficients per firm is 33, on average. The model generally finds higher numbers of non-zero coefficients per firm in $\phi_1$ and $\phi_3$, while $\phi_2$ tends to have a slightly lower number. In terms of the maximum number of non-zero coefficients per firm in a $\phi_\ell$, the maximum in $\phi_3$ is usually at or above the maximum in $\phi_1$; this is likely because longer-term trends are captured in this third lag. For instance, for a rolling window in this date range, the maximum number of non-zero coefficients for a firm is 35 in $\phi_1$, 29 in $\phi_2$, and 37 in $\phi_3$. The median number of non-zero coefficients per firm tends to be higher for $\phi_1$ and often constant between $\phi_2$ and $\phi_3$. For example, for the same window, the median number of non-zero coefficients for a firm is 8 for $\phi_1$, 5 for $\phi_2$, and 5 for $\phi_3$. 

There are persistently some firms in each period which have all zero coefficients in a given $\phi_\ell$. However, examining the non-zero coefficients across $\phi_\ell$ for a given rolling window reveals that firms with all zeros in one $\phi_\ell$ almost always have non-zero coefficients in another $\phi_\ell$. For these firms (of which there are consistently three), company profiles reveal that they are relatively diversified firms who have exposure to real estate, in addition to another sector. This behavior suggests that their price movements are most correlated with their other exposure rather than real estate, as their stock volatilities seem less sensitive to peer firms wholly in the real estate industry. Ultimately, in the network results in Section \ref{sec: results}, these firms are consistently positioned on the periphery by ForceAtlas2, reflecting their weaker linkages with other firms. 

Fast-forwarding to the era of the three red lines and Evergrande's letter to the Guangdong government, I examine $\phi_\ell$ matrices for 29 July-30 November 2020, and find roughly the same pattern. However, as the real estate crisis intensifies, the median number of non-zero coefficients per firm rises. For dates later in the crisis (August 2021 and after), the median number of non-zero counts per firm for $\phi_\ell$ is closer to 10; for instance, for one later rolling window, the median number of non-zero coefficients is 10 for $\phi_1$, 9 for $\phi_2$, and 9 for $\phi_3$. 

I also investigate the correlation matrices of $\phi_\ell$ across time to see how consistent the coefficients selected by the elastic net are across sequential rolling windows. Compared to the pre-COVID and early COVID periods (June/July 2019 and June/July 2020), I find that the selected coefficients for each firm become noticeably smoother day-to-day starting with the three red lines event, with each firm having less fluctuation between consecutive windows. That is, compared to the earlier period, the elastic net seems to be selecting the same lagged data across sequential windows with more consistency.\footnote{In the rolling window estimation, between sequential dates, only two days are different from the previous period as the window shifts over: the first date of the previous period is cut off, the next 99 days are the same, and the last date is new. It is thus a positive sign to see this increased stability of the coefficients selected by the elastic net, as it seems to select most of the same predictors across nearby dates in the window.} This pattern continues for the rest of the data under investigation (until March 2024).

\subsection{Additional plots} \label{Appendix: additional plots}

The following plots are included for reference, with only brief descriptions of what is observed. 

\subsubsection{Network before and after COVID-19 outbreak} \label{appendix: COVID network}

It is first important to briefly note the changes that COVID-19 brought to the network, already shifting the landscape before the first of the policy changes that shook the real estate market. Below in Figure \ref{Covid Figure} is the network shortly before COVID-19 became a serious concern (17 January 2020) and a few months after (14 April 2020).\footnote{17 January 2020 is an appropriate before date given that the first Chinese province to launch a public health warning for COVID-19 did so shortly thereafter, on 22 January 2020 
\begin{CJK*}{UTF8}{gbsn} 
(\cite{noauthor__2020}).
\end{CJK*} }

\begin{figure}[H]
  \caption{Network before and during COVID-19 outbreak}  \centering
  \begin{subfigure}{0.5\textwidth}
    \centering
    \includegraphics[width=\linewidth]{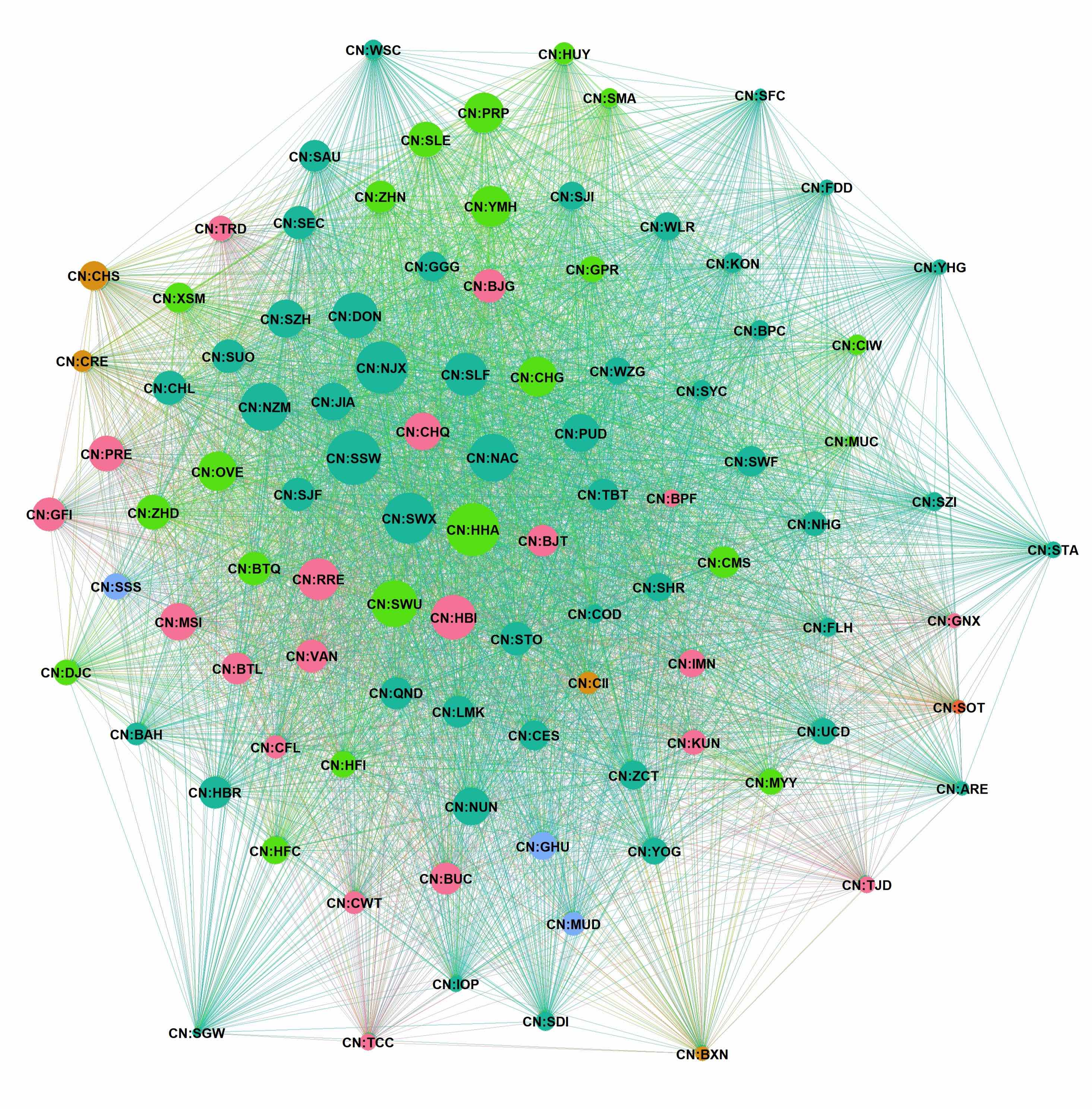}
    \subcaption{17 January 2020}
  \end{subfigure}%
  \begin{subfigure}{0.5\textwidth}
    \centering
    \includegraphics[width=\linewidth]{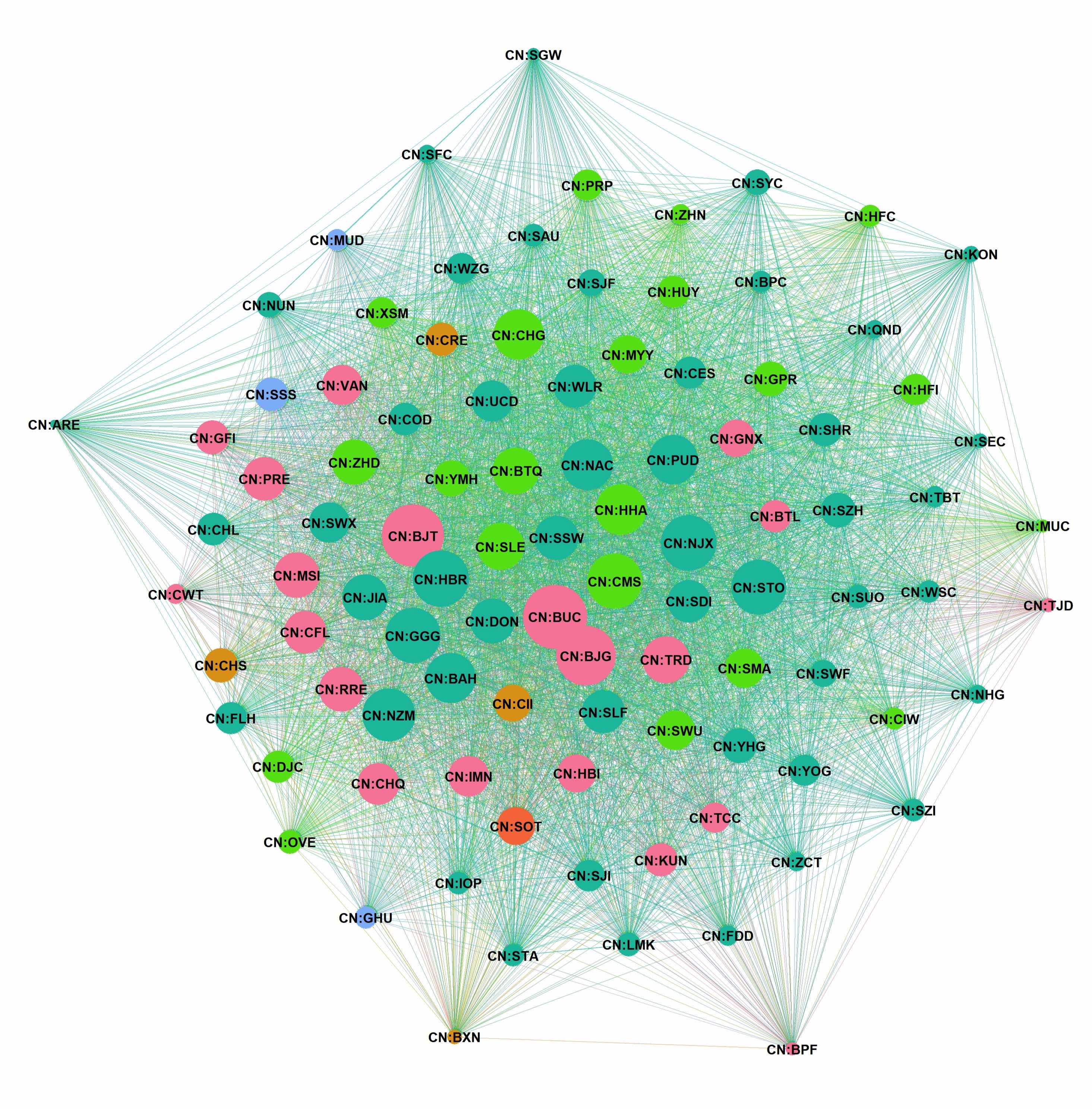}
    \subcaption{14 April 2020}
    \label{covid figure after}
    \end{subfigure}
    \caption*{\footnotesize Note: The left graph shows the network shortly before COVID-19 became a serious concern (17 January 2020) while the right graph shows the network a few months later (14 April 2020). Here, the colors are determined by the region where the developer is primarily focused: pink is the north, light green is the south, teal is the east, bronze is the southwest, light blue is the northwest, and red-orange (of which there is only one node, CN:SOT) is the northeast. Node size is determined by ``to" connectedness; larger node size thus means that the node has a higher level of to connectedness.}
  \label{Covid Figure}
\end{figure}

As is apparent above, the network before COVID-19 is comparatively unstressed: nodes are relatively far apart, the core is very loose and not dense, and those nodes on the periphery have small values of ``to" connectedness.\footnote{While not apparent in Figure \ref{Covid Figure}, the ``from" connectedness of nodes on the periphery is low as well.} ``To" connectedness ranges from 22 to 136 while ``from" connectedness ranges from 43 to 98. 

In contrast, the network in Figure \ref{covid figure after} (depicting 14 April 2020) is much more stressed, with nodes being drawn more tightly together, a clear core forming, and more regional clustering, particularly with pink nodes gathering towards the left side of the network. This 14 April network has ``to" connectedness ranging from 23 to 197 and ``from" connectedness from 63 to 98. While the to connectedness range has grown significantly, the from connectedness range narrows. As discussed in Section \ref{three red lines section}, this indeed confirms the observation that there is a minimum level of connectedness that comes from being a listed developer with some real estate exposure; as time passes and stress increases, this minimum from connectedness increases as well. 

Certainly, there are many more things to analyze here, but I only offer the above as an orientation to the reader for understanding these networks, as well as a comparison of unstressed vs. stressed periods of the network. Indeed, as will be apparent in Section \ref{three red lines section}, a somewhat stressed network will be the base state for the three red lines announcement, one that looks more like the 14 April 2020 network than the 17 January network. Even before the three red lines, the Chinese real estate industry was already under pressure. 

\subsubsection{Network before and after news of the three red lines} \label{appendix: three red lines additional plots}

\begin{figure}[H]
  \caption{Network before and after news of the three red lines: Color by state-ownership}  \centering
  \begin{subfigure}{0.5\textwidth}
    \centering
    \includegraphics[width=\linewidth]{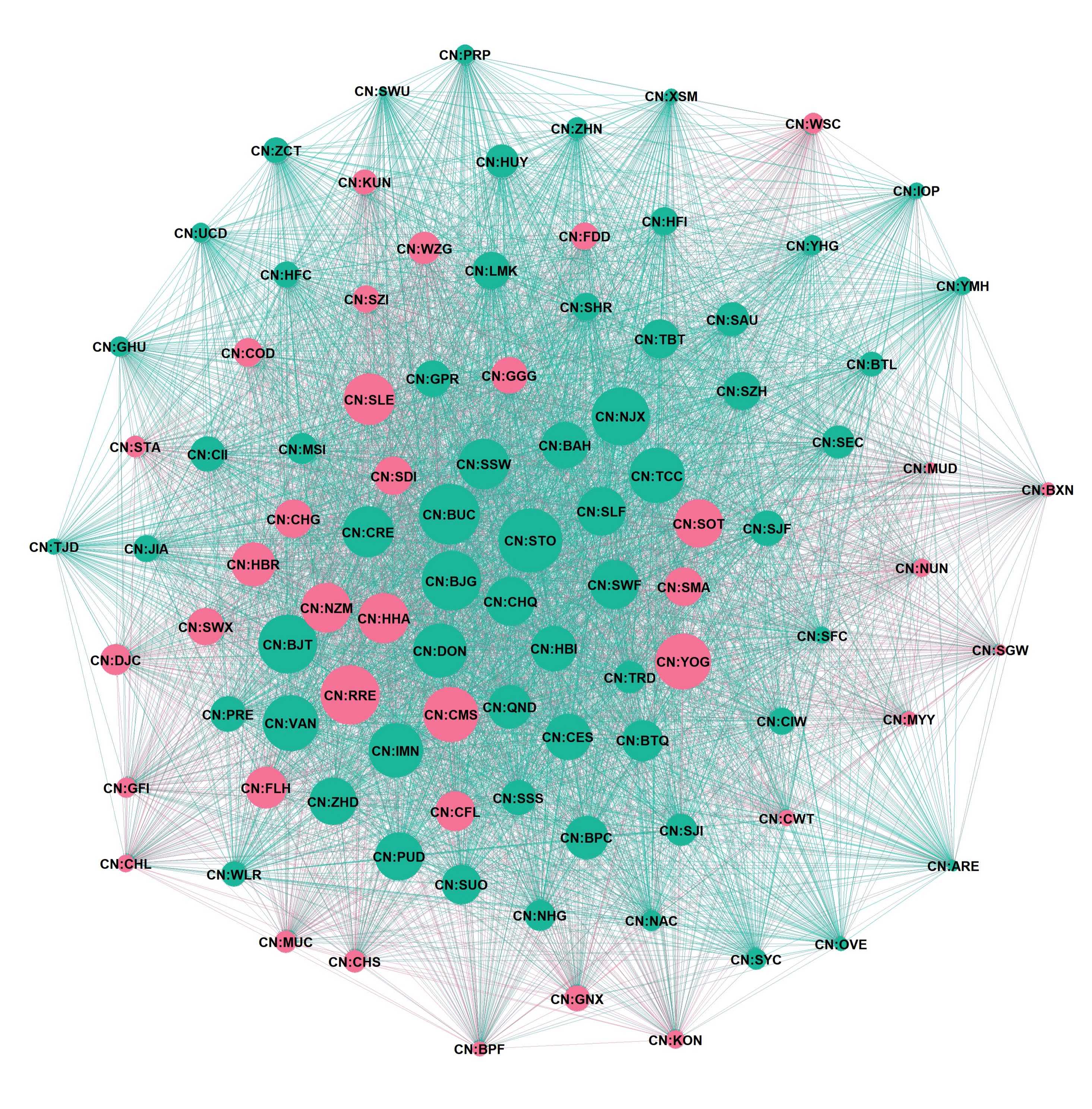}
  \end{subfigure}%
  \begin{subfigure}{0.5\textwidth}
    \centering
    \includegraphics[width=\linewidth]{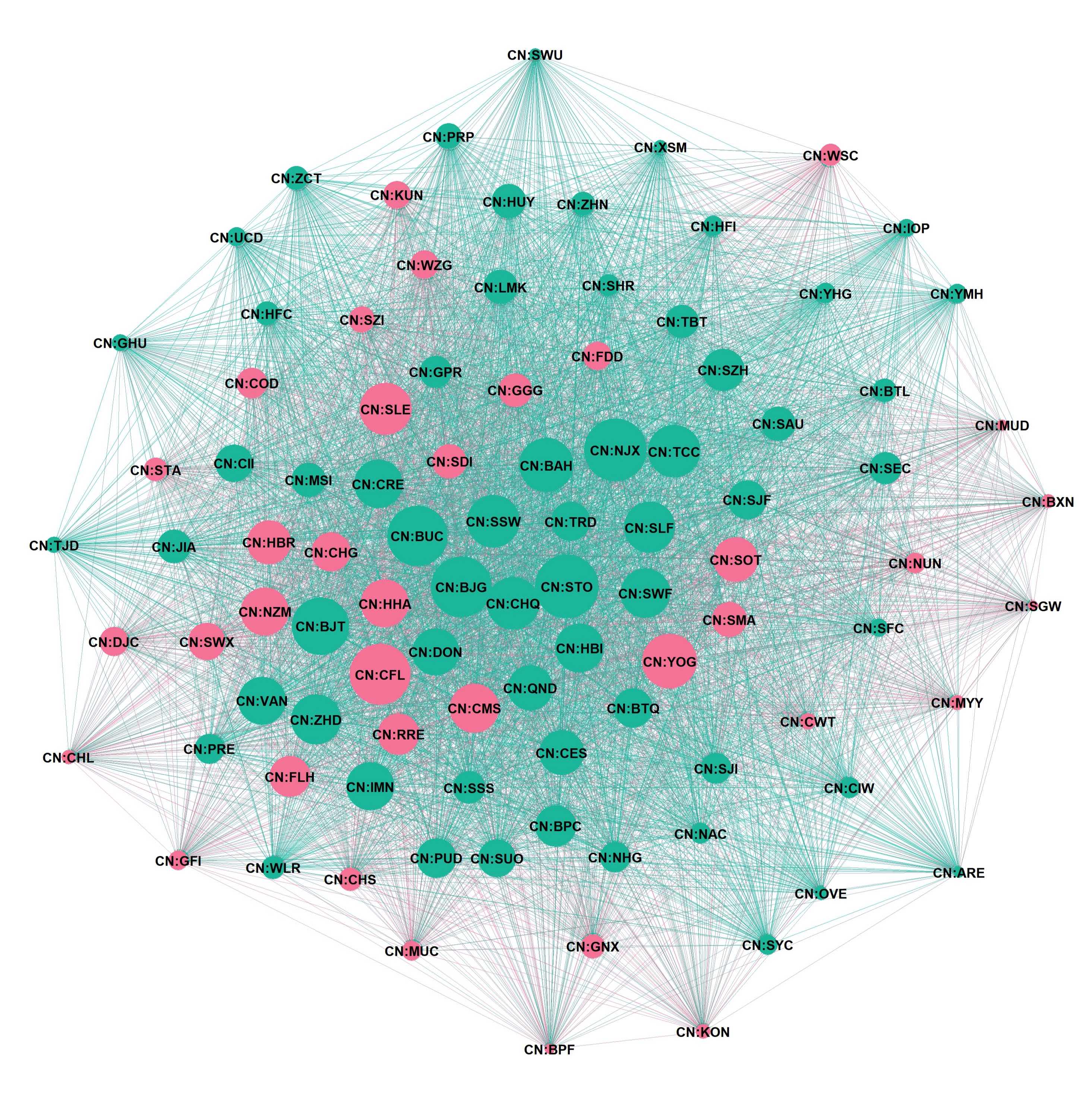}
    \end{subfigure}
    \caption*{\footnotesize Note: The left graph shows the network before news of the three red lines circulated (13 August 2020) while the right graph shows the network the day after (14 August 2020). Here, the colors are determined by state-ownership status: teal nodes are SOEs, and pink nodes are POEs. Node size is determined by ``to" connectedness.}
  \label{State-ownership COVID figure}
\end{figure}

As is apparent in Figure \ref{State-ownership COVID figure}, there does not seem to be different behavior between state-owned and private-owned enterprises. Both SOEs and POEs are pulled in towards the core of the network, and firms of both types have positions on the periphery as well. Additionally, to connectedness does not increase or decrease more for one group than another.

\subsubsection{Network before, during, and after news of Evergrande's letter}

\begin{figure}[H]
  \caption{Network before, during, and after news of Evergrande's letter: Color by region, node size by ``from" connectedness}
  \centering
  \begin{subfigure}{0.5\textwidth}
    \centering
\includegraphics[width=\linewidth]{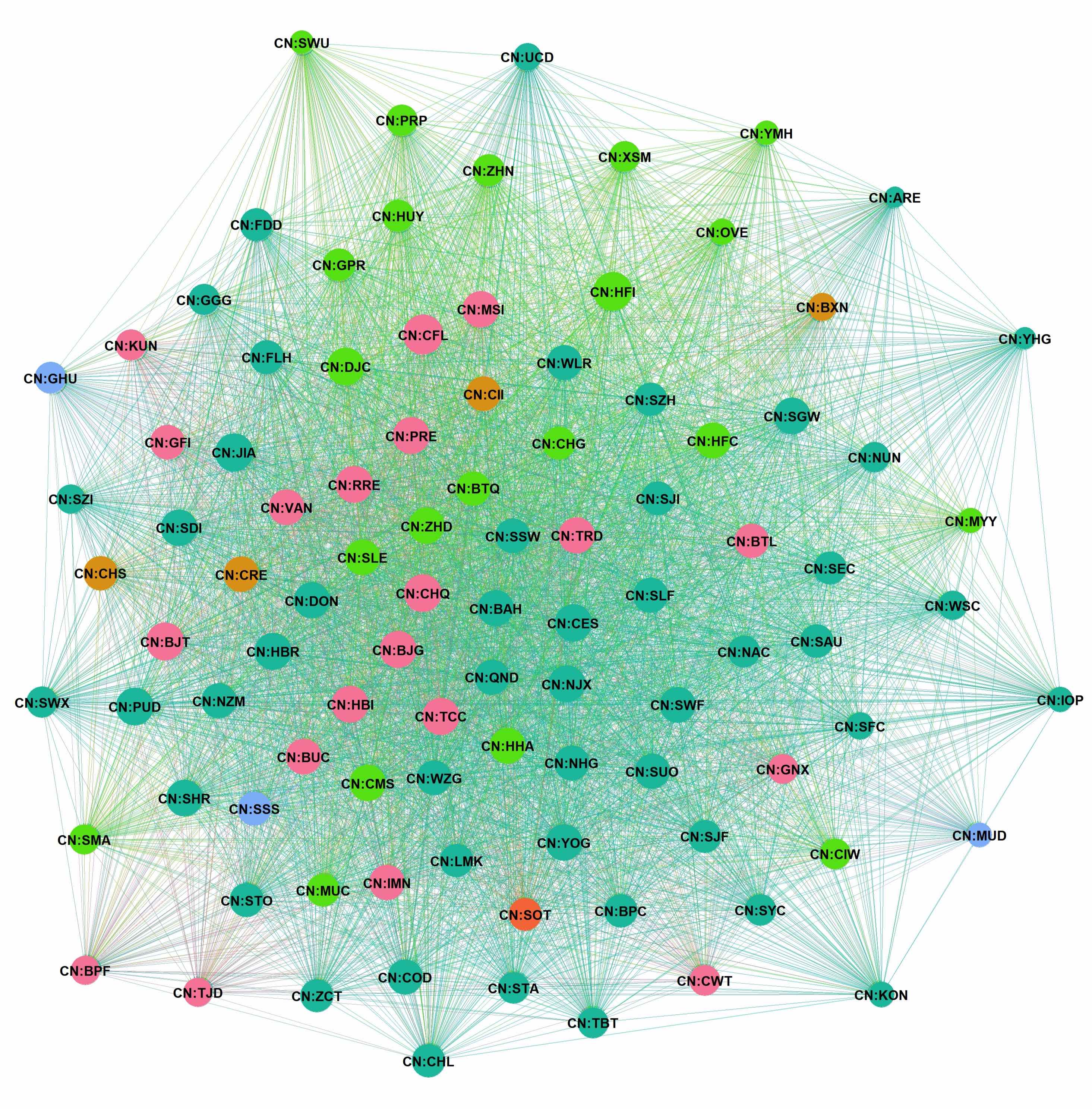}
    \subcaption{Before news circulation (21 September 2020)}
    \label{fig: SO before circ from}
  \end{subfigure}%
  \begin{subfigure}{0.5\textwidth}
    \centering
    \includegraphics[width=\linewidth]{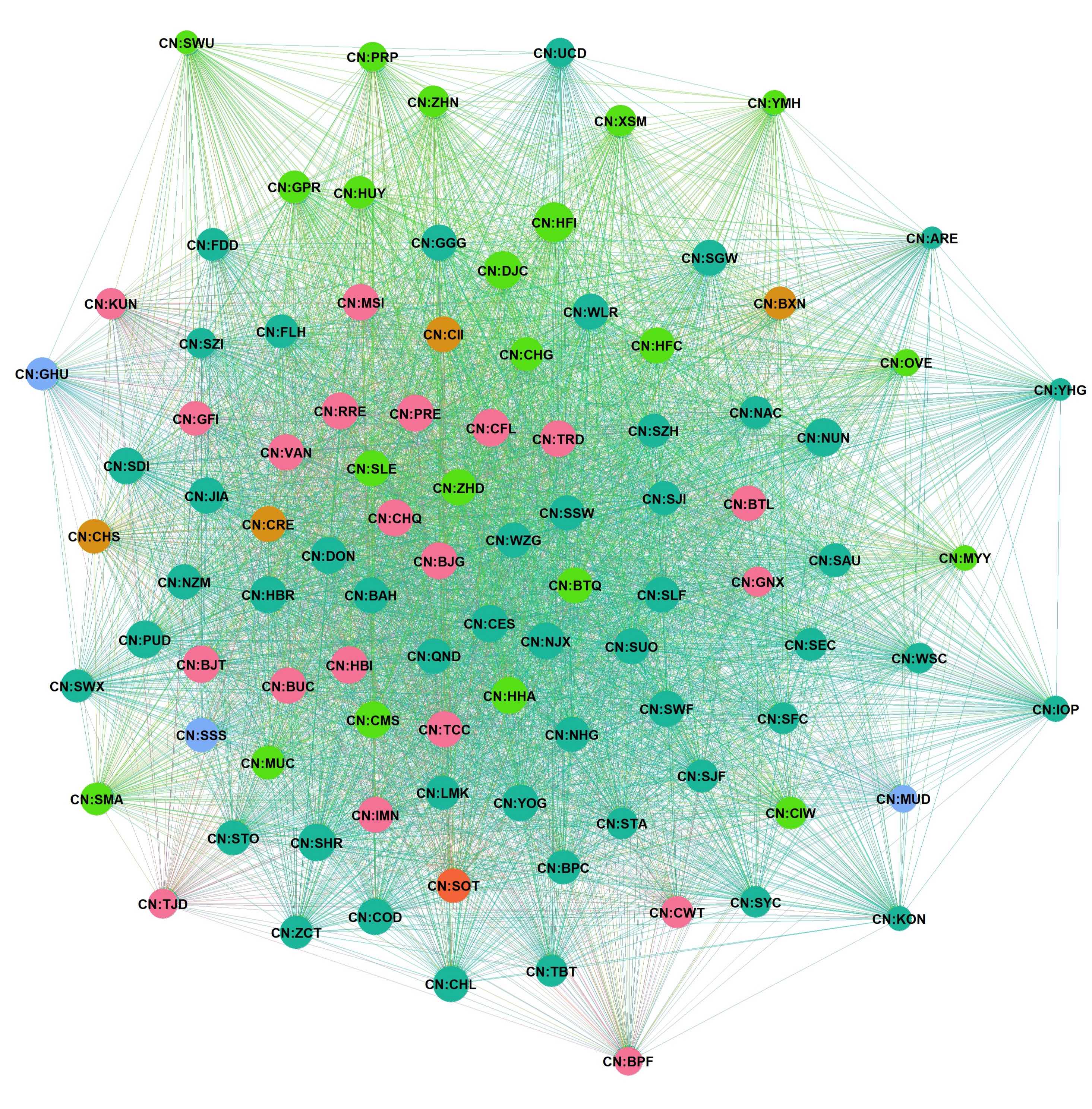}
    \subcaption{Partially circulated (23 September 2020)}
    \label{fig: SO circ partial from}
  \end{subfigure}
  
  \begin{subfigure}{0.5\textwidth}
    \centering
    \includegraphics[width=\linewidth]{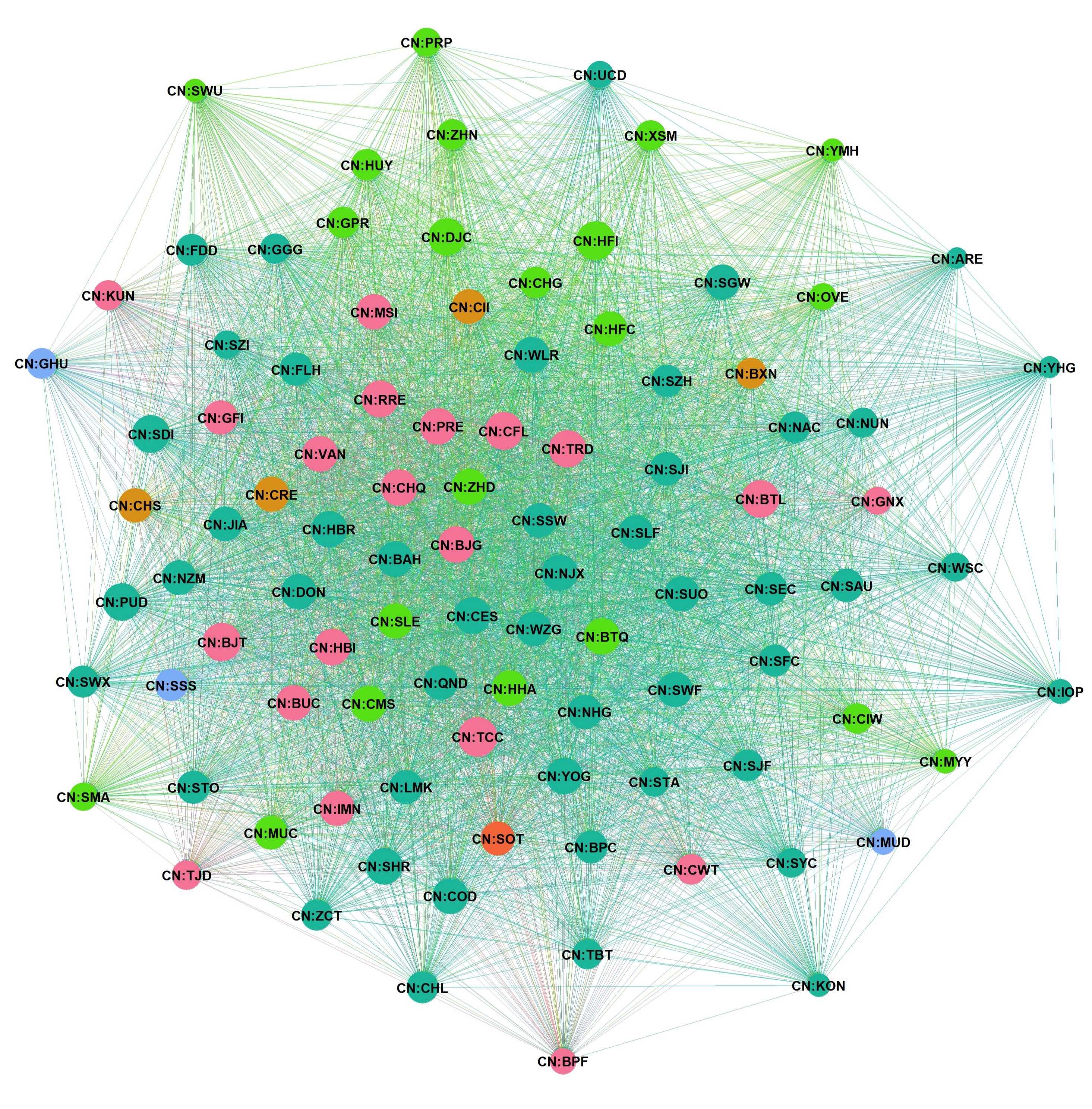}
    \subcaption{Letter goes viral (24 September 2020)}
    \label{fig:SO circ viral from}
  \end{subfigure}%
  \begin{subfigure}{0.5\textwidth}
    \centering
    \includegraphics[width=\linewidth]{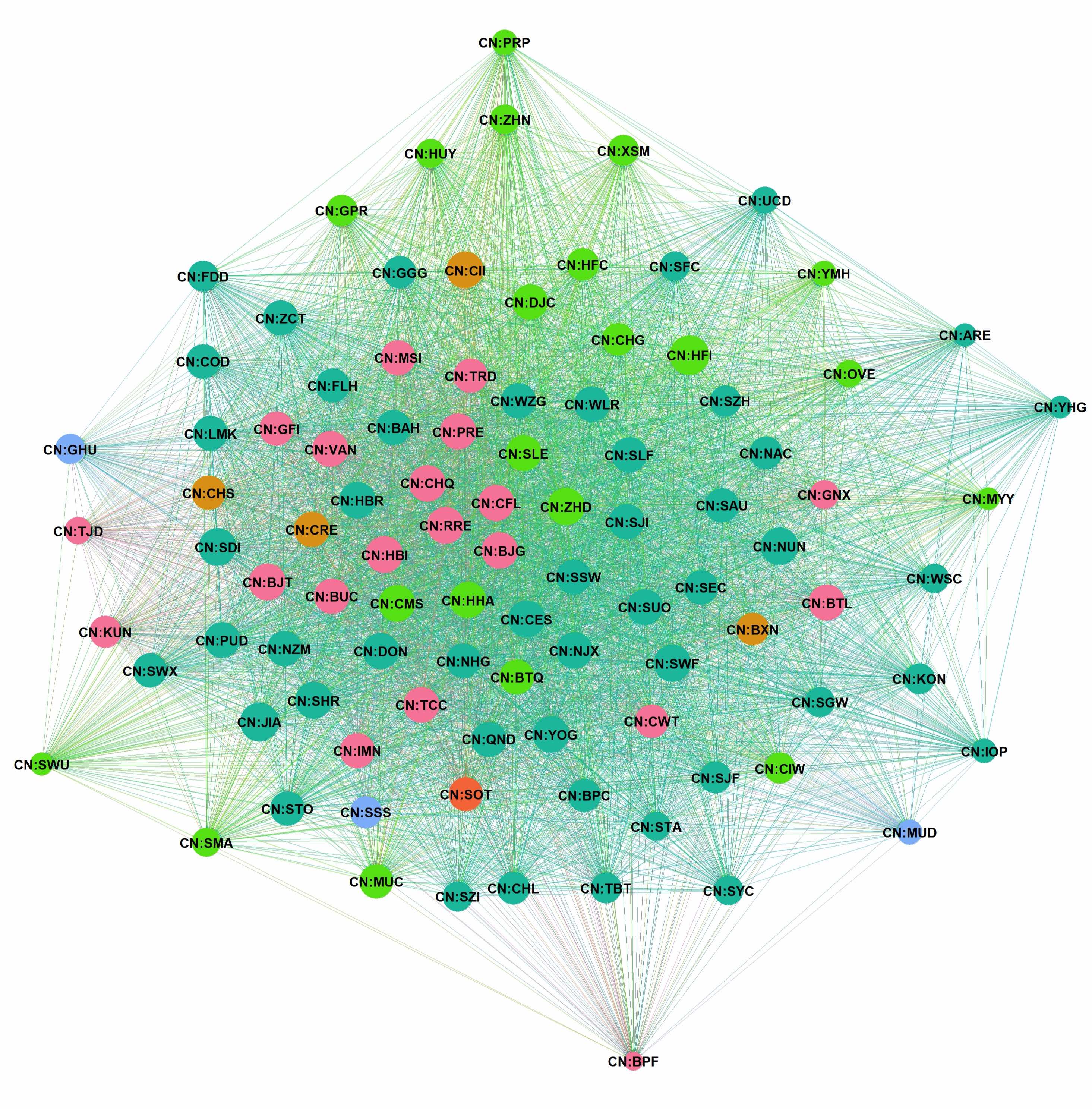}
    \subcaption{Two weeks later (9 October 2020)}
    \label{fig:SO two weeks later from}
  \end{subfigure}
  \label{From Guangdong Gov Letter Network}
\end{figure}
\footnotesize Note: The upper left graph shows the business day before the news (21 September 2020), while the upper right graph shows the network change when the news has been partially circulated (23 September 2020). The bottom left shows the day the news went viral (24 September 2020), and the bottom right shows the network roughly two weeks later (9 October 2020). Node size is determined by from connectedness; larger node size thus implies a higher level of from connectedness for the node. The colors are determined by the region where the developer is primarily focused: pink is the north, green is the south, teal is the east, bronze is the southwest, light blue is the northwest, and red-orange (of which there is only one node, CN:SOT) is the northeast. 

\subsubsection{Network before and after Kaisa suspension} \label{appendix: Kaisa further figures}

\begin{figure}[H]
  \caption{Network before and after Kaisa suspension: Color by region, node size by ``from" connectedness}
  \centering
  \begin{subfigure}{0.5\textwidth}
    \centering
    \includegraphics[width=\linewidth]{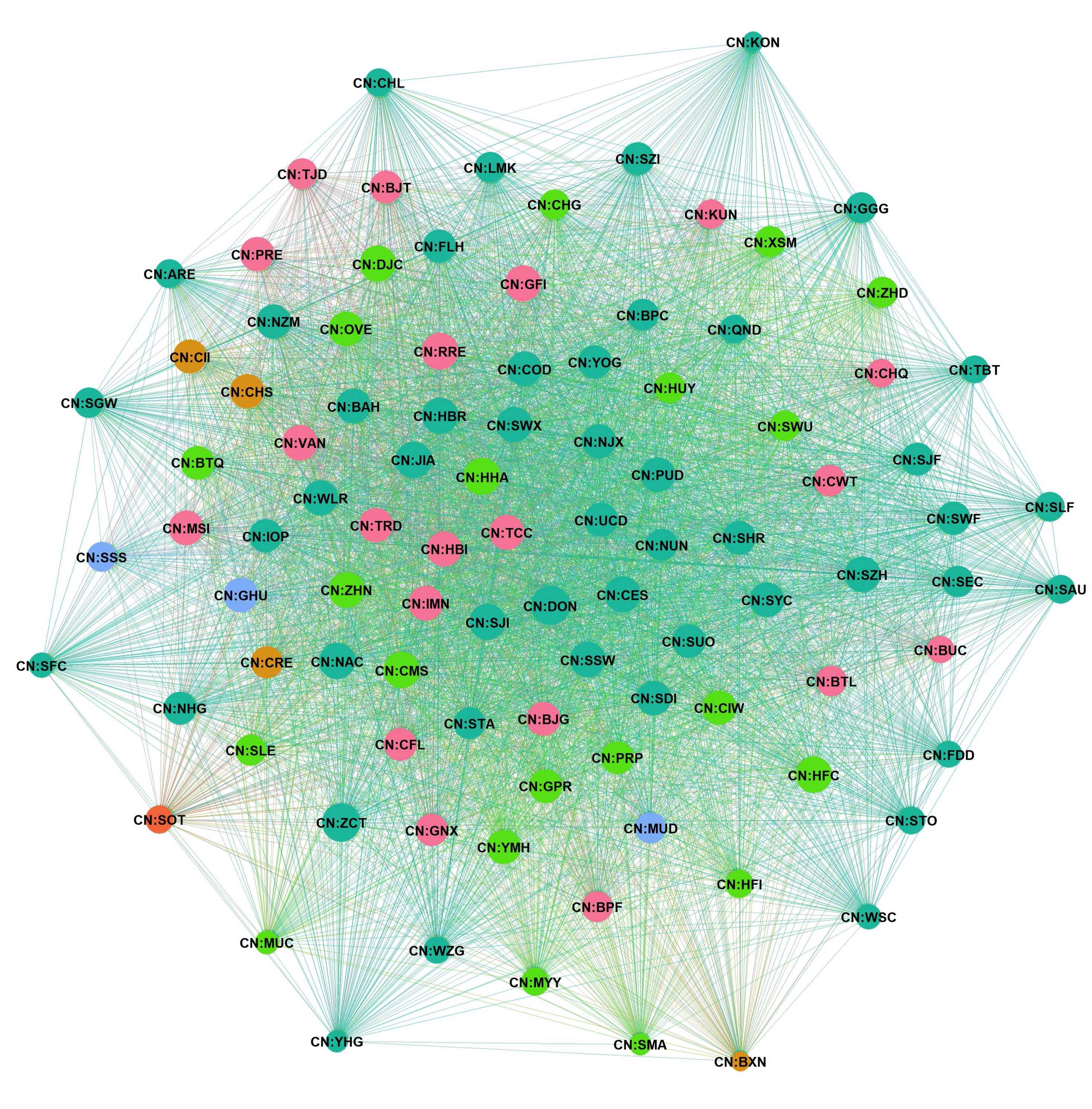}
    \subcaption{Before suspension (3 November 2021)}
    \label{fig: from Kaisa before}
  \end{subfigure}%
  \begin{subfigure}{0.5\textwidth}
    \centering
    \includegraphics[width=\linewidth]{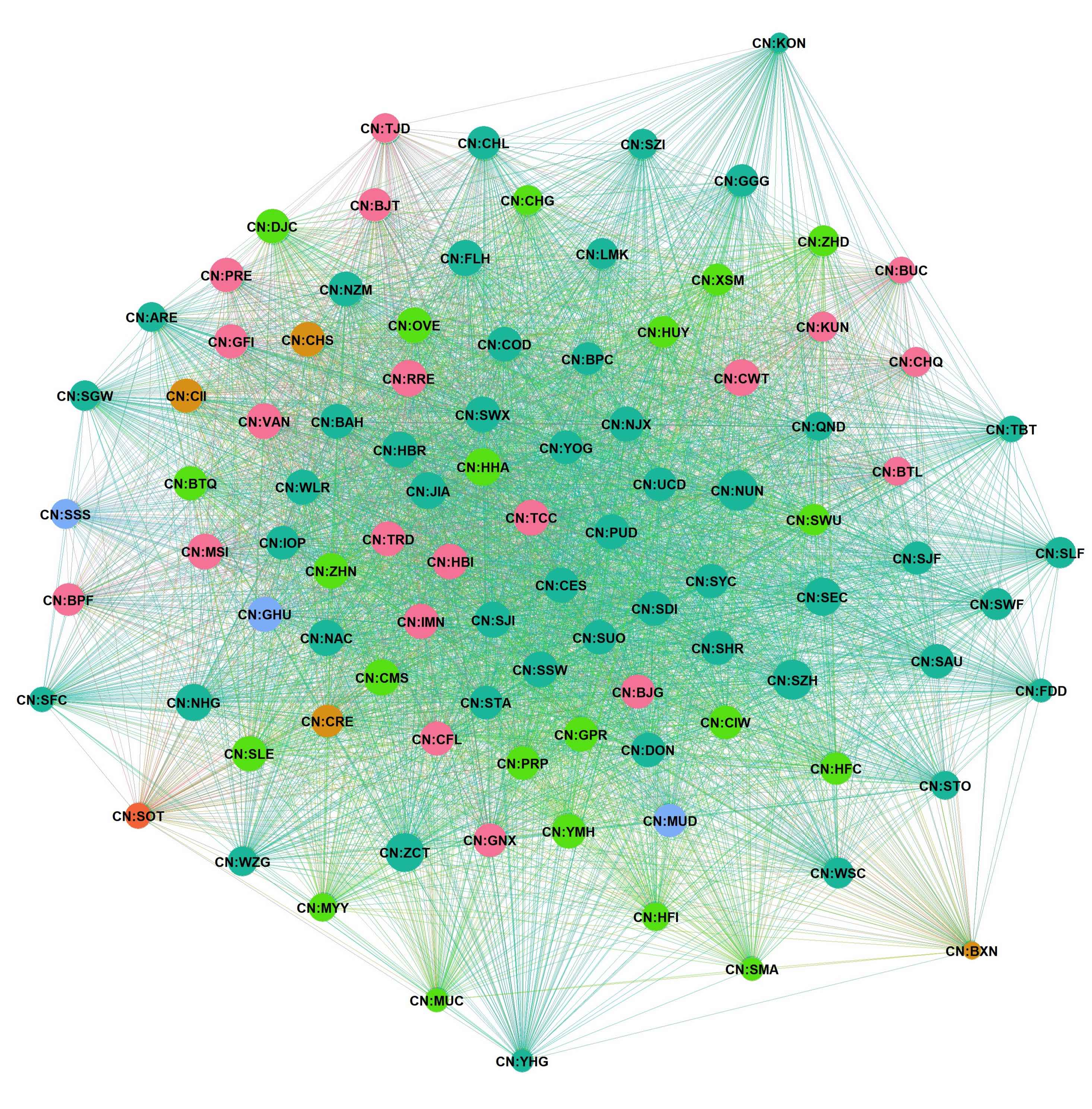}
    \subcaption{After suspension (5 November 2021)}
    \label{fig: from Kaisa after}
  \end{subfigure}
  \label{from Region Kaisa}
      \caption*{\footnotesize Note: The left graph shows the business day before the missed payment to onshore investors, two days before the suspension (3 November 2021). The right graph shows the network change on the day of suspension (5 November 2021), as the stock suspended at 9 am\textemdash the opening of the exchange for the day. The colors are determined by the region where the developer is primarily focused: pink is the north, light green is the south, teal is the east, bronze is the southwest, light blue is the northwest, and red-orange (of which there is only one node, CN:SOT) is the northeast. Node size is determined by from connectedness, meaning larger nodes have higher levels of from connectedness.}

\end{figure}

\begin{figure}[H]
  \caption{Network before and after Kaisa suspension: Color by SOE status and price movement}
  \centering
  \begin{subfigure}{0.5\textwidth}
    \centering
    \includegraphics[width=\linewidth]{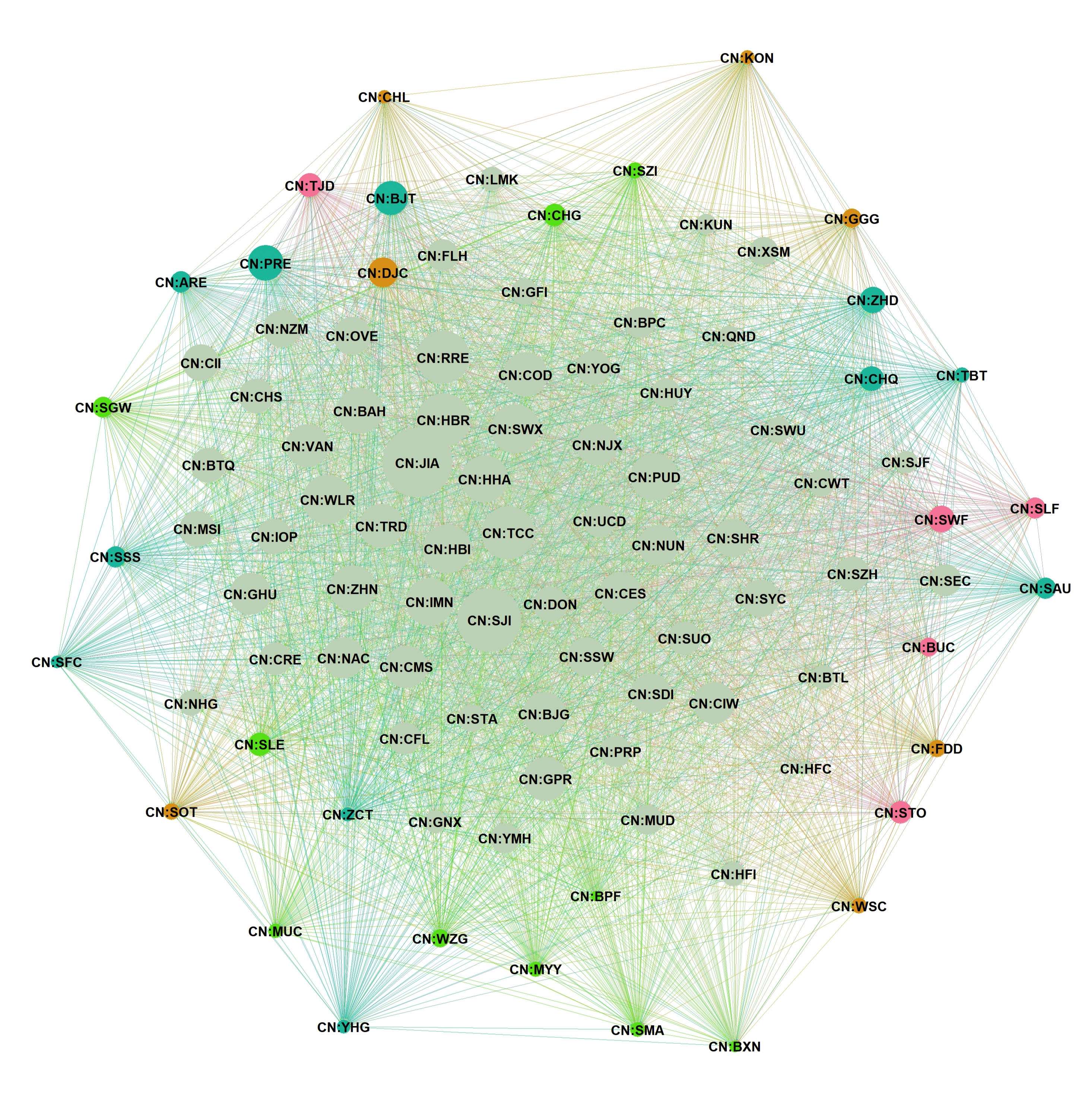}
    \subcaption{Before suspension (3 November 2021)}
    \label{Kaisa before SOExInc}
  \end{subfigure}%
  \begin{subfigure}{0.5\textwidth}
    \centering
    \includegraphics[width=\linewidth]{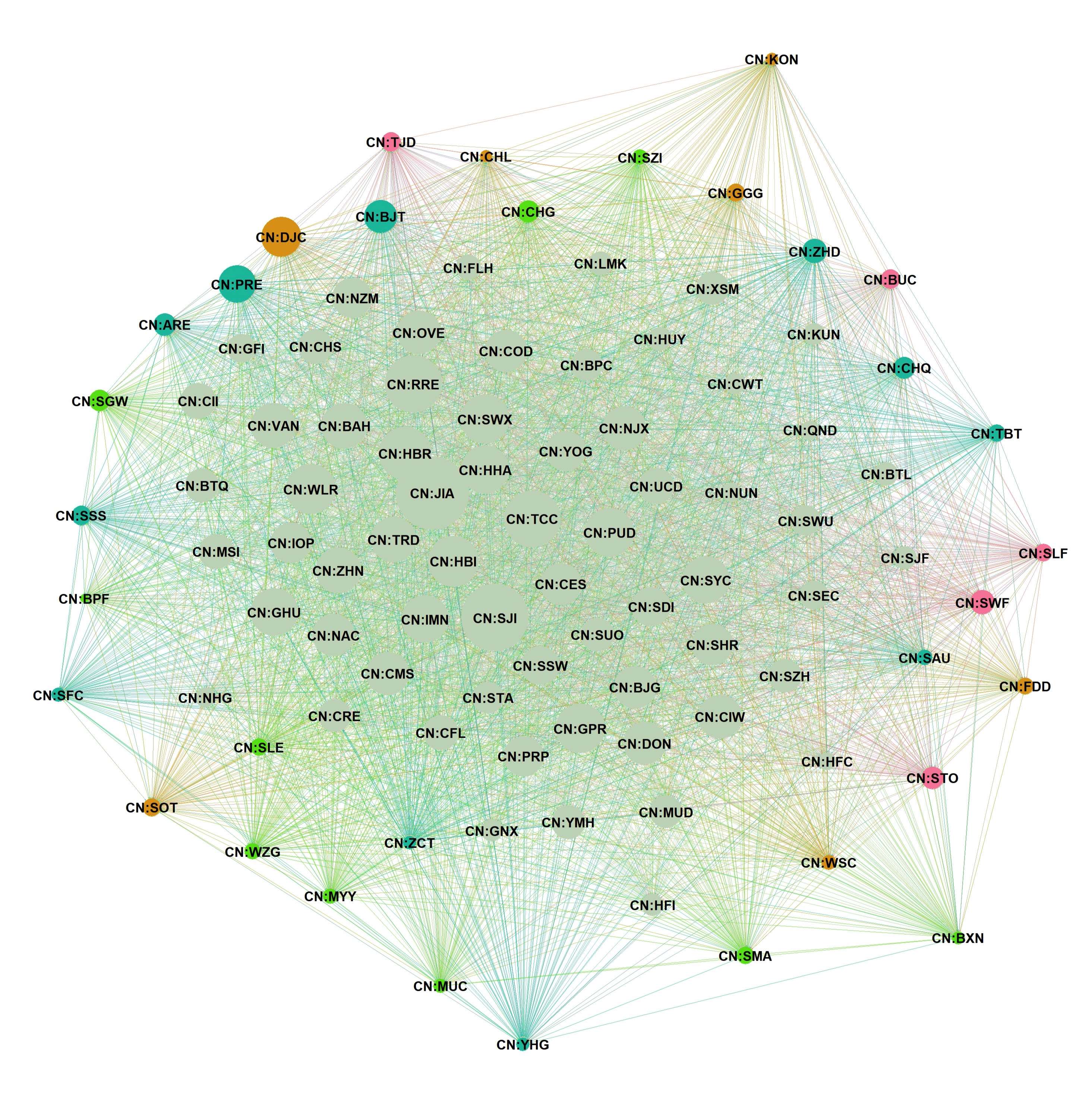}
    \subcaption{After suspension (5 November 2021)}
    \label{Kaisa after SOExInc}
  \end{subfigure}
  \label{Appendix fig Kaisa SOExInc}
    \caption*{\footnotesize Note: The left graph shows the business day before the missed payment to onshore investors, two days before the suspension (3 November 2021). The right graph shows the network change on the day of the suspension (5 November 2021), since the stock suspended at 9 am\textemdash the time the exchange opens for the day. The colors are determined by the state-ownership status of each firm and whether there was a price increase: bronze nodes are firms that are not SOEs and experience a decrease in closing price on 5 November compared to 3 November. Light green nodes are non-SOEs who experience a price increase across the two days. Teal nodes are SOEs who experience a price decrease, and pink nodes are SOEs who experience a price increase between the 3 November and 5 November. The gray nodes in the middle are those not on the periphery; I color the plot this way to allow for an easy visualization of the periphery behavior, as it is much harder to isolate it when examining the full colored plot. Node size is determined by ``to" connectedness. }
\end{figure}

\normalsize As is highlighted in the note below Figure \ref{Appendix fig Kaisa SOExInc}, the node colors are determined by the state-ownership status of each firm and whether there was a price increase: bronze nodes are firms that are privately-owned and experience a decrease in closing price on 5 November, compared to 3 November. Light green nodes are POEs who experience a price increase across the two days. Teal nodes are SOEs who experience a price decrease, and pink nodes are SOEs who experience a price increase between the 3 November and 5 November. The grey nodes in the middle are those not on the periphery (regardless of state ownership or price increase status). The plot is colored this way to allow for an easy visualization of the periphery behavior; it is much harder to isolate it when examining the full colored plot. 

I include both the day before the suspension (3 November 2021) as Figure \ref{Kaisa before SOExInc} and the day after the suspension (5 November 2021) as Figure \ref{Kaisa after SOExInc} for comparison, but the main plot of interest is \ref{Kaisa after SOExInc}. 

As is apparent in Figure \ref{Appendix fig Kaisa SOExInc}, of the nodes on the periphery, the majority of private enterprises experience price increases (10/17), and the majority of state-owned enterprises experience decreases in share price (11/16). There is a small cluster of pink nodes (SOEs with price increases) on the right periphery; given the consistent positions of these nodes on the periphery in these periods and after, as well as their close relative distances to each other, it is likely that these firms are relatively consistently insulated from investor sentiments. Indeed, the three firms clustered together on the left are CN:SLF (Shanghai Lujiazui Finance \& Trade Zone Development), CN: STO (Shanghai Tongji Science \& Technology Industrial), and CN:SWF (Shanghai Waigaoqiao Free Trade Zone Development `A'). All three firms have nearly all of their properties in Shanghai\textemdash one of the largest cities in China and one least likely to experience decreases in demand or property price declines. It thus makes sense that these select firms experience price increases during times of risk, as they are comparatively more stable, but like most other state-owned firms, likely less profitable as well. Ultimately, though, this example helps to illustrate that firms are not homogeneous across their SOE or region-centric status, since there is an inherent degree of variation.  

\subsubsection{Network before and after news of Country Garden’s missed payments} 

As highlighted in Section \ref{COGA body}, there is no serious change in the network between the two days. While the ``to" connectedness of some of the core nodes increases slightly, this is only the case for a few nodes, and it is natural to expect some nodes will change size on any given day in the market. Most nodes experience very small adjustments in position, and the periphery is quite stable. 

\begin{figure}[H]
  \caption{Network before and after news of Country Garden’s missed payment: Color by region}
  \centering
  \begin{subfigure}{0.5\textwidth}
    \centering
    \includegraphics[width=\linewidth]{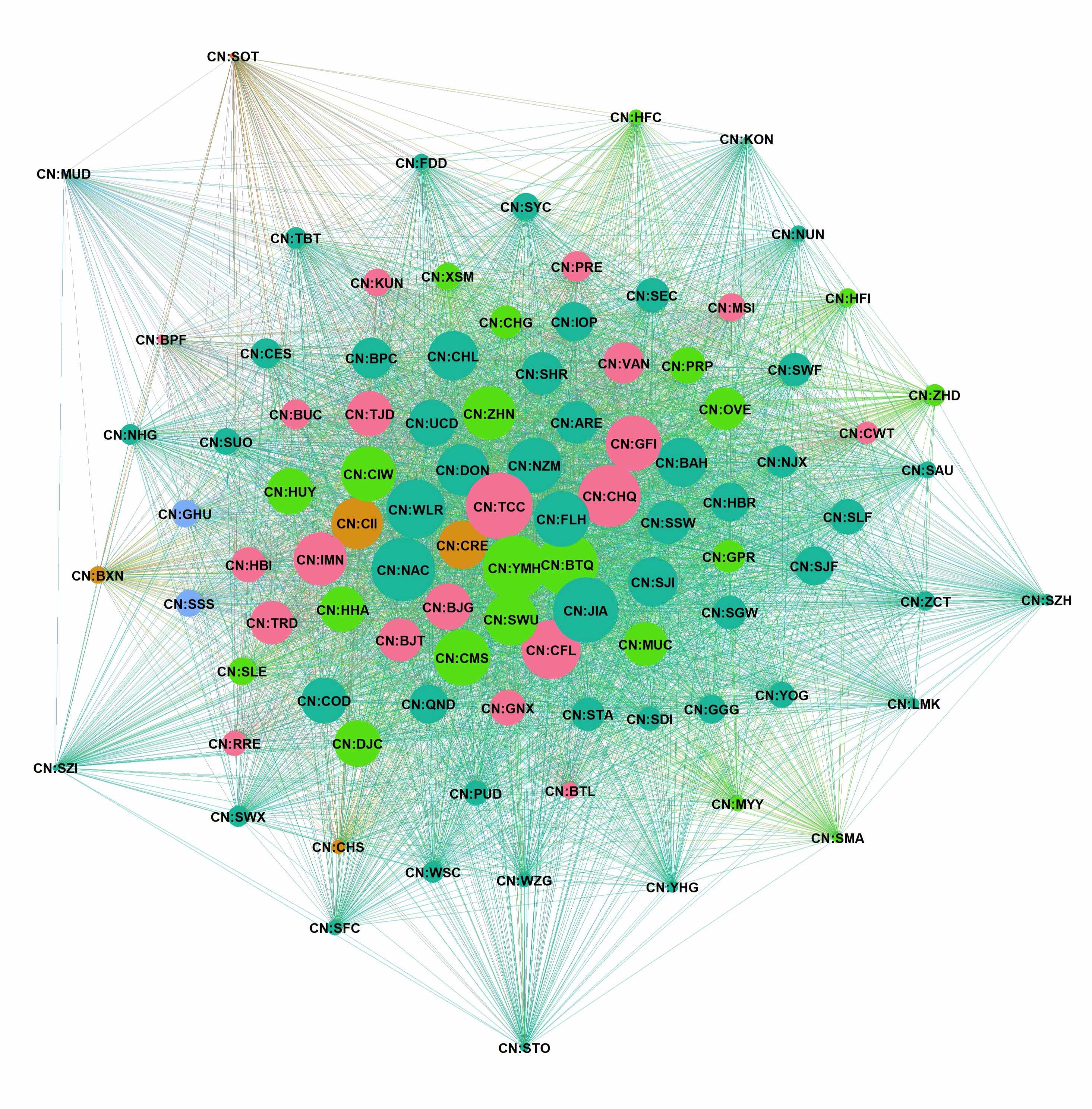}
    \subcaption{Before news circulated (7 August 2023)}
    \label{COGA before missed payment}
  \end{subfigure}%
  \begin{subfigure}{0.5\textwidth}
    \centering
    \includegraphics[width=\linewidth]{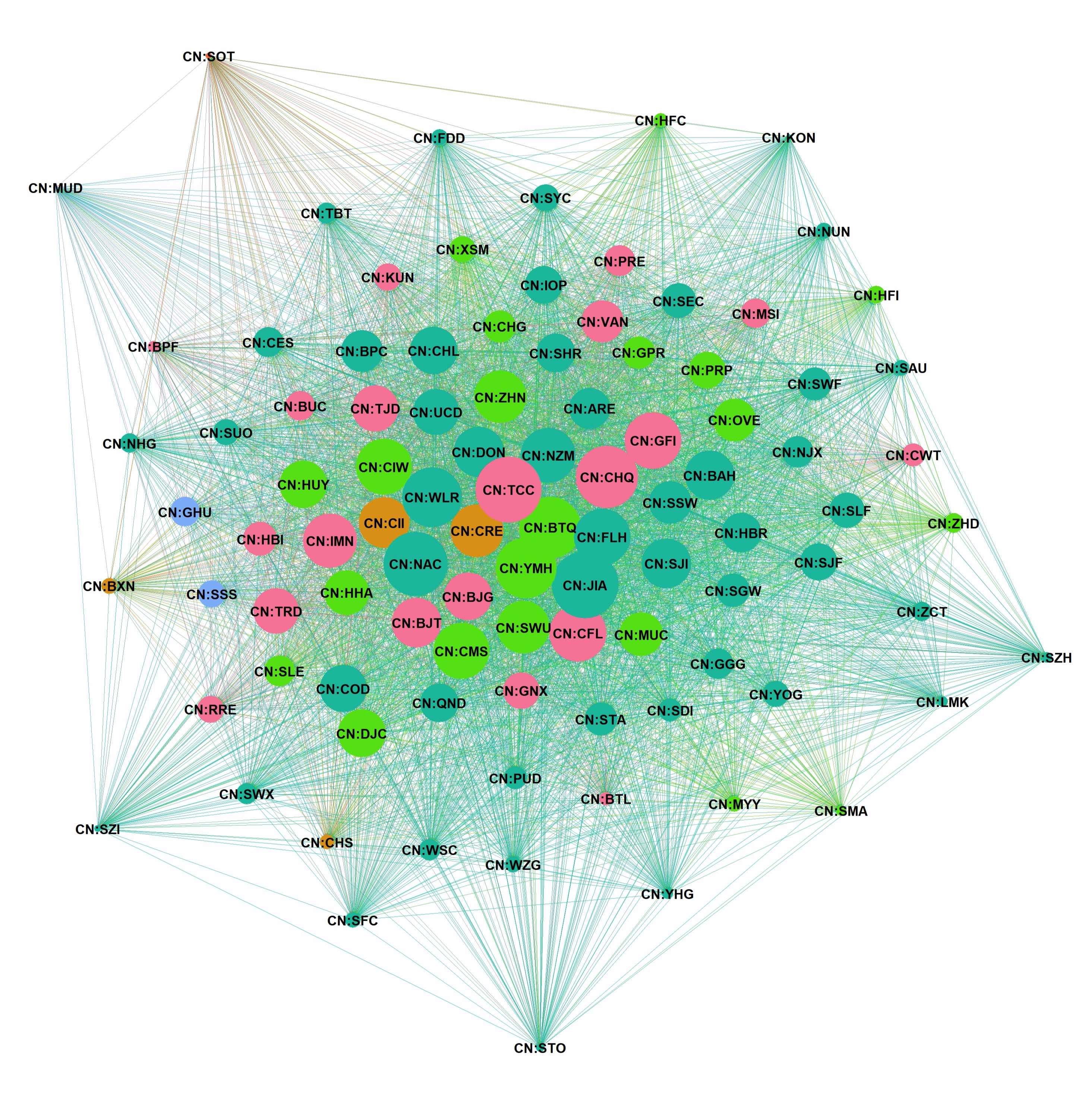}
    \subcaption{After news circulated (8 August 2023)}
    \label{COGA after missed payment}
  \end{subfigure}
  \label{Appendix fig COGA missed payment}
    \caption*{\footnotesize Note: The left graph shows the business day before news of Country Garden's missed debt payment (7 August 2023). The right graph shows the network change on the day that the news of the missed payment circulated (8 August 2023). The colors are determined by the region where the developer is primarily focused: pink is the north, light green is the south, teal is the east, bronze is the southwest, light blue is the northwest, and red-orange (of which there is only one node, CN:SOT) is the northeast. Node size is reflective of ``to" connectedness.}
\end{figure}

\subsubsection{Network before and after news of Evergrande's liquidation} 

As highlighted in Section \ref{COGA body}, there is a much less serious change in the network between the two days than one would expect, given that Evergrande was the first major Chinese developer to be liquidated (and is seen as a bellwether of the Chinese real estate industry). Certainly, many nodes in the core experience increases in to connectedness, but the layout of the network is largely the same: there is no major contraction or expansion. The periphery is also quite stable. This behavior again suggests that the liquidation of Evergrande was largely expected by the market and was already ``priced in," resulting in relatively stable node behavior (``Reactions: Hong Kong court orders liquidation of China Evergrande" 2024). \nocite{noauthor_reactions_2024}

\begin{figure}[H]
  \caption{Network before and after news of Evergrande's liquidation: Color by region}
  \centering
  \begin{subfigure}{0.5\textwidth}
    \centering
    \includegraphics[width=\linewidth]{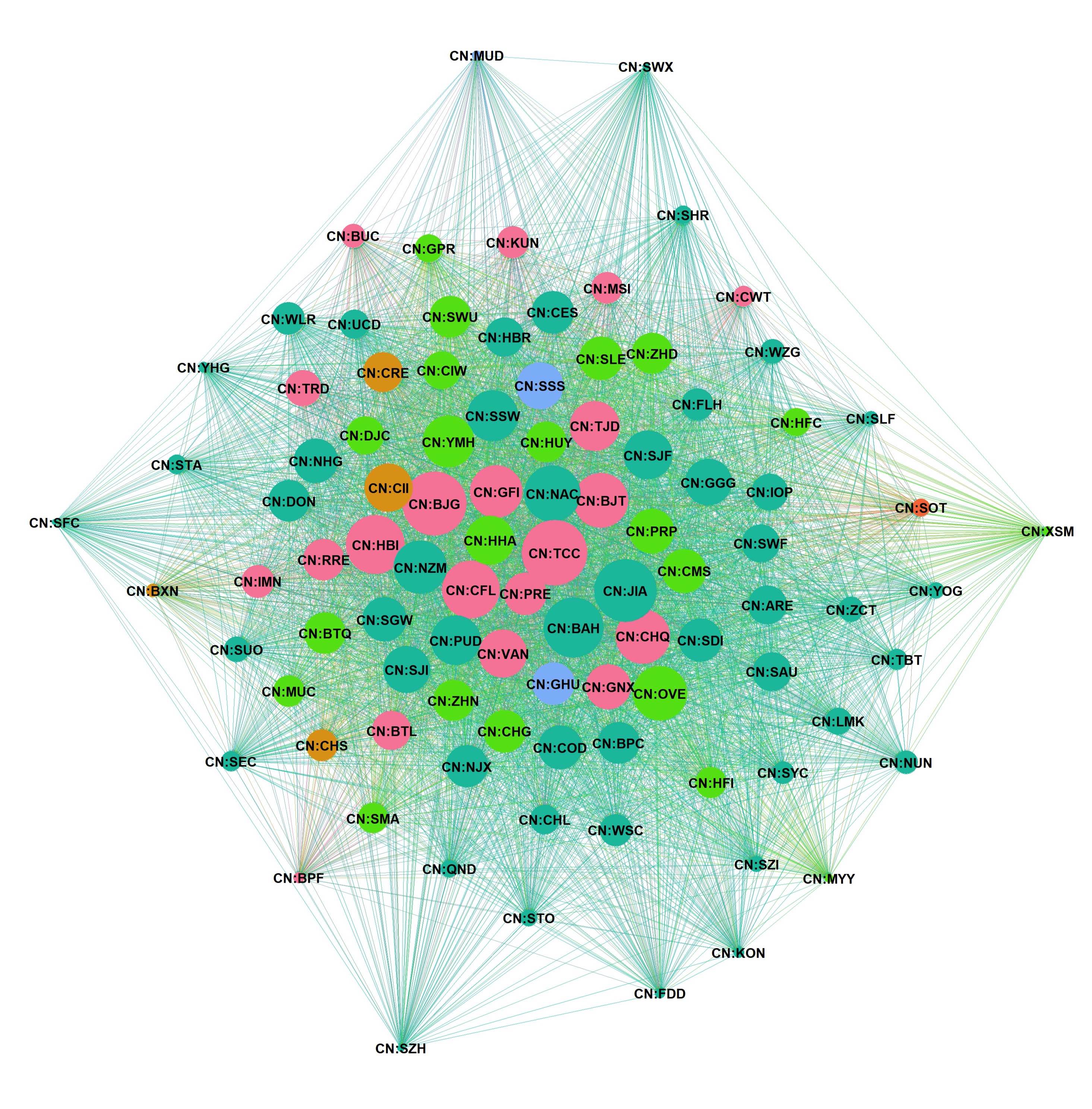}
    \subcaption{Before news circulated (26 January 2024)}
    \label{Evergrande before liquidation}
  \end{subfigure}%
  \begin{subfigure}{0.5\textwidth}
    \centering
    \includegraphics[width=\linewidth]{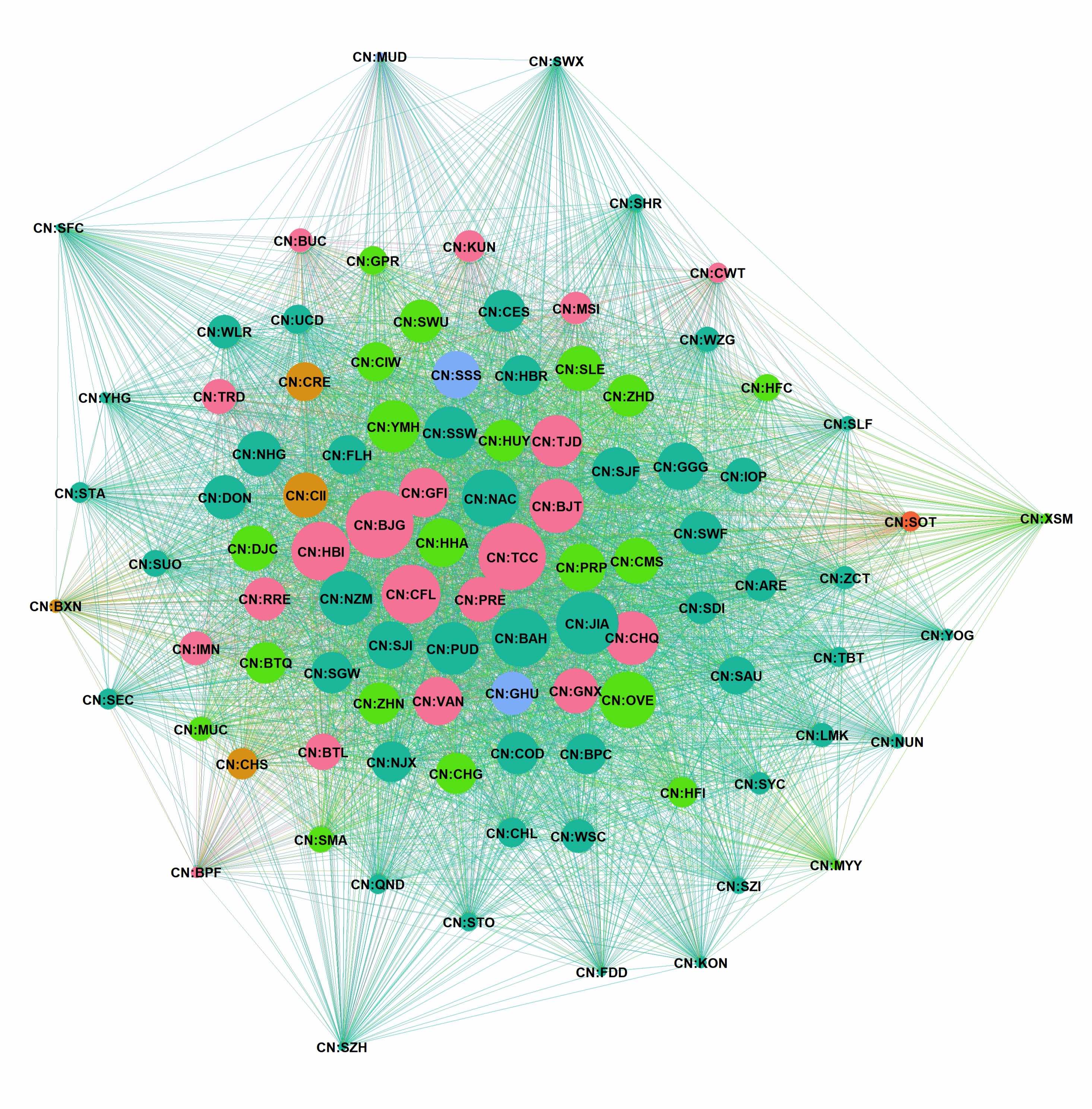}
    \subcaption{After news circulated (29 January 2024)}
    \label{Evergrande after liquidation}
  \end{subfigure}
  \label{Appendix fig Evergrande liquidation}
    \caption*{\footnotesize Note: The left graph shows the business day before news of Evergrande's liquidation (26 January 2024). The right graph shows the network change on the day that the liquidation was announced (29 January 2024). The colors are determined by the region where the developer is primarily focused: pink is the north, light green is the south, teal is the east, bronze is the southwest, light blue is the northwest, and red-orange (of which there is only one node, CN:SOT) is the northeast. Node size is reflective of ``to" connectedness.}
\end{figure}

\subsubsection{Network before and after news of Country Garden's suspension} 

As highlighted in Section \ref{COGA body}, there is a much less serious change in the network between the two days than one would expect, given that Country Garden is one of the largest developers in China. It is also important to note, though, that 29 March and 1 April 2024 were holidays for the Hong Kong exchange, so in accordance with the data cleaning mechanisms described in Section \ref{Appendix: Data Cleaning Procedures}, these days are dropped. Thus, the two windows shown are from 28 March 2024, the business day before the announcement, and 2 April, the first trading day after the announcement included in the sample. This also means that trading still occurred domestically in China on 29 March and 1 April (since the holiday was not celebrated in mainland China), so the comparison is not perfect. 

However, it is even more marked that two trading days after the announcement, the network has changed so little. Certainly, some nodes have experienced increases in to connectedness, but the layout of the network is largely the same: no major contraction or expansion occurs, and the periphery is also quite stable. This behavior again suggests that the suspension of Country Garden was largely expected by the market and was already ``priced in," resulting in relatively stable node behavior, as in the Evergrande liquidation case. 

\begin{figure}[H]
  \caption{Network before and after news of Country Garden's suspension: Color by region}
  \centering
  \begin{subfigure}{0.5\textwidth}
    \centering
    \includegraphics[width=\linewidth]{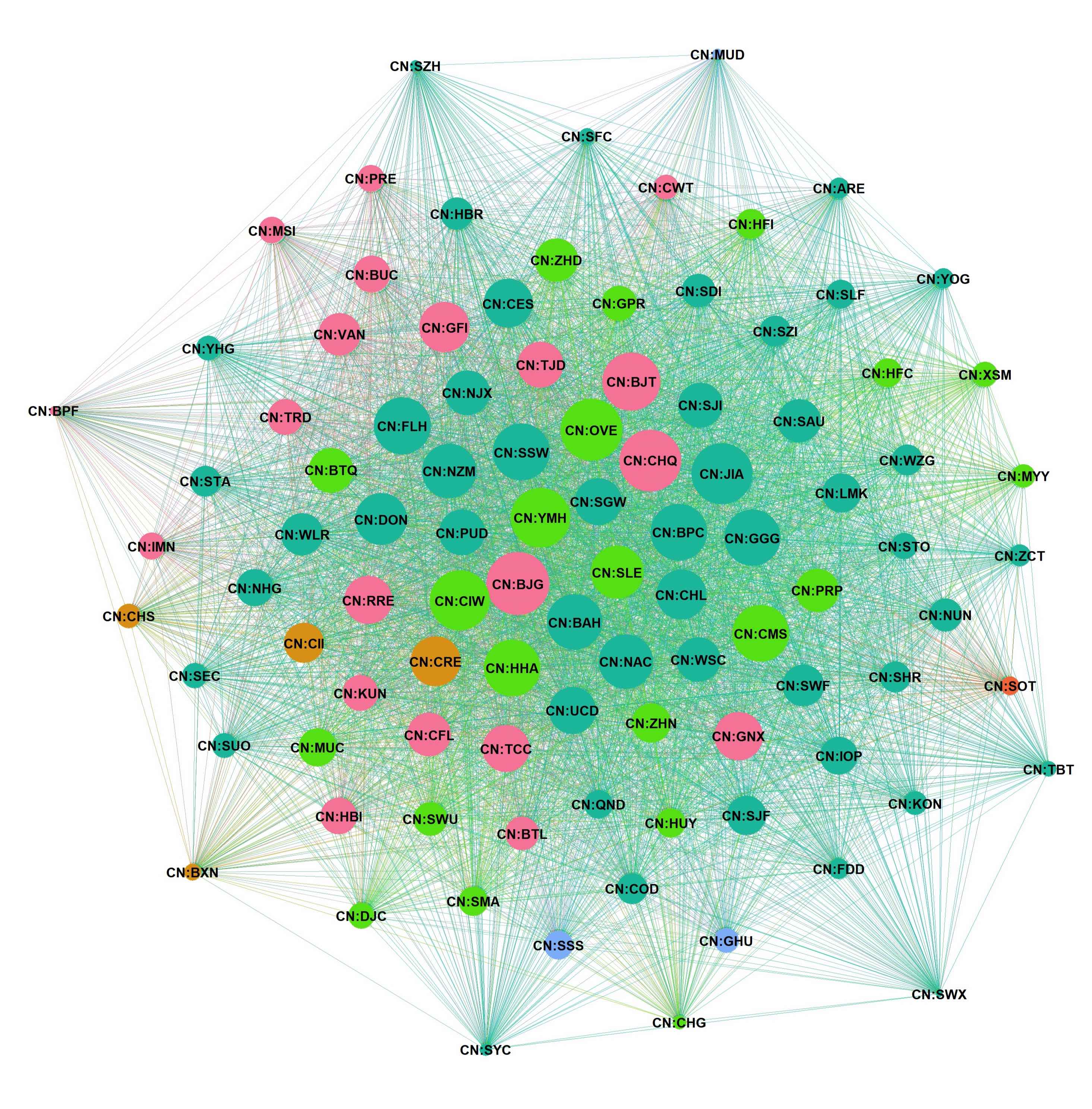}
    \subcaption{Before news circulated (28 March 2024)}
    \label{COGA before suspension}
  \end{subfigure}%
  \begin{subfigure}{0.5\textwidth}
    \centering
    \includegraphics[width=\linewidth]{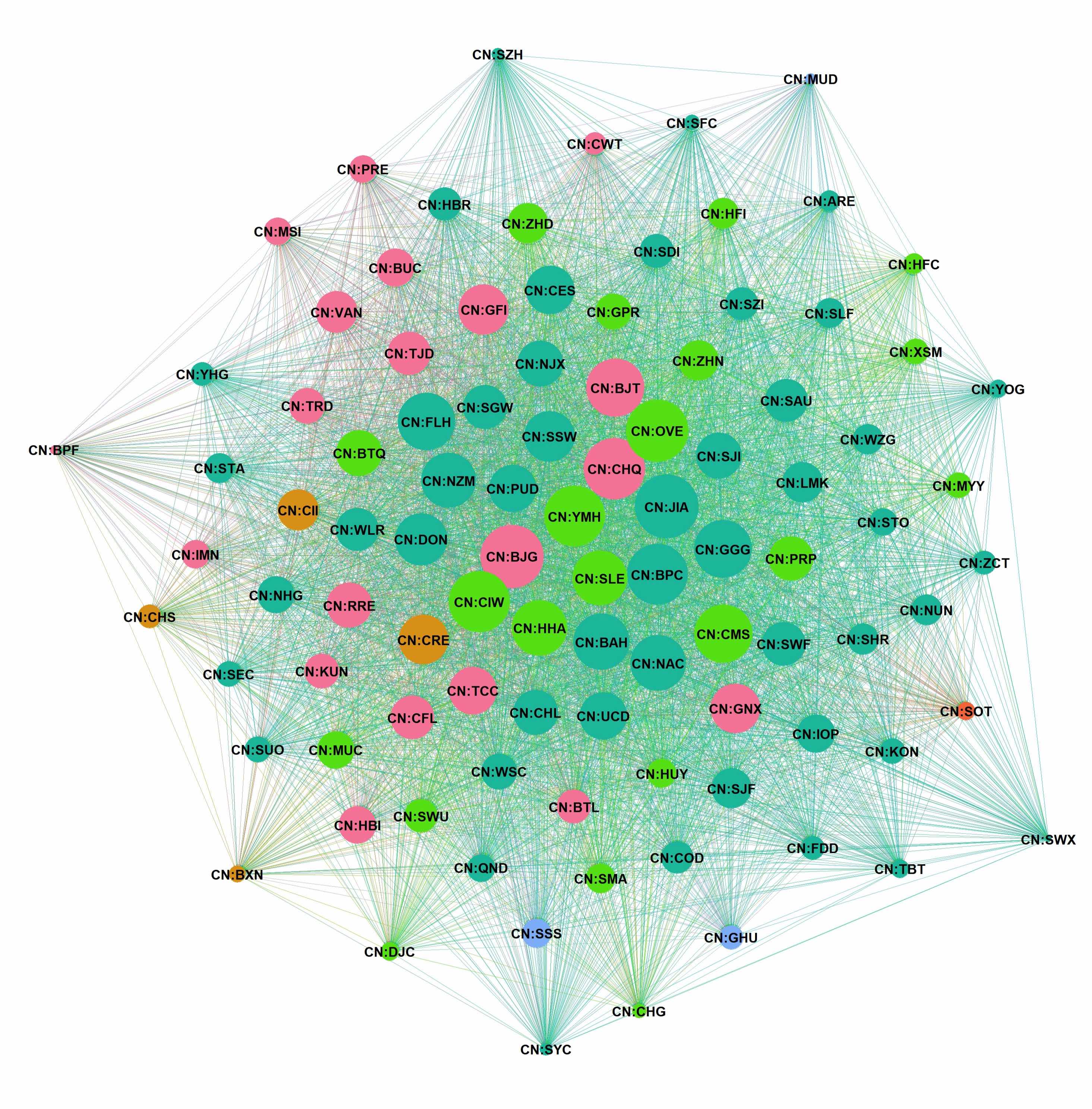}
    \subcaption{After news circulated (2 April 2024)}
    \label{COGA after suspension}
  \end{subfigure}
  \label{Appendix fig COGA suspension}
    \caption*{\footnotesize Note: The left graph shows the business day before Country Garden announced it would not have its financials completed in time for the Stock Exchange of Hong Kong's 31 March deadline and would likely be suspended (28 March 2024). The right graph shows the network change on the next trading day in the sample after the suspension was announced (2 April 2024). The colors are determined by the region where the developer is primarily focused: pink is the north, light green is the south, teal is the east, bronze is the southwest, light blue is the northwest, and red-orange (of which there is only one node, CN:SOT) is the northeast. Node size is reflective of ``to" connectedness.}
\end{figure}

\pagebreak

\begin{CJK*}{UTF8}{gbsn} 
\printbibliography
\end{CJK*} 

\end{document}